\documentclass{aa}

\usepackage{graphics}
\usepackage{graphicx}
\usepackage{multicol}
\usepackage{txfonts}
\usepackage{amsmath}

\usepackage{natbib,twoopt}
\usepackage{amsmath} 
\usepackage{hyperref} 
\makeatletter
\newcommandtwoopt{\citeads}[3][][]{\href{http://adsabs.harvard.edu/abs/#3}%
{\def\hyper@linkstart##1##2{}%
\let\hyper@linkend\@empty\citealp[#1][#2]{#3}}}
\newcommandtwoopt{\citepads}[3][][]{\href{http://adsabs.harvard.edu/abs/#3}%
{\def\hyper@linkstart##1##2{}%
\let\hyper@linkend\@empty\citep[#1][#2]{#3}}}
\newcommandtwoopt{\citetads}[3][][]{\href{http://adsabs.harvard.edu/abs/#3}%
{\def\hyper@linkstart##1##2{}%
\let\hyper@linkend\@empty\citet[#1][#2]{#3}}}
\newcommandtwoopt{\citeyearads}[3][][]%
{\href{http://adsabs.harvard.edu/abs/#3}
{\def\hyper@linkstart##1##2{}%
\let\hyper@linkend\@empty\citeyear[#1][#2]{#3}}}
\makeatother

\usepackage{color}
\definecolor{mygreen}{RGB}{0,128,0}

\hypersetup{colorlinks=true,linkcolor=blue,citecolor=blue,urlcolor=blue}

\def\vsiniSysterr{0.20} 
\def\PASysterr{3.3} 
\def\vsiniSoliderrSystB{0.25}
\def\PASoliderrSystB{3.5}
\def\omegasini{5.6} 
\def\omegasinierr{1.3} 

\def\vsiniSolidA{4.82}

\def\vsysSolidB{+4.52}

\def\vsiniSolidB{5.47}

\def\PASolidB{48.0}

\def\vExpSolidExpB{+0.69}
\def\vExpSolidExperrB{0.08}
\def\rstarB{29.50}
\def\rstarerrB{0.14}
\def\ProtB{36}
\def\ProterrB{8}

\def\vsiniSolidC{5.81}

\def\PASolidC{51.9}

\def\vExpSolidExpC{+1.75}
\def\vExpSolidExperrC{0.08}

\def\PASolidD{45.5}

\begin{document}

\title{The close circumstellar environment of Betelgeuse}
\subtitle{V. Rotation velocity and molecular envelope properties from ALMA}
\titlerunning{Betelgeuse's rotation velocity and molecular envelope properties from ALMA}
\authorrunning{P. Kervella et al.}
\author{
Pierre~Kervella\inst{1,2}
\and
Leen~Decin\inst{3}
\and
Anita~M.~S.~Richards\inst{4}
\and
Graham~M.~Harper\inst{5}
\and
Iain~McDonald\inst{4}
\and
Eamon~O'Gorman\inst{6}
\and
Miguel~Montarg\`es\inst{3}
\and
Ward~Homan\inst{3}
\and
Keiichi~Ohnaka\inst{7}
}
\institute{
Unidad Mixta Internacional Franco-Chilena de Astronom\'{i}a, CNRS/INSU UMI 3386 and Departamento de Astronom\'{i}a, Universidad de Chile, Casilla 36-D, Santiago, Chile, \email{pkervell@das.uchile.cl}.
\and
LESIA, Observatoire de Paris, PSL Research University, CNRS, Sorbonne Universit\'es, UPMC Univ. Paris 06, Univ. Paris Diderot, Sorbonne Paris Cit\'e, 5 Place Jules Janssen, 92195 Meudon, France, \email{pierre.kervella@obspm.fr}.
\and
Institute of Astronomy, KU Leuven, Celestijnenlaan 200D B2401, 3001 Leuven, Belgium
\and
Jodrell Bank Centre for Astrophysics, School of Physics and Astronomy, University of Manchester, Manchester M13 9PL, UK
\and
Center for Astrophysics and Space Astronomy, University of Colorado, Boulder, CO 80309, USA
\and
Dublin Institute for Advanced Studies, Dublin 2, Ireland
\and
Universidad Cat\'olica del Norte, Instituto de Astronom\'ia, Avenida Angamos 0610, Antofagasta, Chile
}
\date{Received ; Accepted}
\abstract
   {
We observed Betelgeuse using ALMA's extended configuration in band 7 ($f \approx 340$\,GHz, $\lambda \approx 0.88$\,mm), resulting in a very high angular resolution of 18\,mas.
Using a solid body rotation model of the $^{28}$SiO($\varv$=2,\,$J$=8-7) line emission, we show that the supergiant is rotating with a projected equatorial velocity of $\varv_\mathrm{eq}\sin i = \vsiniSolidB \pm \vsiniSoliderrSystB$\ km\,s$^{-1}$ at the equivalent continuum angular radius $R_\mathrm{star} = \rstarB \pm \rstarerrB$\ mas.
This corresponds to an angular rotation velocity of $\omega\,\sin i = (\omegasini \pm \omegasinierr) \times 10^{-9}$\ rad\,s$^{-1}$.
The position angle of its north pole is $PA = \PASolidB \pm \PASoliderrSystB^\circ$.
The rotation period of Betelgeuse is estimated to $P / \sin i = \ProtB \pm \ProterrB$\,years.
The combination of our velocity measurement with previous observations in the ultraviolet shows that the chromosphere is co-rotating with the star up to a radius of $\approx 10$\,au (45\,mas or $1.5\times$ the ALMA continuum radius).
The coincidence of the position angle of the polar axis of Betelgeuse with that of the major ALMA continuum hot spot, a molecular plume, and a partial dust shell (from previous observations) suggests that focused mass loss is currently taking place in the polar region of the star.
We propose that this hot spot corresponds to the location of a particularly strong ``rogue'' convection cell, which emits a focused molecular plume that subsequently condenses into dust at a few stellar radii.
Rogue convection cells therefore appear to be an important factor shaping the anisotropic mass loss of red supergiants.
}

\maketitle


\section{Introduction}

The final stages of the evolution of massive stars are complex and still poorly understood, in particular with respect to the physical mechanism of their mass loss.
The modeling efforts reported by \citetads{2013EAS....60...17M}, \citetads{2016ApJ...819....7D}, and \citetads{2017MNRAS.465.2654W} conclude that the nearby red supergiant \object{Betelgeuse} will explode as a Type II-P or II-L supernova, but diverge significantly on its mass and evolutionary state.
A major source of uncertainty of the evolutionary modeling of Betelgeuse is its rotation velocity.
The rotation of red supergiants is expected to be slow, due to the considerable inflation of their radius compared to the main sequence.
However, as shown by \citetads{2012A&A...537A.146E}, this parameter plays a fundamental role in the internal mixing of the stellar material, and therefore in the lifetime of different stages of the star's evolution.
The measurement of the rotation velocity of supergiants is unfortunately difficult, due to their slow rotation compared to the convective broadening of the spectral lines (the macroturbulence is expected at $\approx 10$\,km\,s$^{-1}$ from \citeads{2011A&A...535A..22C}), their pulsation, and the perturbations from the inhomogeneous flux distribution at their surfaces (see, e.g., \citeads{2009A&A...508..923H, 2011A&A...529A.163O} or \citeads{2017Natur.548..310O} for Antares).
We  present here new spatially resolved observations of the close-in molecular envelope of Betelgeuse in the submillimeter domain using the most extended configuration of the ALMA array (Sect.~\ref{observations}), which we interpret using a solid body rotation model (Sect.~\ref{analysis}).
We  discuss in Sect.~\ref{discussion} the rotation velocity profile of Betelgeuse, the geometrical configuration of its rotation axis, and the possible relations between rotation, convection, and mass loss.

\section{Observations \label{observations}}

We observed Betelgeuse with the Atacama Large Millimeter/submillimeter Array (ALMA)
during cycle 3 in band 7 (275--373 GHz) (project code: 2015.1.00206.S, PI: P. Kervella).
Observations were made in the extended configuration (hereafter TE, baseline lengths of 0.056 to 16\,km)
on 6, 7, and 9 Nov 2015, using 47 good antennas, for a total of 112 minutes on target.
We also collected observations on a more compact configuration (hereafter TC, baseline lengths of 0.013
to 1.2\,km) on 16 Aug 2016, using 42 good antennas, for 38 min on target.
The same spectral configuration was used for the TE and TC configurations (Table~\ref{alma-config}).

\begin{figure*}[]
        \centering
        \includegraphics[width=\hsize]{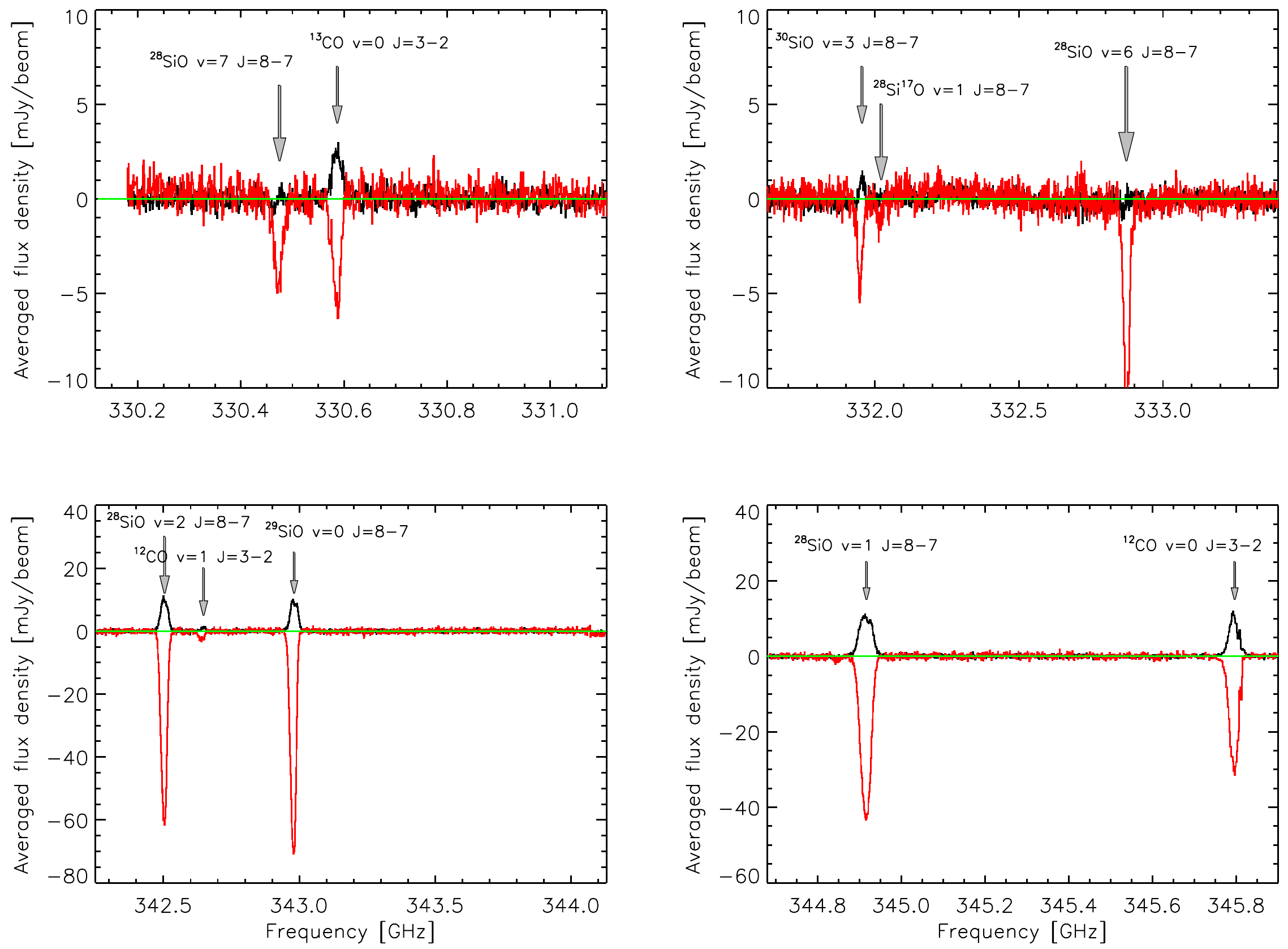}
        \caption{Spectra of Betelgeuse for the spectral windows \emph{spw0} to \emph{spw3} (see Table~\ref{alma-config}) with the identified spectral lines.
        The absorption on the stellar disk (red curve, averaged up to 20\,mas from the star center) and the circumstellar emission (black curve, averaged between 30 and 40\,mas in radius) are represented separately.
        \label{plot_spec}}
\end{figure*}

Standard ALMA procedures were used for instrumentally  derived
calibration (system temperature measurements, water vapor radiometry,
etc.), bandpass, flux scale, and phase reference calibration. The phase
reference source used for the first two TE epochs, \object{J0605+0939}, was
found to be rather faint and solutions from the broader spectral window (hereafter \emph{spw}) were
applied to the narrower.  The brighter \object{J0552+0313} was used for the
remainder of observations (9 Nov 2015, 16 Aug 2016). After applying
these corrections Betelgeuse was split out and all data were adjusted
to constant velocity with respect to the local standard of rest
($\varv_{\mathrm{LSR}}$).
The ALMA frequency scale and velocities are expressed in the local standard of rest kinematic (LSRK), which is the conventional local standard of rest based on the average velocity of stars of the solar neighborhood.
In this frame, the velocity of the Sun is 20.0\,km\,s$^{-1}$ towards $\alpha = 18$h and $\delta = +30^\circ$ at epoch 1900.0.
We made a copy of the data with channels
averaged to 15.625 GHz for speed of continuum imaging. We identified
the line-free channels and used these to make continuum images for
self-calibration, applying the solutions to all data in the same
configuration. For the TE data, the best astrometry was obtained for 9
Nov 2015 and this was used as a starting model for the other two TE
epochs. The 9 Nov data only were the first to be released and were
used for detailed continuum analysis (\citeads{2017A&A...602L..10O};
also see that paper for more observational details) and
astrometric analysis \citepads{2017AJ....154...11H}.

We subtracted the line-free continuum from the
full-spectral-resolution, fully calibrated data and made image cubes of
all spectral windows for the TE, TC, and combined data, both before and after
continuum subtraction. The precise image properties (synthesized
beam, sensitivity, etc.) are a function of frequency and atmospheric
transmission; to within a few percent the parameters are as
follows. The TE-only cubes are sensitive to emission on angular scales
up to 2\,arcseconds; the TC and combined cubes up to 8\,arcseconds.  The typical
synthesized beam sizes and standard deviation of the noise off-source were 18\,mas and 1.3\,mJy,
190\,mas and 1.7\,mJy, 22\,mas and 1\,mJy, for the TE, TC, and TE+TC combined
data, respectively.  The primary beam FWHM is approximately 16\,arcseconds at these
frequencies and we made TC images of comparable size, applying the
primary beam correction. 
However, the results discussed in the present paper use the highest resolution TE
cubes and concentrate on the inner hundreds of milliarcseconds where this is not relevant.

\begin{table}
        \caption{ALMA spectral window (\emph{spw}) configuration. The listed spw width is the extension of the frequency window.
        The channel width is the step in frequency of one spectral element of the ALMA cube, translated into radial velocity.}
        \centering          
        \label{alma-config}
        \begin{tabular}{clcc}
        \hline\hline
        \noalign{\smallskip}
\# & Center & \emph{spw} width& Channel width  \\     
 & (GHz)  &(GHz)    &(km s$^{-1}$) \\
        \noalign{\smallskip}
        \hline    
        \noalign{\smallskip}
0 & 330.65  & 0.94  &   0.89          \\
1 & 332.55  & 1.88  &   0.88          \\
2 & 343.19  & 1.88  &   0.85          \\
3 & 345.15  & 0.94  &   0.85            \\
4 & 345.80  & 0.47   &  1.69            \\
        \hline                      
        \end{tabular}
\end{table}


\section{Analysis}\label{analysis}

\subsection{Detected spectral lines}

\begin{table}
        \caption{Molecular emission lines detected in Betelgeuse, sorted by increasing frequency and spectral window.
        The rest frequencies are taken from the CDMS database \citepads{MULLER2005215, ENDRES201695}. }
        \centering          
        \label{betelgeuse-lines}
        \begin{tabular}{lcccc}
        \hline\hline
        \noalign{\smallskip}
         & \multicolumn{2}{c}{Quantum numbers}  & Rest freq. & Lower state \\
         & Vibrat. & Rotat.  & (GHz) & energy (cm$^{-1}$) \\
        \noalign{\smallskip}
        \hline    
        \noalign{\smallskip}
$^{28}$SiO & $\varv=7$ & $8-7$ & $330.47753$ & 8397.36 \\
$^{13}$CO & $\varv=0$ & $3-2$ & $330.58797$ & 11.03 \\ 
        \hline    
        \noalign{\smallskip}
$^{30}$SiO & $\varv=3$ & $8-7$ & $331.95548$ & 3648.27 \\
$^{28}$Si$^{17}$O & $\varv=1$ & $8-7$ & $332.02199$ & $1245.27$  \\
$^{28}$SiO & $\varv=6$ & $8-7$ & $332.87874$ & 7238.86 \\
        \hline    
        \noalign{\smallskip}
$^{28}$SiO & $\varv=2$ & $8-7$ & $342.50438$ & 2487.33 \\
$^{12}$CO & $\varv=1$ & $3-2$ & $342.64764$ & 2154.70 \\
$^{29}$SiO & $\varv=0$ & $8-7$ & $342.98085$ & 40.05 \\ 
        \hline    
        \noalign{\smallskip}
$^{28}$SiO & $\varv=1$ & $8-7$ & $344.91633$ & 1269.89 \\
$^{12}$CO & $\varv=0$ & $3-2$ & $345.79599$ & 11.54 \\ 
        \hline                      
        \end{tabular}
\end{table}

\begin{figure}[]
        \centering
        \includegraphics[width=\hsize]{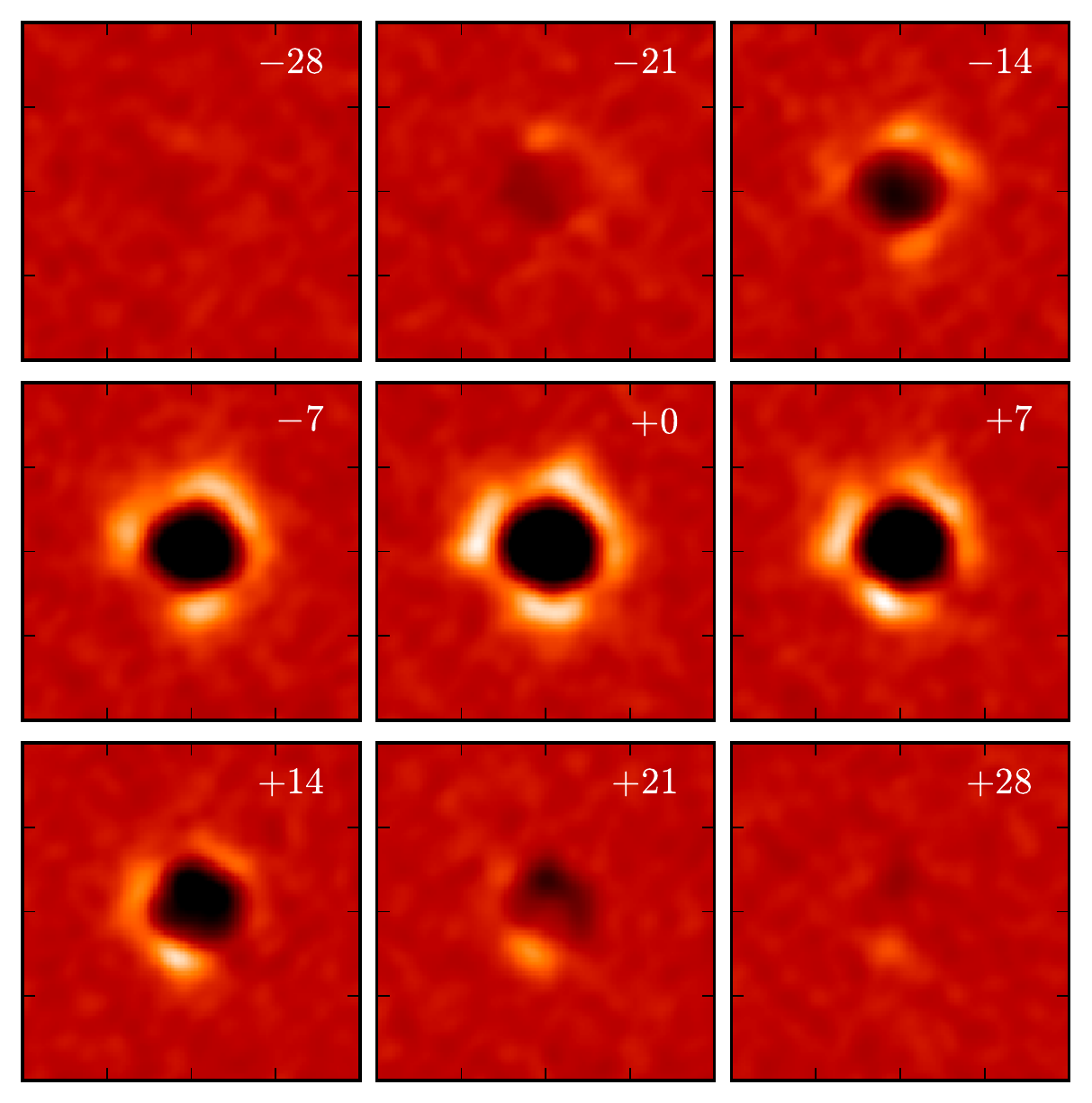}
        \caption{Channel map of the continuum subtracted emission from the $^{28}$SiO($\varv$=2,\,$J$=8-7) line.
        The velocity offset from the LSR systemic velocity of $\varv_\mathrm{sys} = \vsysSolidB$\,km\,s$^{-1}$ is given in the upper right corner of each panel.
        The field of view is $100\,\mathrm{mas} \times 100\,\mathrm{mas}$.
        \label{channel-SiOv2}}
\end{figure}

We detected ten lines of the CO and SiO molecules including several isotopologues, either in absorption only or in  both  emission and  absorption.
The line properties are listed in Table~\ref{betelgeuse-lines}, and the continuum subtracted spectra are shown in Fig.~\ref{plot_spec}.
The velocity structure of the circumstellar emission of Betelgeuse is visible in the $^{28}$SiO($\varv$=2,\,$J$=8-7) line channel map represented in Fig.~\ref{channel-SiOv2}.
It is interesting to note that the circumstellar emission of Betelgeuse exhibits a comparable structure to the ALMA long-baseline observations of the close environment of Mira reported by \citetads{2016A&A...590A.127W}.
Images made at lower resolution and better surface brightness sensitivity from the combined TE and TC data also show extended emission in different molecular emission lines; this will be discussed in a forthcoming paper.

\subsection{Shell model\label{shellmodel}}

\begin{figure}[]
        \centering
        \includegraphics[width=\hsize]{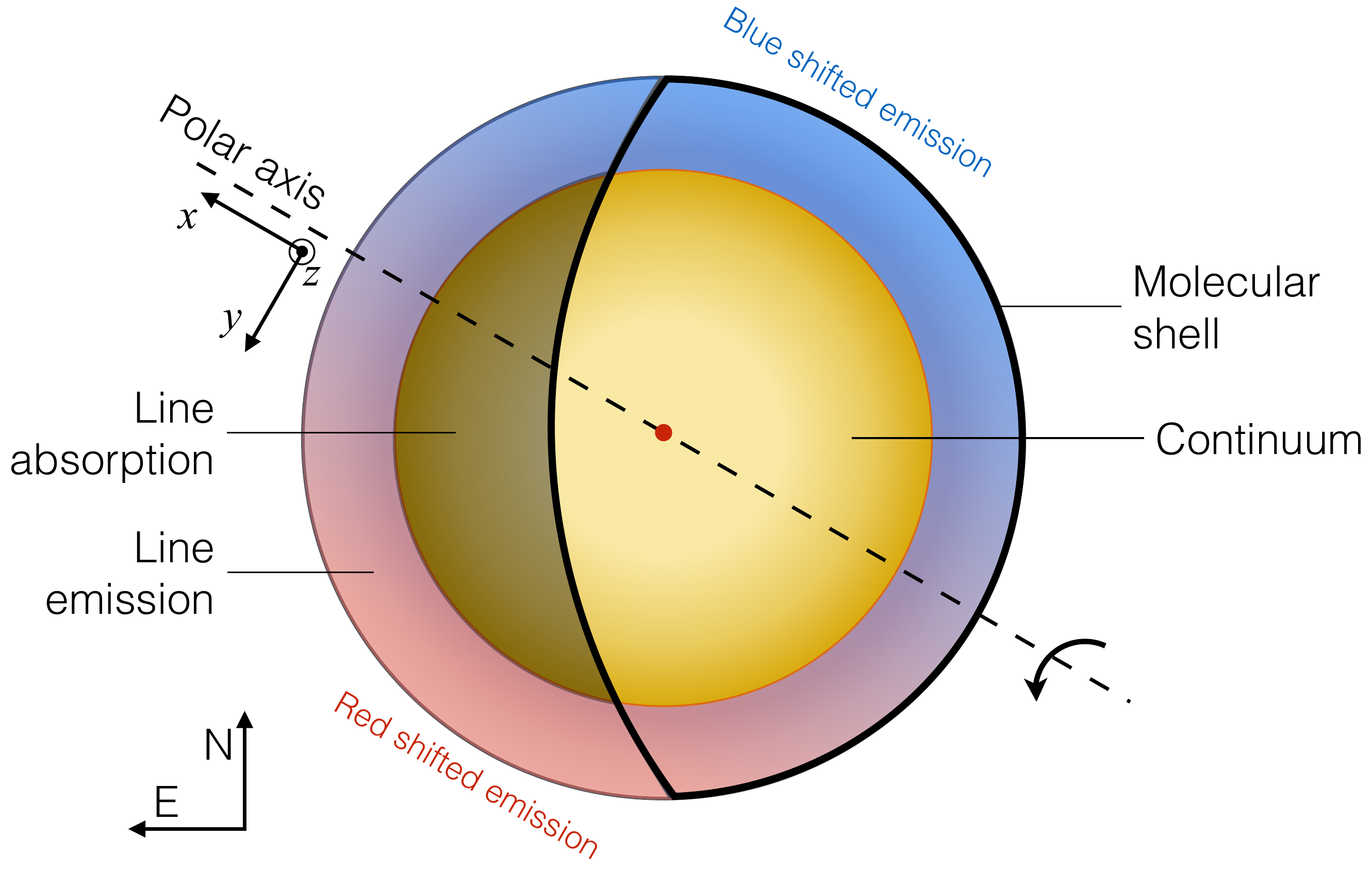}
        \caption{Adopted model of Betelgeuse.
        The molecular shell is shown approximately to scale with respect to the continuum.
        The right part of the shell is omitted to show the continuum without line absorption.
        The cool molecular layer located around the star and shell is not represented.
        The position angle of the polar axis is represented according to the best fit model.
        The $(x,y)$ coordinates are the sky angular coordinates of the reference frame;  the $z$-axis is pointing toward the observer.
        \label{bet-model}}
\end{figure}

The spectral signature of the molecular envelope of Betelgeuse is observed in absorption over the continuum stellar disk, and in emission beyond its limb (see also Sect.~\ref{lineprofile}).
Following \citetads{2004A&A...418..675P}, \citetads{2014A&A...572A..17M}, and  \citetads{2009A&A...503..183O, 2011A&A...529A.163O}
we model our ALMA observations using a spherical thin molecular shell of angular radius $R_\mathrm{shell}$ located above the spherical stellar photosphere of angular radius $R_\mathrm{star}$.
An important difference with the near-infrared domain is that the considered molecular lines are likely to be excited by maser pumping close to the star.
In addition, we expect that a significant absorption from a cool molecular envelope is present in front of both the star and the light-emitting shell.
We therefore dissociate the absorption and emission components in our parametrization of the surface brightness $I(r)$ of the star and its molecular shell observed at an angular radius $r$ from the center of the stellar disk.
This results in the following model:
\begin{itemize}
\item over the stellar continuum disk $(0< r <R_\mathrm{star})$:
\begin{equation}\label{modeloverdisk}
I(r) = \left[I_\mathrm{star}\,e^{-\tau_\mathrm{shell}/G(r)} + I_\mathrm{shell}\,\,\left( 1 - e^{-\tau_\mathrm{shell}/G(r)} \right) \right] \,e^{-\tau_\mathrm{cool}}
;\end{equation}
with $\tau_\mathrm{cool}$ the optical depth of the cooler material located in front of the star and shell, $\tau_\mathrm{shell}$ the optical depth of the light-emitting thin shell, and $G(r)$ the geometrical projection factor:
\begin{equation}
G(r) =  \sqrt{1-r^2/R_\mathrm{shell}^2}
;\end{equation}
\item between the stellar limb and the shell $(R_\mathrm{star}<r<R_\mathrm{shell})$:
\begin{equation}\label{eq2}
I(r) = I_\mathrm{shell}\,\left[ 1 - e^{-2\,\tau_\mathrm{shell}/G(r)} \right] \,e^{-\tau_\mathrm{cool}}
;\end{equation}
\item outside of the shell $(r>R_\mathrm{shell})$: $I(r) = 0$,
\end{itemize}
where $I_\mathrm{star}$ and $I_\mathrm{shell}$ are the star's continuum and molecular shell surface brightnesses, respectively.
An overview of the model geometry is presented in Fig.~\ref{bet-model}.
We make  the assumption here that the light-emitting layer is optically thin ($\tau_\mathrm{shell} \ll 1$), which enables us to simplify Eq.~\ref{modeloverdisk} (valid between $0< r <R_\mathrm{star}$),
\begin{equation}\label{eq4}
I(r) = \left(I_\mathrm{star} + \frac{I_\mathrm{shell}\,\tau_\mathrm{shell}}{G(r)}\right)\,e^{-\tau_\mathrm{cool}}
,\end{equation}
and Eq.~\ref{eq2} $(R_\mathrm{star}<r<R_\mathrm{shell})$ becomes
\begin{equation}
I(r) = \left( \frac{2\,I_\mathrm{shell}\,\tau_\mathrm{shell}}{G(r)} \right) \,e^{-\tau_\mathrm{cool}}
.\end{equation}\label{eq5}
This static shell model is defined by five parameters: $R_\mathrm{star}$, $I_\mathrm{star}$, $R_\mathrm{shell}$, $\left(I_\mathrm{shell}\,\tau_\mathrm{shell}\right)$, and $\tau_\mathrm{cool}$.
We note that we consider the product $I_\mathrm{shell}\,\tau_\mathrm{shell}$ as a single parameter, as the optical depth and surface brightness of the emitting shell are degenerate quantities.

The absorption and emission is modulated by the profile $P(f)$ in frequency of the molecular emission line, assumed to be Gaussian (see Sect.~\ref{modelparameters}) normalized with a unit amplitude.
The solid body rotation of the shell is introduced through a position-dependent Doppler shift $\Delta f(\delta)$ of the molecular line profile
\begin{equation}
\Delta f(\delta) = f \left( \frac{\varv_\mathrm{eq} \sin i}{c} \right) \left(\frac{\delta}{R_\mathrm{shell}} \right),
\end{equation}
where $f$ is the frequency, $\varv_\mathrm{eq}$ the equatorial rotation velocity, $i$ the inclination of the polar axis on the line of sight, and $\delta$ the distance of the considered point to the polar axis.
More complex, non-solid body rotation can be introduced by changing the $\Delta f$ law, for instance using latitude-dependent rotation velocity
\citepads{2004A&A...418..781D}.

\subsection{Static shell model parameters \label{modelparameters}}

In this section, we analyze the $^{28}$SiO($\varv$=2,\ $J$=8-7) emission line.
We chose this line as its intensity profile is the best match to our thin shell model (see Sect.~\ref{shellbrightness}) out of the four intense emission lines visible in Fig.~\ref{plot_spec}.
A possible reason for the behavior is that this line has the highest upper state energy and therefore traces  the regions closest to the star.
The lower excitation lines are excited farther away, and this is the reason why the thin shell model is less applicable.

In addition, the $^{28}$SiO($\varv$=2,\ $J$=8-7) line is also probably excited through a maser process, which facilitates its observation.
The channel width of our data cubes is $\approx 0.85$\,km\,s$^{-1}$ so we cannot look for maser spikes in the spectrum and the total line width is greater than 10\,km\,s$^{-1}$.
The broad line width by itself does not rule out saturated masers, but higher spectral and spatial resolution is needed to identify whether high brightness temperature maser components are present or if the emission is mostly thermal.
However, the $\varv=2$ (respectively $\varv=1$) lines are of similar intensity to the $\varv=0$ line, which suggests that the brightness of the excited states is enhanced by the maser process;
otherwise, they would be only about 1/6 (respectively 1/2) as bright, although the spectral resolution is not fine enough to allow us to confirm this.
This supports the conclusion that masing is present, but it is not definitive.
The analyses of the other detected emission lines are presented in Appendices~\ref{12COv0line} to \ref{29SiOv0line}.

\subsubsection{Stellar radius $R_\mathrm{star}$ and surface brightness $I_\mathrm{star}$}

\begin{figure}[]
        \centering
        \includegraphics[height=7.5cm]{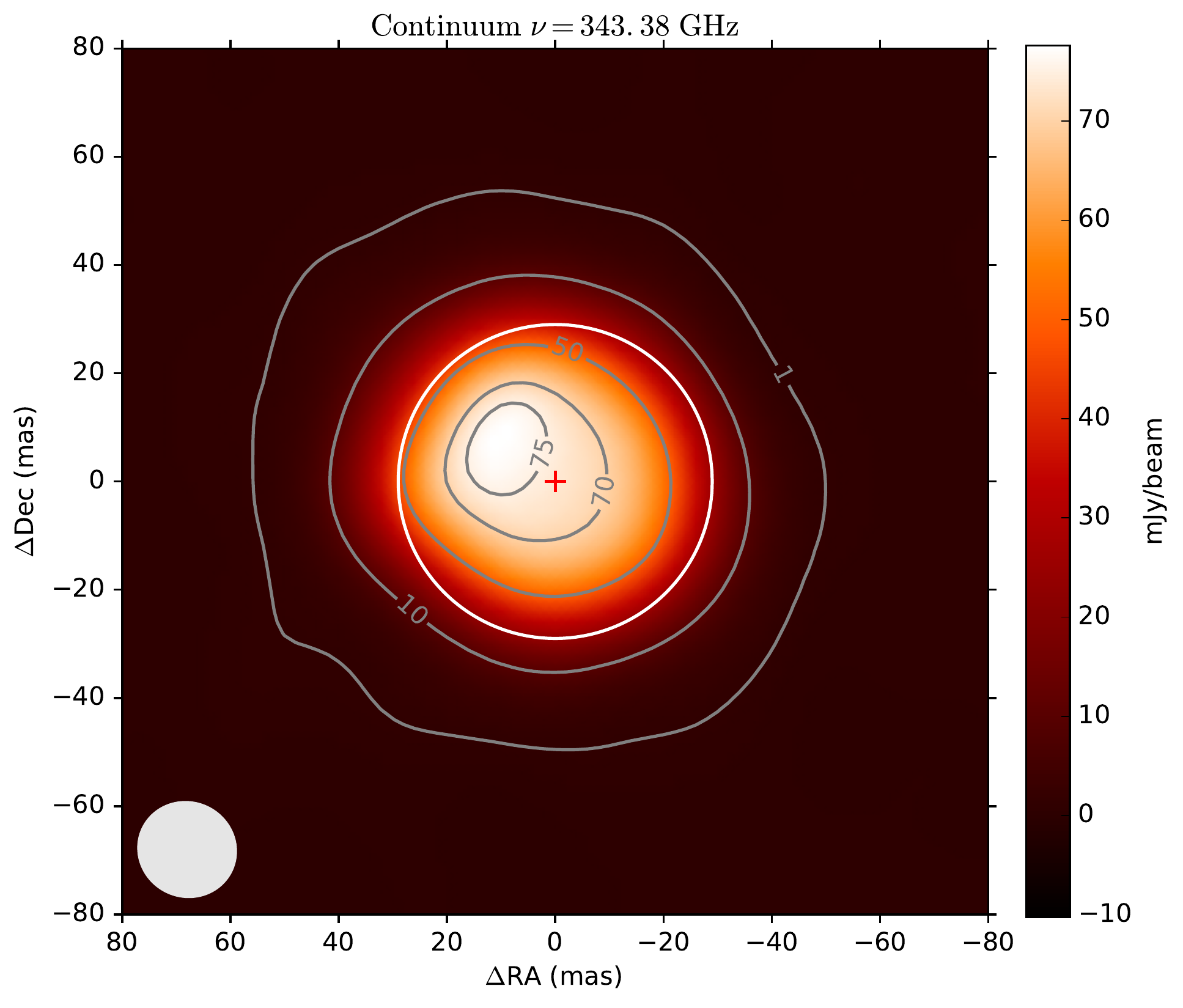}
        \caption{Image of Betelgeuse in the continuum for the spectral window \emph{spw2} associated with the $^{28}$SiO($\varv$=2,\,$J$=8-7) and $^{29}$SiO($\varv$=0,\ $J$=8-7) emission lines.
        The white circle represents the equivalent uniform disk size of the star ($R_\mathrm{star}=29.5$\,mas), and the red cross marks the centroid of the intensity distribution taken as the center of the stellar disk.
        The gray disk at the bottom left of the plot is the ALMA beam size, and the contour levels are labeled in mJy\,beam$^{-1}$.
        \label{bet-spw2-cont}}
\end{figure}

We  determine the $R_\mathrm{star}$ and $I_\mathrm{star}$ parameters of the model from the image of Betelgeuse in the continuum computed around the considered molecular line.
The continuum image for \emph{spw2} is presented in Fig.~\ref{bet-spw2-cont}.
The surface brightness $I_\mathrm{star}$ is measured at the center of the stellar disk.
We define $R_\mathrm{star}$ as the equivalent uniform disk angular radius of the star in the ALMA continuum, that is, at a wavelength of $\lambda \approx 0.9$\,mm.
As the stellar disk is resolved by ALMA, we estimate $R_\mathrm{star}$ from the integrated continuum flux $I(r,\theta)$ and the surface brightness measured at the center of the disk $I_\mathrm{star}$ through
\begin{equation}
R_\mathrm{star} = \sqrt{\frac{1}{\pi} \frac{\int_\mathrm{star}{I(r,\theta)\,r\,dr\,d\theta}}{I_\mathrm{star}}}
\end{equation}
For \emph{spw2}, we measure $I_\mathrm{star} = 74.81 \pm 0.14$\,mJy\,beam$^{-1}$ and we obtain $R_\mathrm{star} = \rstarB \pm \rstarerrB$\,mas.
The model parameters for the four studied emission lines are summarized in Table~\ref{betelgeuse-line-params}.

\begin{table*}
        \caption{Continuum and line emission parameters of the thin shell models for the four studied lines. The intensities $I_\mathrm{star}$ and $I_\mathrm{shell}$ are expressed in mJy\,beam$^{-1}$, the angular radii $R_\mathrm{star}$ and $R_\mathrm{shell}$ in milliarcseconds, and the velocities (expressed in the LSR) and line widths in km\,s$^{-1}$. The uncertainties are listed in the subscript of each value.}
        \centering          
        \label{betelgeuse-line-params}
        \begin{tabular}{lcccccccccc}
        \hline\hline
        \noalign{\smallskip}
         Line & $I_\mathrm{star}$ & $R_\mathrm{star}$ & $I_\mathrm{shell}\tau_\mathrm{shell}$ & $R_\mathrm{shell}$ & $\tau_\mathrm{cool}$ & $\varv_\mathrm{emi}$ & $\sigma_\mathrm{emi}$ & $\varv_\mathrm{abs}$ & $\sigma_\mathrm{abs}$ & $\chi^2_\mathrm{red}$ \\
        \noalign{\smallskip}
        \hline    
        \noalign{\smallskip}
 $^{28}$SiO($\varv$=2,$J$=8-7) & $74.8_{0.4}$ & $29.50_{0.14}$ & $27_{3}$ & $37.8_{2.5}$ & $1.08_{0.03}$ & $4.52_{0.09}$ & $10.20_{0.08}$ & $4.80_{0.14}$ & $10.26_{0.13}$ & $0.16$ \\ 
$^{12}$CO($\varv$=0,$J$=3-2) & $76.6_{0.4}$ & $29.97_{0.12}$ & $34_{3}$ & $39.5_{2.9}$ & $1.12_{0.03}$ & $5.56_{0.15}$ & $8.80_{0.13}$ & $4.93_{0.17}$ & $8.69_{0.15}$ & $1.46$ \\ 
$^{28}$SiO($\varv$=1,$J$=8-7) & $75.0_{0.4}$ & $29.60_{0.13}$ & $71_{11}$ & $39.6_{2.9}$ & $1.83_{0.09}$ & $4.74_{0.10}$ & $11.37_{0.09}$ & $4.34_{0.11}$ & $10.45_{0.10}$ & $0.36$ \\
$^{29}$SiO($\varv$=0,$J$=8-7) & $74.8_{0.4}$ & $29.50_{0.14}$ & $31_{3}$ & $39.5_{2.9}$ & $1.28_{0.03}$ & $4.43_{0.09}$ & $10.29_{0.08}$ & $5.78_{0.11}$ & $9.62_{0.10}$ & $0.32$ \\  
        \hline                      
        \end{tabular}
\end{table*}

\subsubsection{Line profile $P(f)$\label{lineprofile}}

\begin{figure}[]
        \centering
        \includegraphics[height=7.5cm]{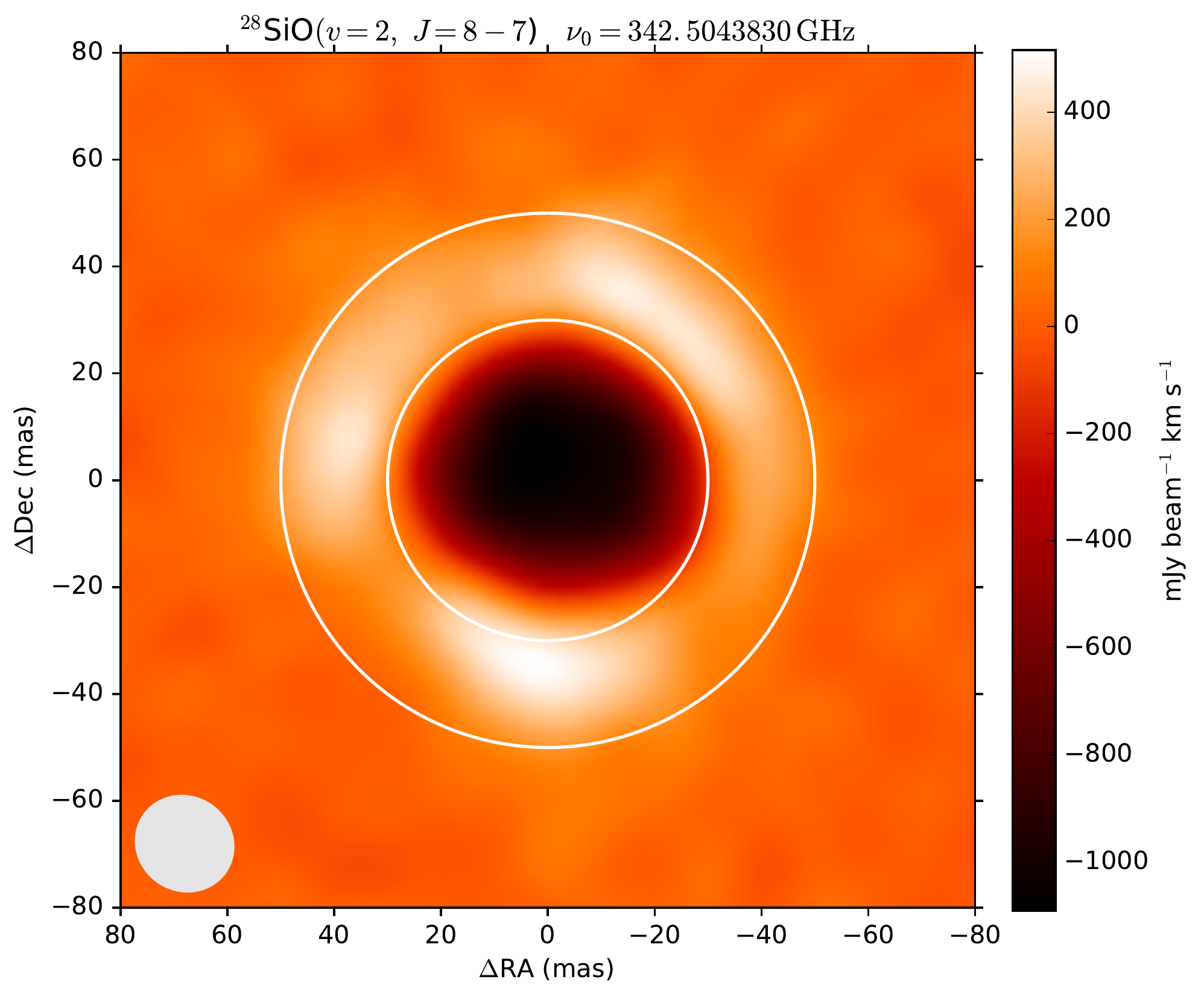}
        \caption{Continuum subtracted emission intensity in the $^{28}$SiO$(v=2,\ J=8-7)$ line, integrated along the frequency axis.
        The ring used to estimate the average emission line profile presented in Fig.~\ref{profiles} is located between the two white circles of angular radii 30 and 50\,mas.
        The absorption profile is measured within the 30\,mas inner circle.
        \label{totalemission}}
\end{figure}

\begin{figure}[]
        \centering
        \includegraphics[width=\hsize]{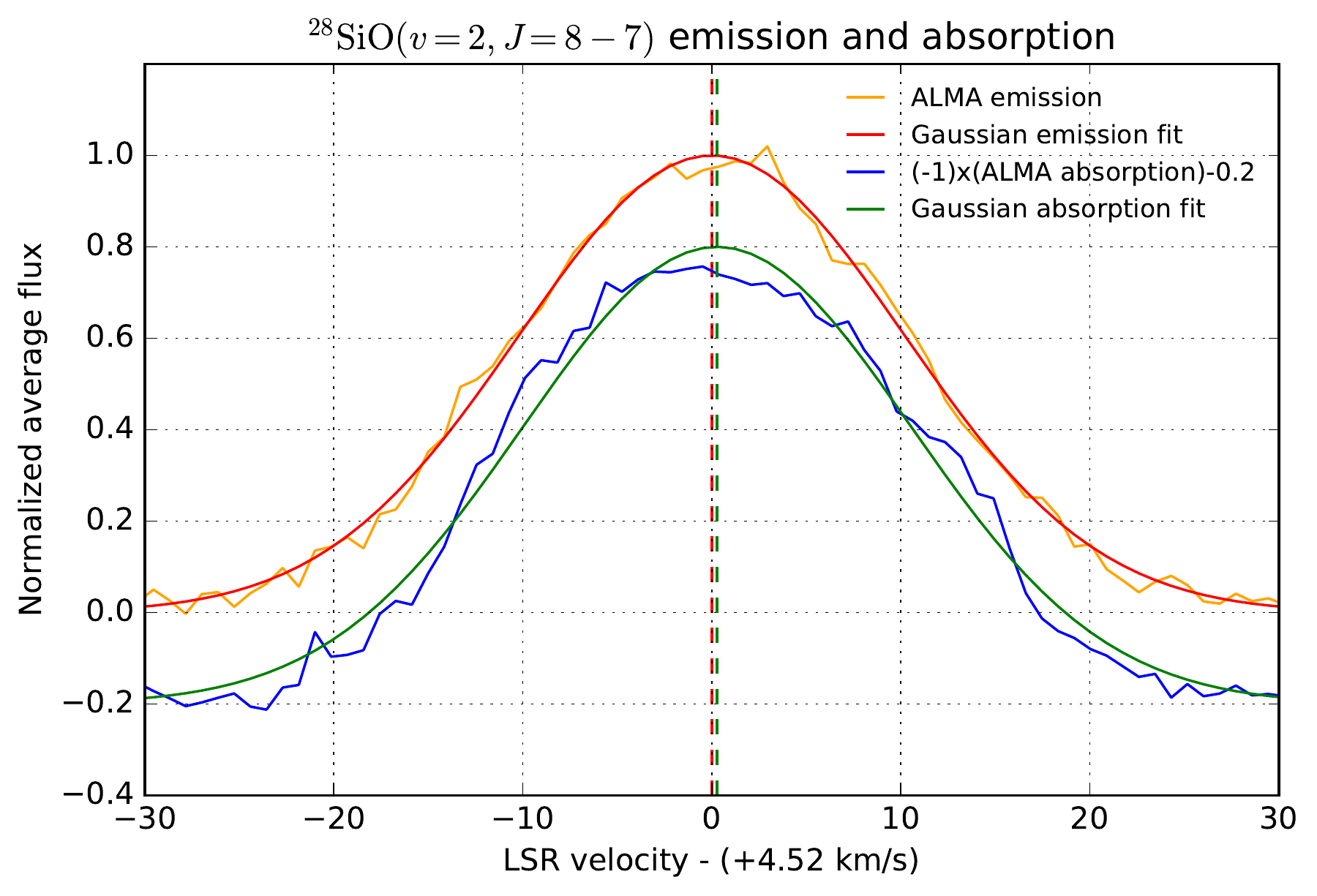}
        \caption{Absorption and emission normalized profiles $P(f)$ of the $^{28}$SiO($\varv$=2,\ $J$=8-7) line.
        The sign of the absorption profile is inverted and shifted by $-0.2$ to facilitate the visual comparison with the emission profile.
        The absorption profile is estimated within a radius of 30\,mas, and the emission profile between 30 and 50\,mas (Fig.~\ref{totalemission}).
        \label{profiles}}
\end{figure}

We determine the line profile $P(f)$ from the continuum subtracted data cubes using two observables: the absorption at the center of the stellar disk and the emission in the circumstellar shell ring (Fig.~\ref{totalemission}).
For a homogeneous non-expanding molecular envelope, the two profiles are identical.
In the case of a global net spherical outflow (or inflow) of molecular material, the absorption profile will be blueshifted (respectively redshifted)  compared to that of the emission ring.
The reason is that the absorption component is essentially on the line of sight between us and the photosphere, hence with a minimal radial velocity projection effect.
On the other hand, the emission ring is located, on average, in the plane of the sky (perpendicular to the line of sight)  for which the projection effect results, for a  homogeneous shell, in a zero radial velocity.
A comparison of the emission and absorption line profiles for the $^{28}$SiO($\varv$=2,\,$J$=8-7) line is presented in Fig.~\ref{profiles}.
Both profiles have very similar dynamical characteristics, with velocities of $\varv_\mathrm{abs} = +4.80 \pm 0.14$\,km\,s$^{-1}$ and $\varv_\mathrm{emi} = +4.52 \pm 0.09$\,km\,s$^{-1}$ and line widths of $\sigma_\mathrm{abs} = 10.20 \pm 0.08$~km\,s$^{-1}$ and $\sigma_\mathrm{emi} = 10.26 \pm 0.12$~km\,s$^{-1}$.
The velocities of the absorption and emission components are close but do not overlap, probably because of the inhomogeneities that are present in the molecular shell and  visible in Fig.~\ref{totalemission} (see also Fig.~\ref{channel-SiOv2}).

We note that we choose to define in this article the line width $\sigma$ as the standard deviation of the adjusted Gaussian line profile,
which is related to the full width at half maximum (FWHM) through $\mathrm{FWHM} \simeq 2.355\, \sigma$.
For the $^{28}$SiO($\varv$=2,\,$J$=8-7) absorption and emission line profiles presented in Fig.~\ref{profiles}, the FWHM is therefore $24.0 \pm 0.2$\,km\,s$^{-1}$, and it is similar for the other ALMA SiO lines.
The FWHM of the ALMA $^{12}$CO emission line), which extends to larger radii than the SiO emission, is $20.7 \pm 0.3$\,km\,s$^{-1}$.
The detected ALMA molecular lines are therefore broader than the FWHM = 18.3\,km\,s$^{-1}$ value determined by \citetads{2000ApJ...532..487G} in the optical domain.
\citetads{2000ApJ...545..454L} measured a macroturbulent and rotational broadening of FWHM = 20\,km\,s$^{-1}$ on unblended metallic lines in the near-infrared.
In the mid-infrared domain, a narrower FWHM of $\approx 14.5$\,km\,s$^{-1}$ was found by \citetads{2006ApJ...637.1040R} for water vapor lines.
The ALMA molecular lines thus appear significantly broader than the visible, near- and mid-infrared lines.
 
We  adopted the emission line velocity as the systemic velocity $\varv_\mathrm{sys}$ as it is in principle less sensitive to the presence of a radially expanding velocity than the photospheric absorption line. The difference between the two velocities $\Delta \varv = \varv_\mathrm{abs} - \varv_\mathrm{emi} = +0.28 \pm 0.17$~km\,s$^{-1}$ is  small, however,  and does not affect the rest of the analysis.
We therefore model the line profile $P(f)$ as a Gaussian of mean LSR velocity $\varv_\mathrm{sys\,LSR} = +4.52 \pm 0.09$~km\,s$^{-1}$ and width $\sigma = 10.20 \pm 0.08$~km\,s$^{-1}$ (red curve in Fig.~\ref{profiles}).
For the emission components of the three studied SiO emission lines, the systemic LSR velocity of Betelgeuse is consistently estimated between +4.43 and +4.74\,km\,s$^{-1}$ (Table~\ref{table-models}).
It is slightly higher for the $^{12}$CO($\varv$=0,\ $J$=3-2) line, at $+5.56$\,km\,s$^{-1}$.
The weighted average of all emission and absorption component velocities listed in Table~\ref{table-models} (including the CO line) is $+4.79$\,km\,s$^{-1}$, close to the adopted $^{28}$SiO($\varv$=2,\,$J$=8-7) emission line value.

The adopted emission velocity in the LSR derived for Betelgeuse corresponds to a heliocentric velocity of $\varv_\mathrm{helio}= 20.4 \pm 0.1$\,km\,s$^{-1}$, which is lower by $1.6\sigma$ than the velocity of $20.9 \pm 0.3$\,km\,s$^{-1}$ found by \citetads{2017ApJ...836...22H} from a weighted mean of four different estimates.
The latter were a 100-year average of photospheric radial velocities, the centroid velocity of $4.6\,\mu$m scattered CO emission, the central velocity of CARMA CO mm-radio emission, and the mean velocity of the co-rotating chromospheric emission as measured by the HST.
Given that each velocity estimate is based on a different diagnostic, each with its own systematic uncertainty, and the differences in the emission velocities for the different ALMA lines, we regard the ALMA value as consistent with \citetads{2017ApJ...836...22H}.

\subsubsection{Shell radius $R_\mathrm{shell}$ and surface brightness $I_\mathrm{shell}\,\tau_\mathrm{shell}$ \label{shellbrightness}}

\begin{figure}[]
        \centering
        \includegraphics[width=\hsize]{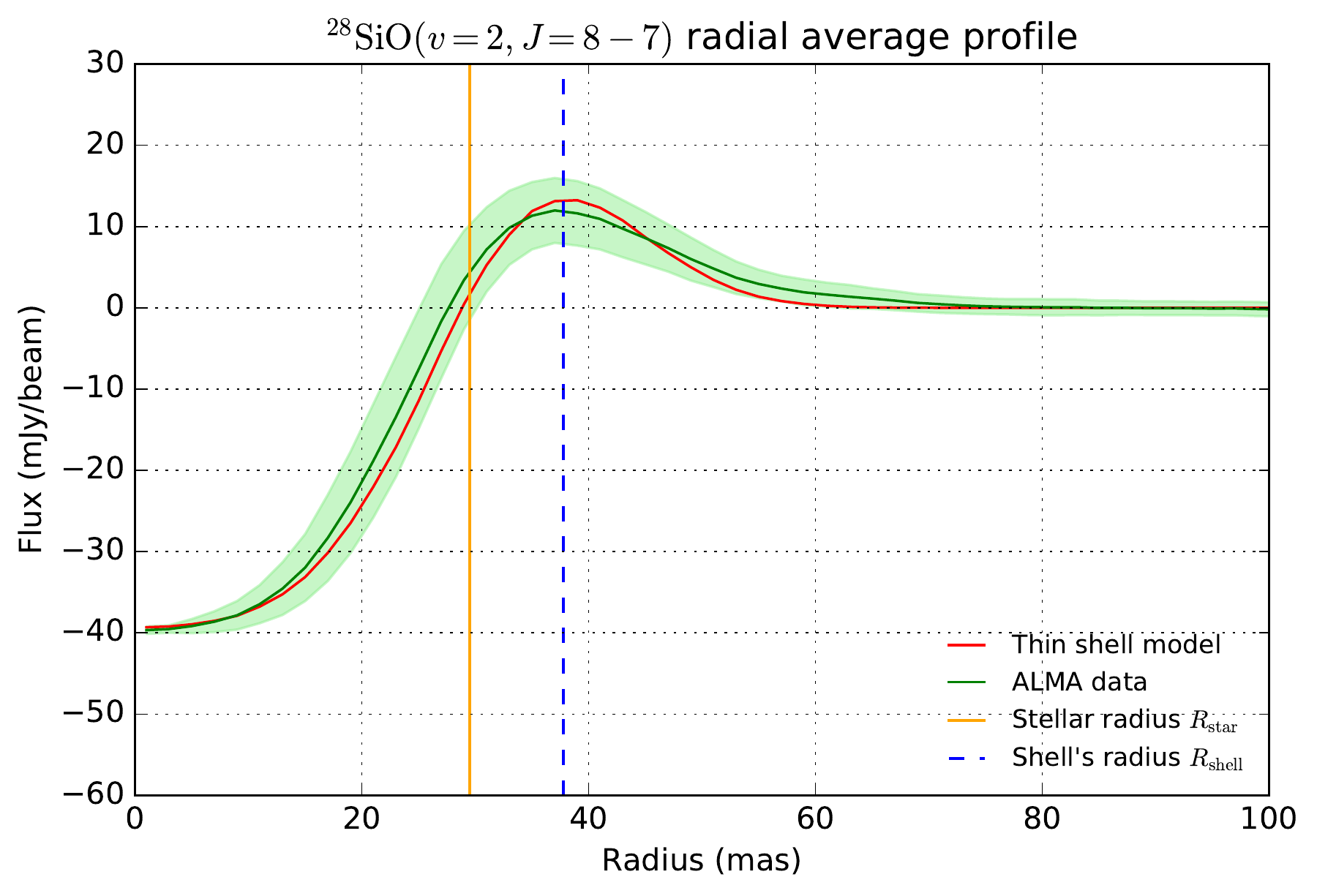}
        \caption{Azimuthally averaged radial profile of the $^{28}$SiO($\varv$=2,\,$J$=8-7) line emission (green curve) and best fit model (red curve).
        It is computed from the sum along the frequency axis of the continuum-subtracted data cube, scaled to the flux level measured at the sky frequency of the line.
        The shaded green area is the standard deviation of the shell over the considered ring radius.
        The equivalent uniform disk radius of the star is shown with a solid orange line, and the radius of the shell is represented with a dashed blue line.
        \label{radprofiles}}
\end{figure}

We determine the radius of the thin shell from the location of the maximum flux in the azimuthally averaged radial profile (Fig.~\ref{radprofiles}).
As shown in Fig.~\ref{radprofiles}, we obtain a value of $R_\mathrm{shell} = 37.8 \pm 2.5$\,mas (equivalent to $1.28 \pm 0.08\,R_\mathrm{star}$).
The uncertainty is estimated as the standard deviation of the radius of the peak emission measured as a function of the position angle.

\begin{figure*}[]
        \sidecaption
        \includegraphics[width=12.5cm]{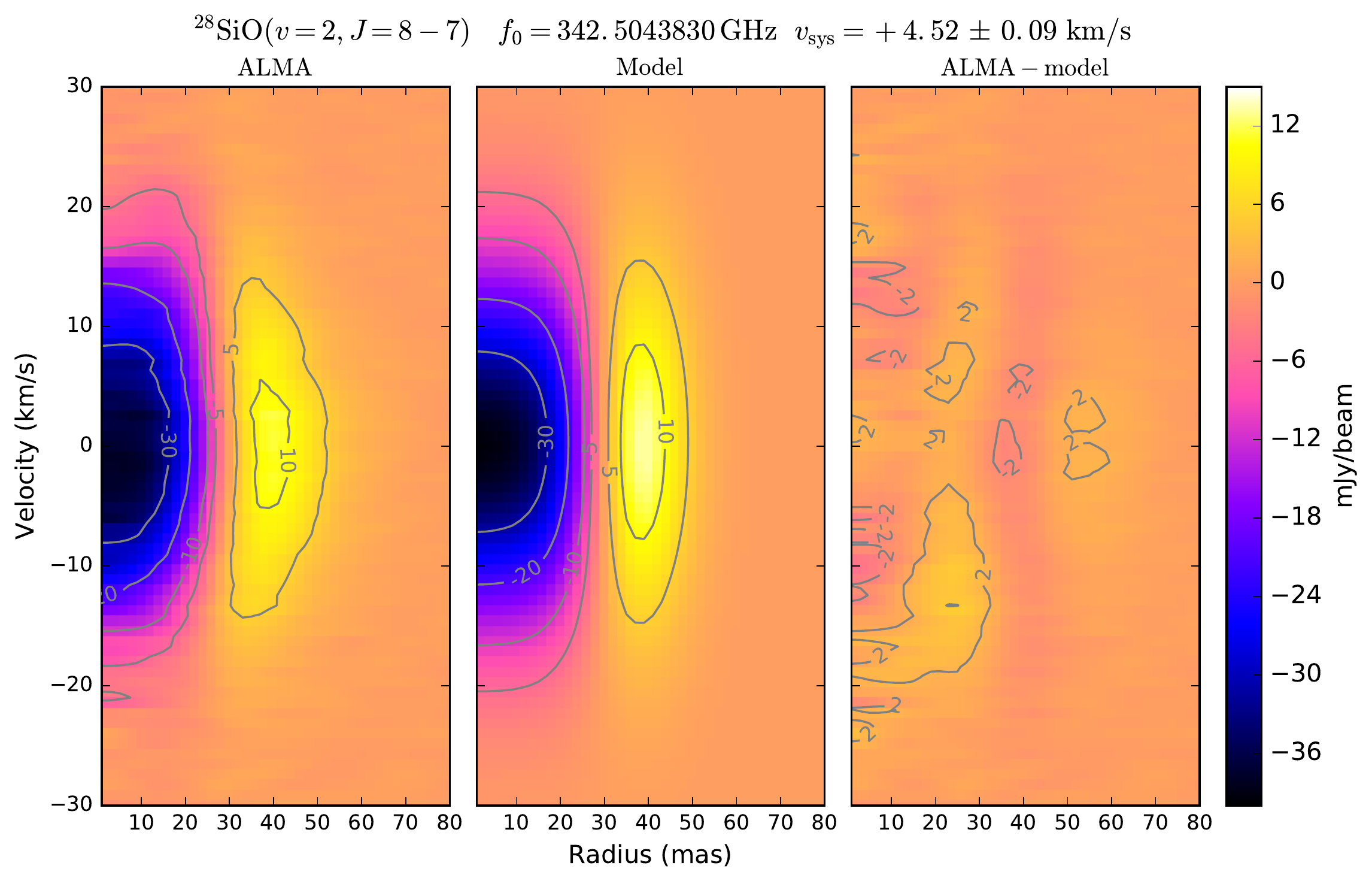}
        \caption{Azimuthally averaged radial profile of the ALMA continuum subtracted observation of the $^{28}$SiO($\varv$=2,\,$J$=8-7) line as a function of velocity offset (left panel), best fit model (center panel), and residuals of their subtraction (right panel).
        \label{profileresiduals}}
\end{figure*}

To determine the shell's surface brightness $I_\mathrm{shell}\,\tau_\mathrm{shell}$ we compute the
azimuthally averaged radial profile $I(r)$ of the continuum subtracted surface brightness map.
We adopt the model described in Sect.~\ref{shellmodel}, convolved with the instrumental beam, to derive the model radial profile.
The model profile is adjusted to the ALMA radial intensity profile $I(r)$ (Fig.~\ref{radprofiles}), using as variables $I_\mathrm{shell}\,\tau_\mathrm{shell}$ and the optical depth of the cooler absorbing layer $\tau_\mathrm{cool}$.
For the $\chi^2$ minimization, we used a classical Levenberg--Marquardt algorithm based on the SciPy routine \texttt{scipy.optimize.leastsq} \citepads{Jones:2001aa}.
The other parameters $R_\mathrm{star}$, $I_\mathrm{star}$, $R_\mathrm{shell}$, and $P(f)$ were determined independently as described previously.
The resulting parameters are listed in Table~\ref{betelgeuse-line-params}.
The observed and best fit radial profiles are presented in Fig.~\ref{radprofiles}.
Despite the simplicity of our model (single thin molecular layer over a spherical continuum), the quality of the fit is good up to the angular radius of the shell  $R_\mathrm{shell}$.
We note, however,  that our thin shell model does not reproduce well the tail of the $^{12}$CO line emission at larger radii (Fig.~\ref{profileresiduals12COv0}), indicating that the light-emitting envelope of this molecular species is extended in radius.
We attempted to fit a thick shell model, adding one parameter in the model (the shell thickness), but the results are unstable as the ALMA beam insufficiently resolves the thickness of the shell.
In the three panels of Fig.~\ref{profileresiduals} we show the azimuthally averaged ALMA radial profiles as a function of wavelength, the best fit model, and the residuals of the subtraction.

\subsection{Molecular emission velocity map\label{molecvelocity}}

\begin{figure*}[ht!]
        \centering
        \includegraphics[width=6cm]{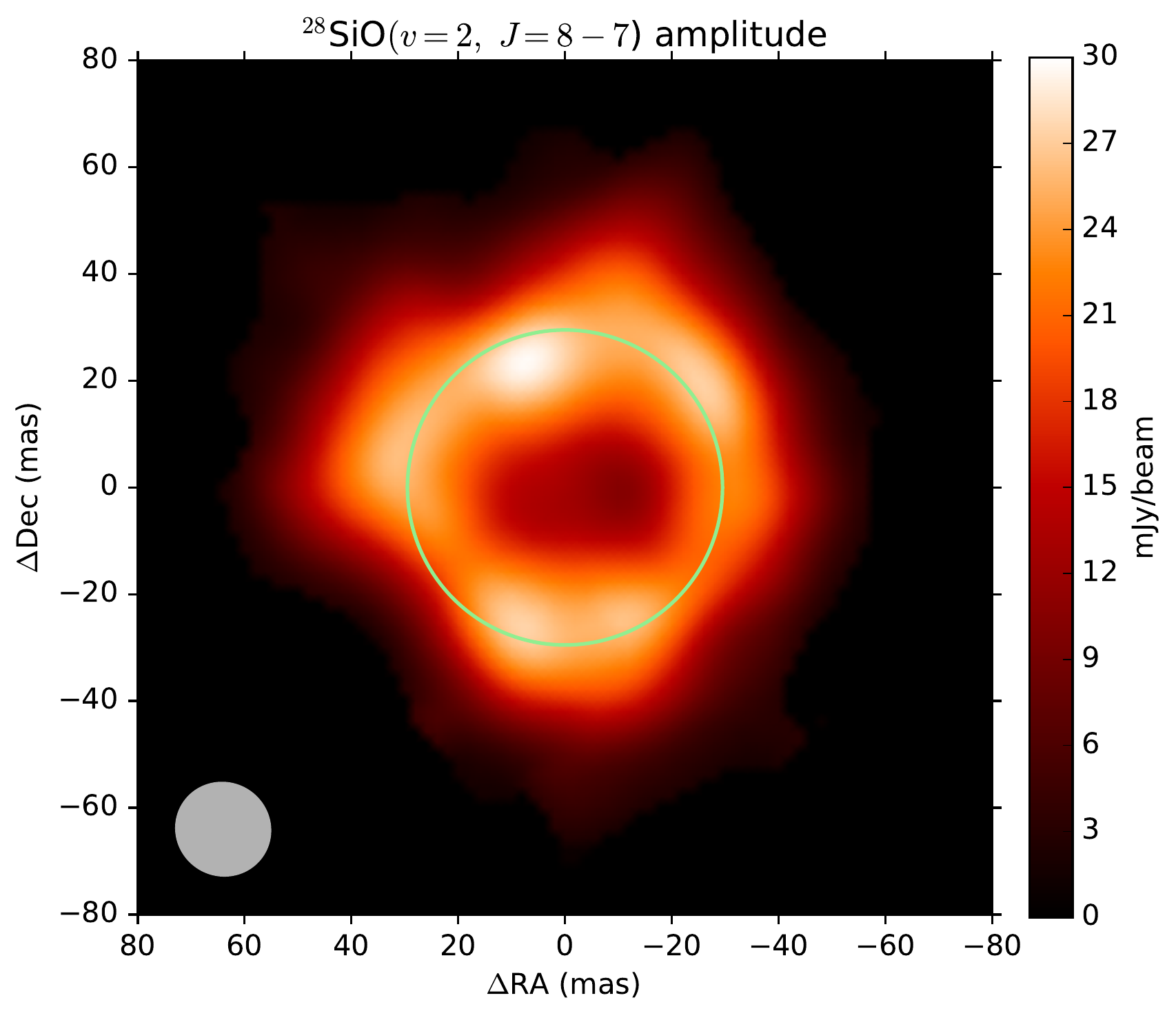}
        \includegraphics[width=6cm]{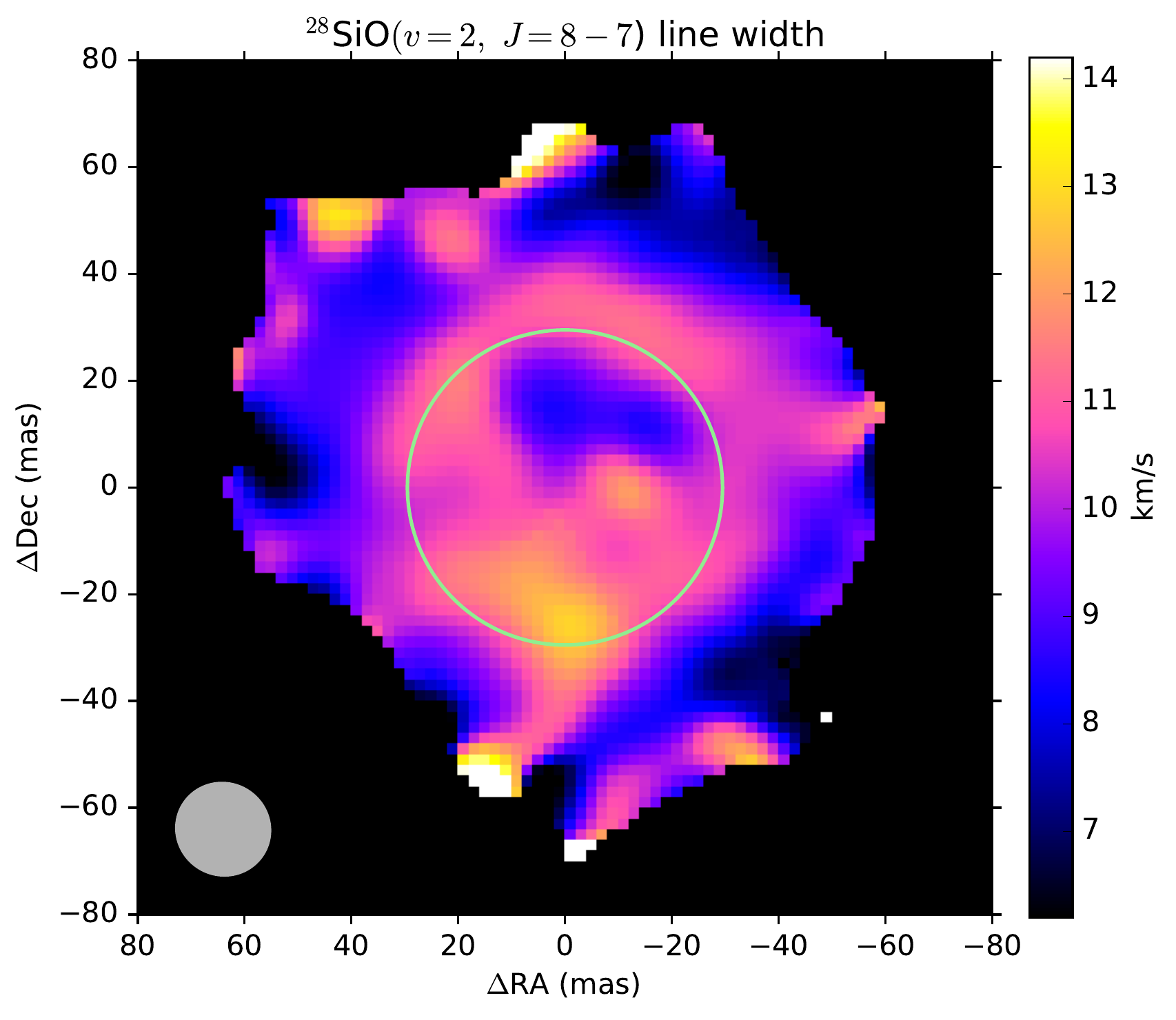}
        \includegraphics[width=6cm]{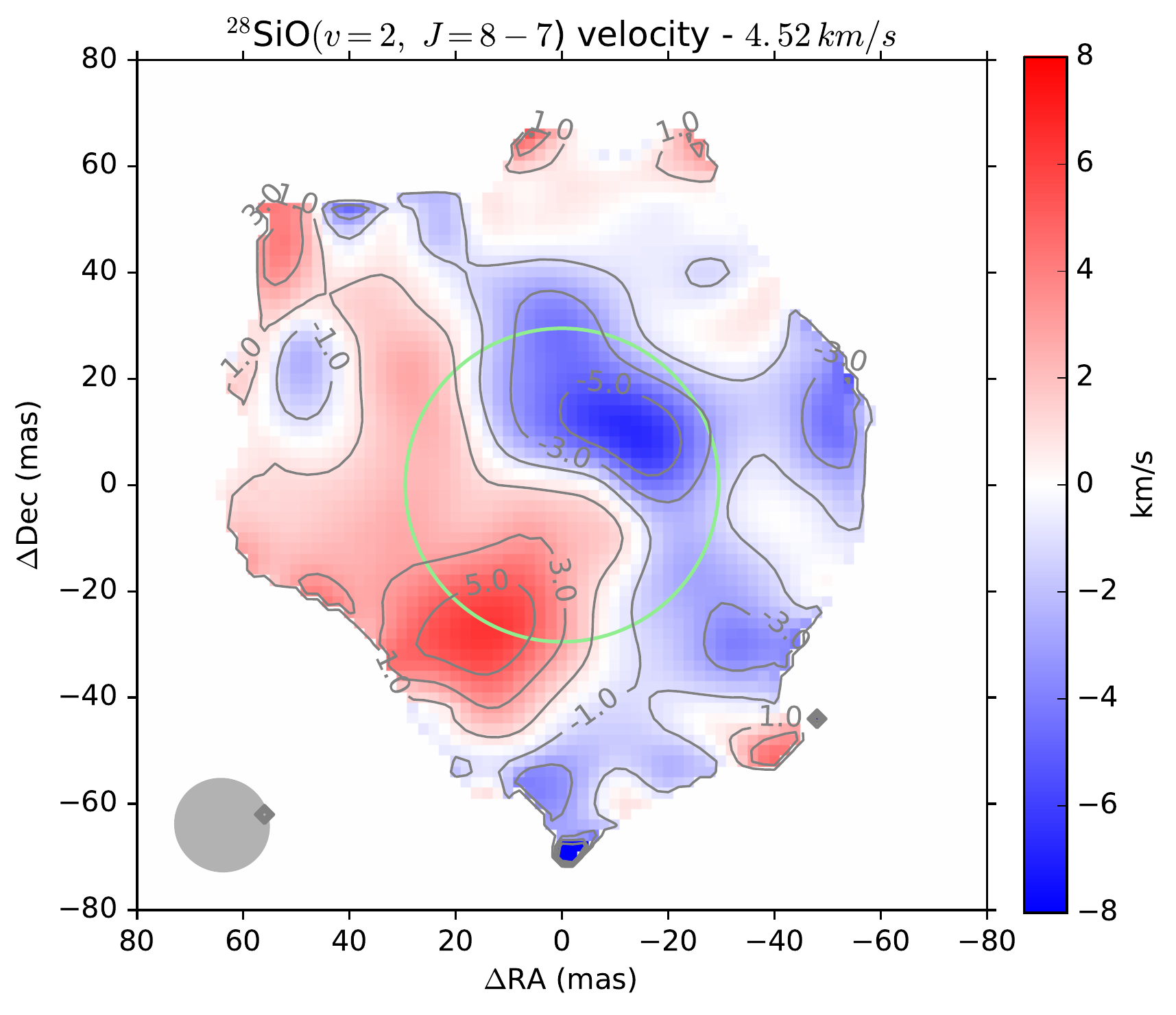}
        \caption{Emission line profile maps for the $^{28}$SiO($\varv$=2,\,$J$=8-7) line computed from the absorption corrected data cube.
        The left panel is the map of the amplitude of the best fit Gaussian, the middle panel the width of the emission line, and the right panel is the radial velocity.
        The systemic velocity determined from the line profile fitting is subtracted from the velocity map.
        The equivalent uniform disk size of the ALMA continuum emission is represented with a green circle, and the beam is shown in the lower left corner of each panel.
        \label{linemaps}}
\end{figure*}

The principle of our approach for the measurement of the rotational velocity of Betelgeuse is to map the Doppler shift of the emission line from the light-emitting shell, and fit a rotation model to the resulting velocity map.
To isolate the shell's emission over the stellar disk, we must compensate for the absorption component, that is, the first term in Eq.~\ref{eq4}.
In Sect.~\ref{modelparameters}, we determined all the relevant parameters, that is, the continuum surface brightness of the star $I_\mathrm{star}$ and the optical depth of the cool absorbing molecular shell $\tau_\mathrm{cool}$.
The resulting subtracted data cube shows only the Doppler shifted emission from the shell.
At each position of the absorption compensated data cube, we fit a Gaussian profile on the emission line with three parameters: the emission amplitude $A_\mathrm{emi}$, the line velocity width $\sigma_\mathrm{emi}$, and the mean velocity $\varv_\mathrm{emi}$.
%
The amplitude of the line emission (Fig.~\ref{linemaps}, left panel) is inhomogeneous and exhibits stronger emission at the limb along three preferential sectors, to the northeast, south, and northwest  of the star.
A strong unresolved emission peak is located close to the northern limb of the star, with a surface brightness of $A_\mathrm{emi}=30$\,mJy\,beam$^{-1}$.
The width of the line is  relatively homogeneous over the full map (Fig.~\ref{linemaps}, center panel),  with a broader velocity dispersion close to the limb.
The line velocity map (Fig.~\ref{linemaps}, right panel) exhibits a redshifted component to the southeast of the star, and a blueshifted component to the northwest. This is the classical signature of rotation, and we use this map in Sect.~\ref{rotationvelocity} to derive the rotation velocity of Betelgeuse.
The presence of low order, non-radial pulsations  (see, e.g., \citeads{2001ApJ...558..815L}) could in principle mimic the velocity signature of rotation.
As is discussed in Sect.~\ref{positionanglediscussion}, we determine a position angle of the symmetry axis of the velocity map that is closely consistent with the value reported by \citetads{1998AJ....116.2501U}.
This stability is well explained by rotation, but the temporal coherence of non-radial oscillations over a period of 20 years appears more unlikely.
In addition, we obtain a projected rotational velocity amplitude ($v_\mathrm{eq}\,\sin i \approx 5.5$\,km\,s$^{-1}$, Sect.~\ref{rotationvelocitydiscussion}) that is comparable to previous measurements.
Another indication in favor of a rotational origin for the Doppler shift is that we also observe co-rotating material outside of the stellar photosphere, as shown in Fig.~\ref{linemaps}.
This  behavior is difficult to explain by  large-scale non-radial pulsations of the photosphere.
For this reason, we interpret the present ALMA observations in terms of the rotation of Betelgeuse.
We foresee additional epochs of ALMA observations in the coming years to test the persistence of the velocity distribution in the close environment of the star.
As the time scale of the velocity variations observed in the UV domain with the HST is relatively short (a few months; \citeads{2001ApJ...558..815L}), a stability of the velocity map over several years would confirm the rotational origin of the observed Doppler wavelength shifts.

\subsection{Rotation velocity\label{rotationvelocity}}
We adopt the classical solid body rotation model, that we define in the $(x,y,z)$ cartesian coordinate system represented in Fig.~\ref{bet-model}.
This simple model assumes that the angular rotation velocity is independent of latitude.
%
In this framework, the projected velocity varies linearly as a function of the distance $y$ to the polar rotation axis:
\begin{equation}
\varv_\mathrm{solid}(x,y) = \frac{-y}{R_\mathrm{star}}\,(\varv_\mathrm{eq} \sin i) + \varv_\mathrm{sys}.
\end{equation}
The three model parameters are the systemic velocity $\varv_\mathrm{sys}$ in the local standard of rest (LSR), the projected rotation velocity $\varv_\mathrm{eq}\,\sin i$, and the position angle of the polar axis $PA$.
The inclination $i$ of the polar axis on the line of sight is degenerate with the linear equatorial velocity $\varv_\mathrm{eq}$ and thus cannot be estimated separately from the radial velocity alone.

In addition, we test for the presence of a spherical expansion velocity component $\varv_\mathrm{exp}$ resulting in the following supplementary radial velocity component $\varv_\mathrm{rad}$:
\begin{equation}
\varv_\mathrm{rad}(x,y) = - \varv_\mathrm{exp} \left( \frac{\sqrt{x^2 + y^2}}{R_\mathrm{star}} \right).
\end{equation}


\begin{figure*}[]
\centering
        \includegraphics[height=6.5cm]{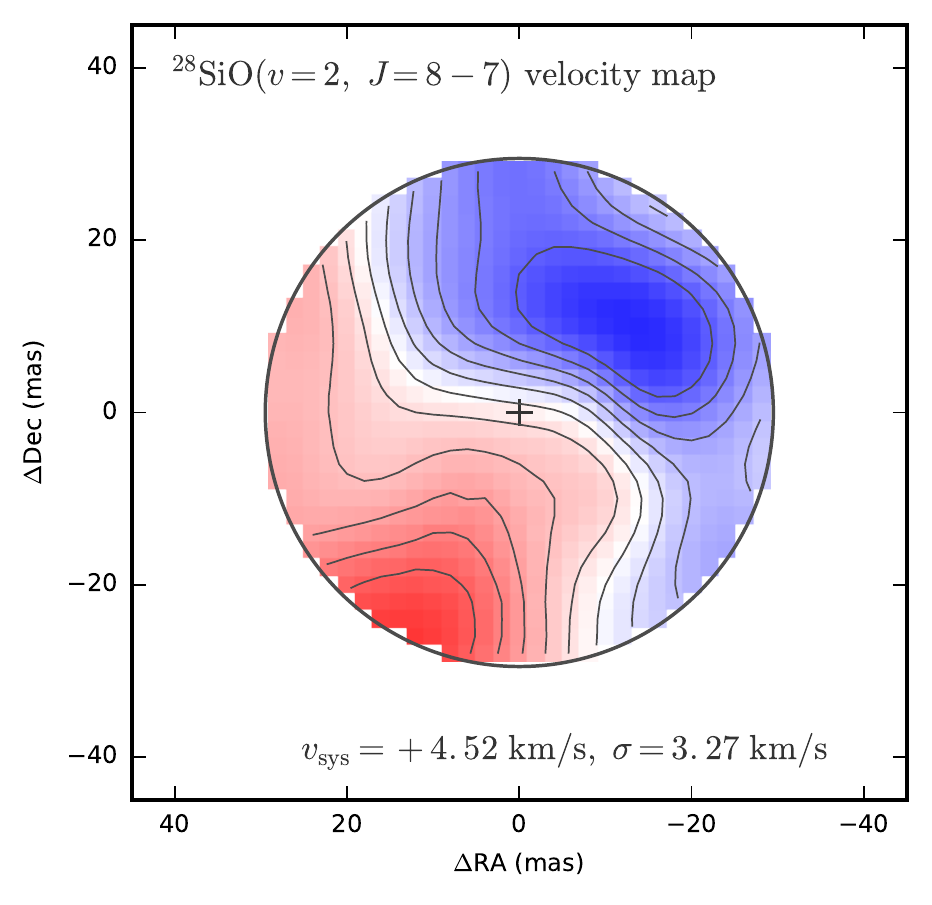}\\
        \includegraphics[height=6.5cm]{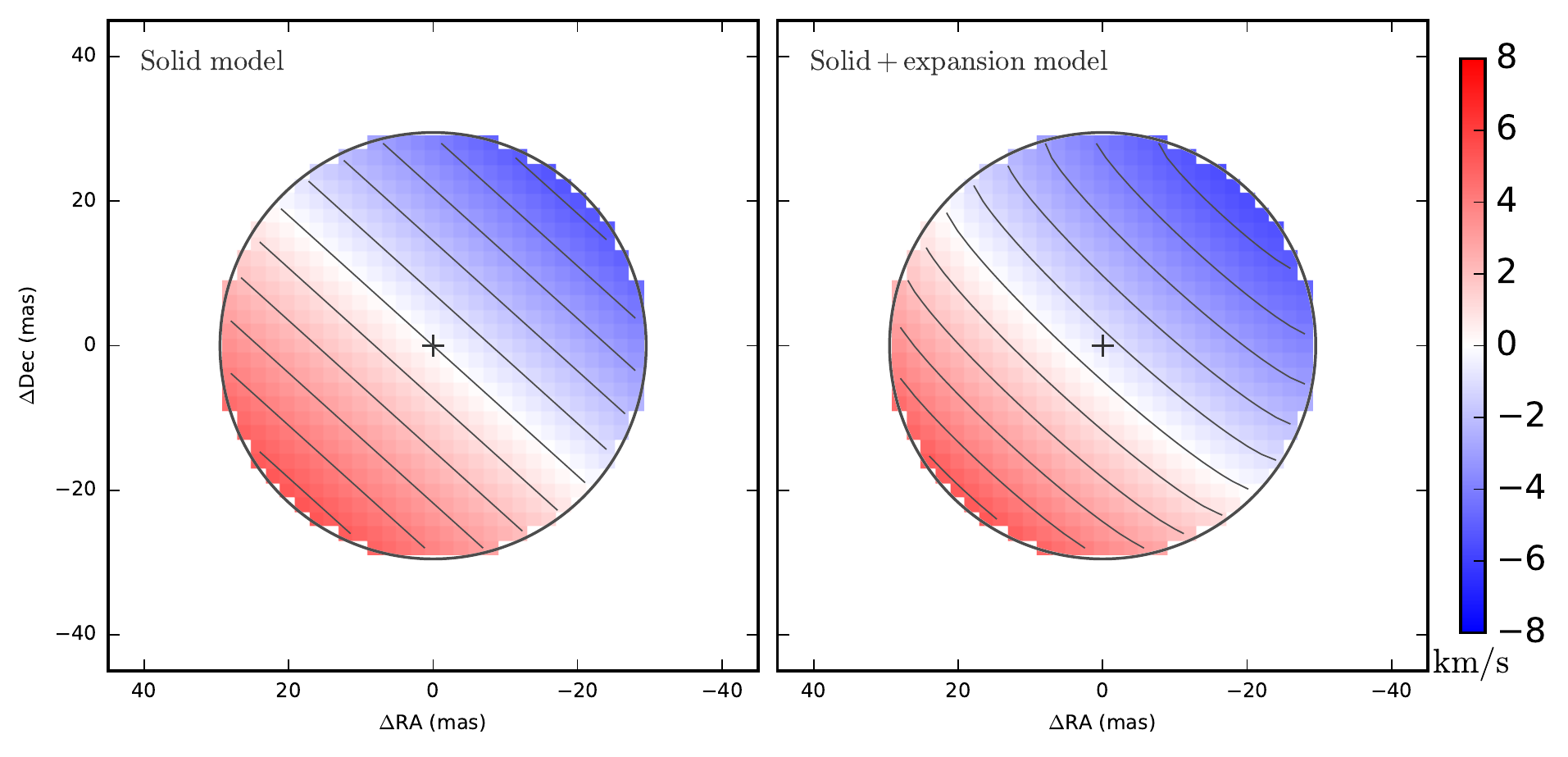}\\
        \includegraphics[height=6.5cm]{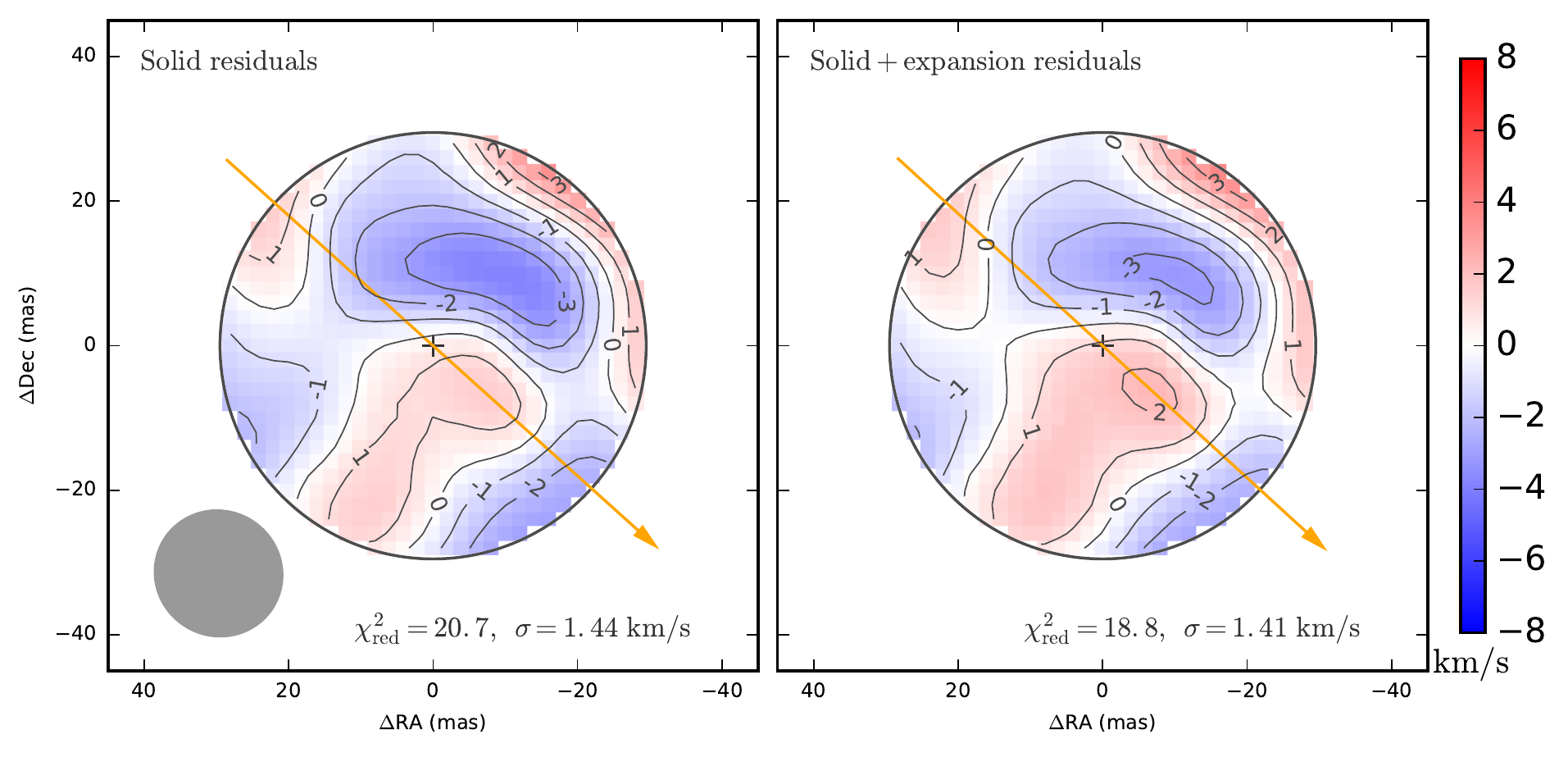}
        \caption{Velocity map of Betelgeuse in the $^{28}$SiO($\varv$=2,\,$J$=8-7) emission line measured over the equivalent continuum disk of the star (top panel)
        and best fit velocity models of Betelgeuse (center row).
        The corresponding model parameters are listed in Table~\ref{table-models}.
        The residuals of the fit are plotted in the bottom row, and the beam is shown in the lower left corner.
        The contour levels are separated by 1\,km\,s$^{-1}$ intervals.
        The polar axis is represented with an arrow pointing toward the direction of the angular momentum vector
        for a right-handed rotating coordinate system.
        \label{2models}}
\end{figure*}

The angular resolution of our ALMA observations provides approximately four resolution elements over the stellar continuum diameter of Betelgeuse.
This is sufficient to adjust the solid body rotation model as the correlation between the rotation velocity and the position angle of the polar axis is low.
The additional parameter of the expansion velocity $\varv_\mathrm{exp}$ is, however, largely correlated with the systemic velocity $\varv_\mathrm{sys}$ and its determination is therefore uncertain.
The reason for this correlation is that the average value of the spherical expansion velocity over the stellar disk is equivalent to a change in the systemic velocity.
The differentiation between the spherical expansion and systemic velocities is only possible at the limb of the star where perturbations from random velocity fluctuations degrade the quality of the fit.

The rotation models for the $^{28}$SiO($\varv$=2,\,$J$=8-7) line and the corresponding residuals are presented in Fig.~\ref{2models}.
We set the systemic LSR velocity of the model to the emission ring value $\varv_\mathrm{emi}$ listed in Table~\ref{betelgeuse-line-params}.
The best fit rotation model parameters are listed in Table~\ref{table-models} for the four analyzed molecular lines (treated independently), with and without a spherical expansion velocity component.
The adjustment of a spherical expansion parameter to the velocity map gives a radial expansion velocity of $\varv_\mathrm{exp} = \vExpSolidExpB \pm \vExpSolidExperrB$\,km\,s$^{-1}$, which is also very low compared to the velocities observed by \citetads{2011A&A...529A.163O}.
The expansion velocities of the other analyzed ALMA emission lines are less than 1\,km\,s$^{-1}$ in absolute value, except for the $^{28}$SiO($\varv$=1,\,$J$=8-7) line that reaches $\vExpSolidExpC \pm \vExpSolidExperrC$\,km\,s$^{-1}$.
The listed uncertainties of the model parameters have been scaled by $(\chi^2_\mathrm{red})^{1/2}$ and are therefore independent of the $\chi^2$ of the fit (that is, they take into account the residual velocity dispersion after subtraction of the model).
The maps and figures related to the $^{12}$CO($\varv$=0,\,$J$=3-2), $^{28}$SiO($\varv$=1,\,$J$=8-7) and $^{29}$SiO($\varv$=0,\,$J$=8-7) lines are presented in Appendices~\ref{12COv0line} to \ref{29SiOv0line}.

The correction of the continuum absorption (using the optical depth parameter $\tau_\mathrm{cool}$) induces a systematic uncertainty of $\pm 0.10$\,km\,s$^{-1}$ on the projected rotation velocity.
We add this systematic contribution quadratically to the error bars.
In addition, the molecular shell surrounding Betelgeuse is an evolving and dynamical environment.
The simple rotation model that we adopted does not include a provision for the presence of perturbing bulk motions in the atmosphere, which vary both temporally and spatially.
To estimate the associated systematic uncertainties on the adjusted model parameters, we chose the approach to modulate the radius over which the fit is computed by $\pm 15\%$ around the continuum radius of the star $R_\mathrm{star}$.
This modulation amplitude of 30\% corresponds approximately to the ratio of the angular resolution of our ALMA observations ($\approx 18$\,mas) to the continuum diameter of the star ($\approx 60$\,mas).
The standard deviation of the model parameters that we obtain through this approach are $\sigma(PA) = \PASysterr^\circ$ for the position angle of the polar axis and $\sigma(\varv_\mathrm{eq}\sin i) = \vsiniSysterr$\,km\,s$^{-1}$ for the projected rotation velocity.
In the following discussion, we adopt these values as systematic uncertainties and add them quadratically to the statistical error bars.

\begin{table*}
        \caption{Rotation model parameters of Betelgeuse. The position angle of the polar axis $PA$ is in degrees east of north.
        The listed error bars are scaled by $(\chi^2_\mathrm{red})^{1/2}$ to account for the residual dispersion of the fit.
        They do not comprise the systematic uncertainty that we estimate to $\pm \PASysterr^\circ$ for the position angle of the polar axis $PA$ and $\pm \vsiniSysterr$\,km\,s$^{-1}$ for the projected rotation velocity $\varv_\mathrm{eq}\sin i$ (see Sect.~\ref{rotationvelocity} for details).}
        \centering          
        \label{table-models}
        \begin{tabular}{lccccc}
        \hline\hline
        \noalign{\smallskip}
Model & $PA$ & $\varv_\mathrm{eq}\sin i$ & $\varv_\mathrm{exp}$ & $\chi^2_\mathrm{red}$ & $\sigma_\mathrm{res}$ \\
        & ($^\circ$) & (km\,s$^{-1}$) & (km\,s$^{-1}$) &  & (km\,s$^{-1}$) \\
        \noalign{\smallskip}
        \hline    
        \noalign{\smallskip}
\multicolumn{6}{c}{$^{28}$SiO($\varv$=2,\ $J$=8-7)} \\
        \noalign{\smallskip}
Solid & $48.0 \pm 1.1$ & $5.47 \pm 0.10$ & $-$ & $20.7$ & $1.44$ \\
Solid+Exp & $47.6 \pm 1.1$ & $5.56 \pm 0.10$ & $+0.69 \pm 0.08$ & $18.8$ & $1.41$ \\
        \hline
        \noalign{\medskip}
\multicolumn{6}{c}{$^{12}$CO($\varv$=0,\ $J$=3-2)} \\
        \noalign{\smallskip}
Solid & $49.5 \pm 1.0$ & $4.82 \pm 0.08$ & $-$ & $12.5$ & $1.13$ \\
Solid+Exp & $47.9 \pm 1.0$ & $4.76 \pm 0.08$ & $+0.60 \pm 0.06$ & $11.1$ & $1.12$ \\
        \hline
        \noalign{\medskip}
\multicolumn{6}{c}{$^{28}$SiO($\varv$=1,\,$J$=8-7)} \\
        \noalign{\smallskip}
Solid & $51.9 \pm 1.3$ & $5.81 \pm 0.13$ & $-$ & $35.8$ & $1.39$ \\
Solid+Exp & $50.6 \pm 1.0$ & $5.73 \pm 0.10$ & $+1.75 \pm 0.08$ & $20.6$ & $1.34$ \\
        \hline
        \noalign{\medskip}
\multicolumn{6}{c}{$^{29}$SiO($\varv$=0,\,$J$=8-7)} \\
        \noalign{\smallskip}
Solid & $45.5 \pm 1.0$ & $5.68 \pm 0.09$ & $-$ & $18.9$ & $1.34$ \\
Solid+Exp & $45.8 \pm 1.0$ & $5.68 \pm 0.09$ & $-0.21 \pm 0.08$ & $18.7$ & $1.34$ \\
        \hline
      \end{tabular}
\end{table*}

\section{Discussion\label{discussion}}

\subsection{Radial structure of the envelope and stellar wind}

The continuum angular radius of Betelgeuse in the near-infrared is $\approx 21$\,mas (see \citeads{2016A&A...588A.130M} and references therein), and the MOLsphere radius is 25 to 30\,mas \citepads{2014A&A...572A..17M, 2011A&A...529A.163O}.
The MOLsphere term, introduced by \citetads{2000ApJ...540L..99T}, designates a non-photospheric molecular layer.
The stellar continuum radius that we measure with ALMA is close to 30\,mas, and the radius of the molecular emission is $\approx 40$\,mas.
 The near-infrared MOLsphere thus closely corresponds to the ALMA continuum radius $R_\mathrm{star}$, while the ALMA line emission reaches approximately twice the near-infrared continuum radius of the star.
\citetads{2015A&A...580A.101O, 2017A&A...602L..10O} established that the apparent continuum radius of Betelgeuse
decreases as a power law with wavelength between 6.1 and 0.7\,cm, and breaks down at $\lambda = 0.9$\,mm, the wavelength of the present ALMA observations.

The fact that the radial velocities that we measure on the stellar disk ($\varv_\mathrm{abs}$ in Table~\ref{betelgeuse-line-params}) and in the circumstellar emission ring ($\varv_\mathrm{emi}$) are close to identical (Sect.~\ref{lineprofile}) implies that there is no significant global wind or large-scale inflow or outflow of material
through the molecular layer located at a radius of 40\,mas, at the angular resolution of our observations ($\approx 18$\,mas).
This conclusion is in agreement with \citetads{2006ApJ...646.1179H}.
This overall static molecular layer therefore seems to act as a buffer absorbing the kinetic energy of the regular underlying convection cells,
but it cannot confine the outflows of the strongest convective cells, hence the formation of focused molecular plumes \citepads{2009A&A...504..115K}.
In this scenario, the MOLsphere within $\approx 40$\,mas therefore includes a globally static spherical component and localized emerging flows of material (see Sect.~\ref{massloss} for a more detailed discussion).
It should be noted that the deeper molecular layer imaged by \citetads{2011A&A...529A.163O} is very turbulent and clumpy (see also \citeads{2017Natur.548..310O} for \object{Antares}).
The ALMA molecular shell is likely also strongly turbulent, considering the observed emission line widths.

\subsection{Rotation velocity\label{rotationvelocitydiscussion}}

From our analysis of the $^{28}$SiO($\varv$=2,\,$J$=8-7) line, we measure a projected rotation velocity of $\varv_\mathrm{eq}\,\sin i = \vsiniSolidB \pm \vsiniSoliderrSystB$\,km\,s$^{-1}$ at an angular radius of $\rstarB \pm \rstarerrB$\,mas.
As listed in Table~\ref{table-models}, the solid body projected rotation velocities from the three SiO emission lines range from \vsiniSolidB\,km\,s$^{-1}$ to \vsiniSolidC\,km\,s$^{-1}$.
The extreme values are therefore only $1.4\,\sigma$ apart from each other.
The $^{12}$CO($\varv$=0,\,$J$=3-2) line velocity appears slower at \vsiniSolidA\,km\,s$^{-1}$.
We note, however, that the CO emission radial profile deviates significantly from our thin shell model as shown in Fig.~\ref{profileresiduals12COv0} (left panel).
The CO emission is more extended radially than the high excitation SiO lines, in particular around the systemic velocity.
This is expected from previous observations by \citetads{2012AJ....144...36O} of the $^{12}$CO($\varv$=0,\,$J$=2-1) molecular emission of Betelgeuse ($\lambda = 1.3$\,mm).
As the measured shell emission includes material located at larger radii from the star, this results in an underestimation of the stellar rotation velocity due to the de-coupling of the chromosphere rotation from the stellar rotation beyond 10\,au (see Sect.~\ref{rotationprofile}).
The vibrationally excited, high energy level SiO is confined to the inner regions, and although SiO masing is possible, especially from vibrationally excited states \citepads{2009MNRAS.394...51G}, this will not affect our kinematic analysis.
For this reason, we adopt the velocity determined using the $^{28}$SiO($\varv$=2,\,$J$=8-7) line.
It corresponds to a projected angular rotation velocity of $\omega\,\sin i = (\omegasini \pm \omegasinierr) \times 10^{-9}$\,rad\,s$^{-1}$.

\subsection{Radial rotation velocity profile\label{rotationprofile}}

To evaluate whether the MOLsphere is co-rotating with the star or is in detached orbit around it, here we  compare the measured projected rotation velocity to the Keplerian velocity.
The mass of Betelgeuse is poorly known, and significantly different estimates can be found in the literature (see, e.g., \citeads{2013EAS....60...17M}, \citeads{2011ASPC..451..117N}, \citeads{2016ApJ...830..103N} and \citeads{2016ApJ...819....7D}) from approximately 10 to $20\,M_\odot$.
Assuming a mass of $m_\mathrm{Bet} = 15 \pm 5\ M_\odot$ and a radius of $R_\mathrm{star} = 1400 \pm 250\ R_\odot$ (from our equivalent ALMA continuum angular radius and the parallax $\pi = 4.51 \pm 0.80$\,mas by \citeads{2017AJ....154...11H}), the Keplerian rotation velocity at the surface of Betelgeuse is $\varv_K = 45 \pm 28$\,km\,s$^{-1}$.
This is approximately a factor of 8 higher than the velocity that we observe in the MOLsphere emission.
This is an indication that the molecular shell is corotating with the star, and that, although measured at $\approx 1.3\,R_\mathrm{star}$ the shell's rotation velocity is a reliable proxy for the rotation velocity of the star itself.
We note that the Keplerian velocity $\varv_K$ at the surface of the star is also the critical rotation velocity.
We therefore conclude that the surface rotation velocity of Betelgeuse is far from being critical, which is consistent with its essentially spherical geometry measured by \citetads{2017A&A...602L..10O}.
The evolution models including rotation by \citetads{2012A&A...537A.146E} indicate that a $15\,M_\odot$ star starting on the main sequence with a rotation velocity of 250\,km\,s$^{-1}$ (close to critical) reaches the red supergiant stage with an equatorial velocity below 10\,km\,s$^{-1}$, which is consistent with our observations.

\begin{figure}[]
        \centering
        \includegraphics[width=\hsize]{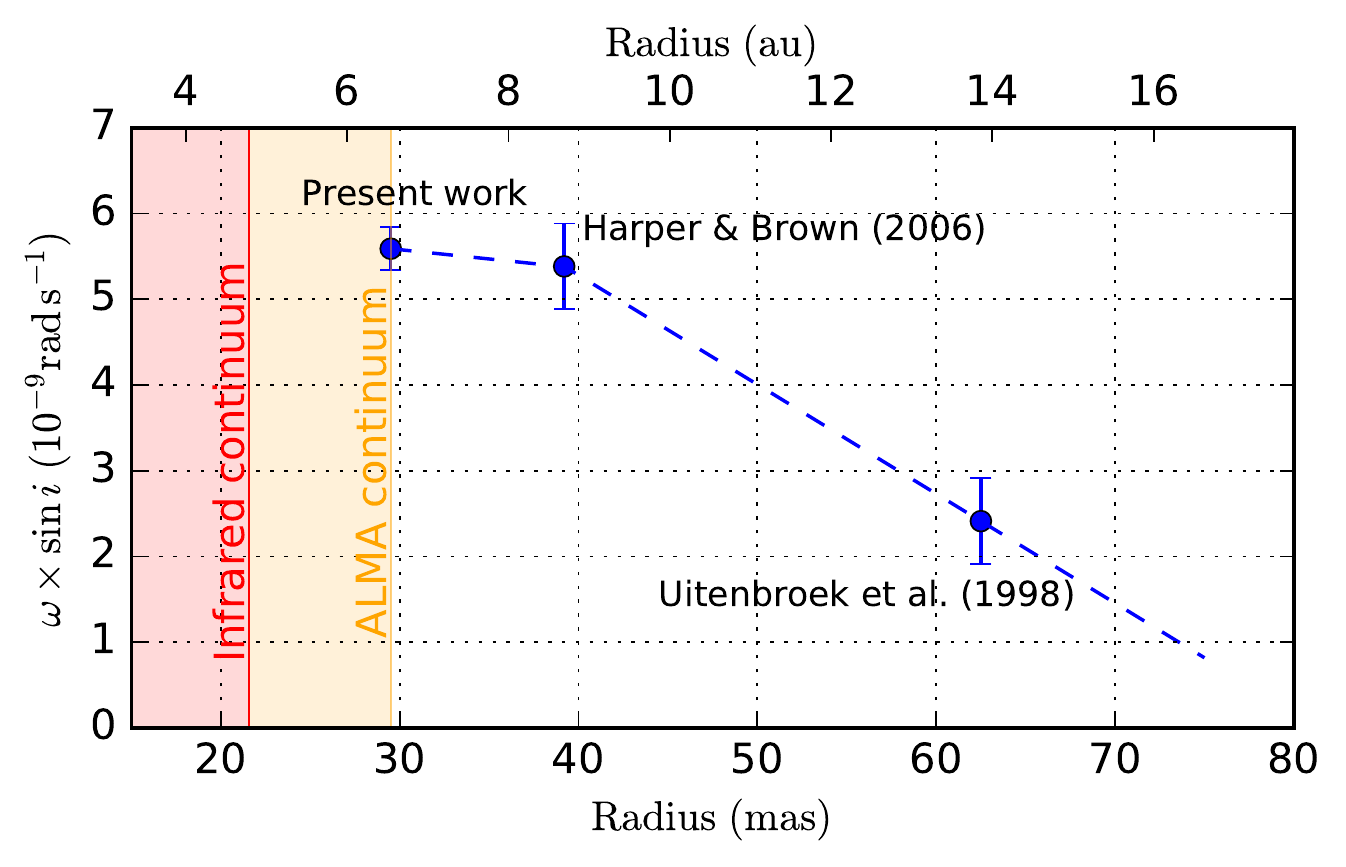}
        \caption{Angular rotation velocity of the envelope of Betelgeuse as a function of radius.
        The infrared continuum radius is the November 2014 epoch from \citetads{2016A&A...588A.130M}.
        \label{rotprofile}}
\end{figure}

From spatially scanned HST/STIS spectra, \citetads{2006ApJ...646.1179H} measured a projected rotation velocity of $\varv_\mathrm{eq} \sin i = 7.0$\,km\,s$^{-1}$ at a radius of 39.2\,mas ($1.3\times R_\mathrm{star}$) in the UV domain ($\lambda = 232.5$\,nm).
This measurement is consistent with a co-rotation with the ALMA molecular shell as the scaled velocity $7.0 \times 29.5 / 39.2 = 5.3$\,km\,s$^{-1}$ is identical to the value we measure within its error bar.
\citetads{1998AJ....116.2501U} measured a velocity of $\varv_\mathrm{eq} \sin i = 5.0$\,km\,s$^{-1}$ at an angular radius or 62.5\,mas and at a wavelength of $\lambda \approx 280$\,nm \citepads{1996ApJ...463L..29G}.
This corresponds to an angular rotation velocity of $\omega \sin i = 2.4 \ 10^{-9}$\,rad\,s$^{-1}$, less than half the value we measure with ALMA.
It is interesting to note that this layer is actually rotating at a very similar linear velocity to that of the SiO molecular shell we observed with ALMA.
It is even closer to the rotation velocity of the $^{12}$CO emission line (Appendix~\ref{12COv0line}) that forms, on average, at larger angular radii from the star than the SiO lines.

The distribution of the measured angular rotation velocities as a function of radius is presented in Fig.~\ref{rotprofile}.
Following \citetads{2006ApJ...646.1179H}, we propose that the chromospheric material is co-rotating with the star up to a radius of $\approx 40$ to 50\,mas ($1.5\times\,R_\mathrm{star} \approx 9$ to 11\,au) and for larger radii, it approximately preserves its linear rotation velocity $\varv_\mathrm{eq} \sin i$, thus decreasing its angular velocity $\omega \sin i$.
A possible explanation for this velocity profile is that the chromosphere of Betelgeuse is magnetically coupled to the star in rigid rotation up to a radius of  10\,au.
At larger distances from the star, the weaker magnetic coupling is not efficient enough to drag the charged chromospheric material, and it continues its rotation with the linear velocity acquired at the de-coupling radius.
Such an extended and rotationally coherent volume may be structured by the presence of a global, possibly dipolar magnetic field.

\subsection{Position angle of the polar axis \label{positionanglediscussion}}

For the solid body rotation model that we adopt (Sect.~\ref{rotationvelocity}), the determination of the position angle of the axis of rotation  of Betelgeuse is robust as its determination relies on the spatial symmetry of the velocity field.
In agreement with the IAU recommendation, we define the north pole as ``that pole of rotation that lies on the north side of the invariable plane of the solar system'' \citepads{Archinal2011}.
The position angles for the four analyzed emission lines vary between $\PASolidD^\circ$ and $\PASolidC^\circ$ (Table~\ref{table-models}).
The position angle from the $^{28}$SiO($\varv$=2,\,$J$=8-7) line is close to the middle of this range at $PA = \PASolidB \pm \PASoliderrSystB^\circ$ and we adopt this value as our reference.
The rotation of Betelgeuse is therefore retrograde as its right-handed angular momentum vector is pointing to the southwest quadrant.

From HST/GHRS spatially scanned spectroscopy of chromospheric absorption lines obtained through a $200\times 200$~mas aperture, \citetads{1998AJ....116.2501U} estimated the position angle of the rotation axis to be $55^\circ$, which was confirmed by \citeads{2006ApJ...646.1179H} using archival HST/STIS $25\times 100$~mas aperture scans of co-rotating low opacity chromospheric emission lines.
This is in good agreement with our determination.
The near-infrared interferometric observation by \citetads{2011A&A...529A.163O} was recorded using an essentially linear configuration of the VLTI/AMBER interferometer, at a position angle of $73^\circ$. It was therefore roughly aligned with the direction of the rotation axis and the AMBER spectro-interferometry, thus does not show the rotational Doppler line shift.

\subsection{Inclination of the polar axis, rotation period}

The inclination $i$ of the rotation axis of Betelgeuse on the line of sight is particularly difficult to determine from observations, due to it slow rotation and the absence of persistent and traceable surface features.
\citetads{1998AJ....116.2501U} proposed that $i\approx 20^\circ$, based on the hypothesis that the hot spot they observed in the southwest quadrant of the UV chromosphere corresponds to the position of the pole of the star, but  this association between a UV spot and the polar axis appears fragile.
\citetads{2013EAS....60...77D} reported several epochs of HST/FOC UV imaging observations, where the persistence of the UV hot spot over several years in not demonstrated.
In addition, we do not confirm the presence of a hot spot in the southwest quadrant, but rather at a nearly centrally symmetric position angle in the northeast quadrant.
The coincidence of the hot spot detected by \citetads{2017A&A...602L..10O} with the position angle of the rotation axis may indicate that this spot is located at the pole.
This hypothesis, already employed by \citetads{1998AJ....116.2501U}, is difficult to justify unambiguously.
One difficulty is that we do not observe in the ALMA data a symmetrical hot spot or molecular plume at the south pole of Betelgeuse.
This could be due, for instance, to a random fluctuation of the polar hot spot activity over the 400-day oscillation period of the star.
It could also be caused by a latitudinal offset of the hot spots with respect to the polar axis, which would modulate their position with rotation,
or to a very long-term cyclic modulation of the polar convective flow.
Under the assumption that the northeast spot is located at the pole, as it appears very close to the limb of the star, the inclination of the rotation axis would be $i \gtrsim 60^\circ$.
Considering the measured $\varv_\mathrm{eq} \sin i = \vsiniSolidB \pm \vsiniSoliderrSystB$\,km\,s$^{-1}$, the linear equatorial velocity would therefore be $\varv_\mathrm{eq} \lesssim 6.3$\,km\,s$^{-1}$, and the angular rotation velocity $\omega \lesssim 6.5 \times 10^{-9}$\,rad\,s$^{-1}$.
As a side note, from purely geometrical considerations, the a priori probability that a star's rotation axis has an inclination greater than $i_0$ (between 0 and $90^\circ$) is equal to $P = \cos i_0$.
As a consequence, and in absence of direct constraints, it is equally probable that the inclination of the rotation axis of Betelgeuse is higher  or lower than $60^\circ$.
The 68\% probability interval of $i$ can be represented by $i = 60_{-27}^{+21} \deg$, which corresponds to a probable equatorial velocity of $\varv_\mathrm{eq} = 6.5^{+4.1}_{-0.8}$\,km\,s$^{-1}$.

We attempted the adjustment of a latitudinal angular differential rotation model (see, e.g., \citeads{2004A&A...418..781D}) to the ALMA emission line velocity maps.
In the presence of differential rotation, the degeneracy between the latitude and the radial velocity is removed, and it is possible to determine the inclination $i$ of the polar axis on the line of sight.
However, the adjustment of this model does not bring a significant improvement of the $\chi^2$ or residuals of the fit compared to the simpler solid body rotation model.
This could be due either to the absence of differential rotation or to the perturbation from the random fluctuations of the velocity field ($\sigma \approx 1$\,km\,s$^{-1}$) superimposed on the rotation velocity field.
A higher angular resolution would be required to  efficiently constrain the differential rotation latitudinal profile and the inclination of the polar axis $i$.
For this reason, we cannot conclude firmly on the inclination of the polar axis on the line of sight.

Combining the newly determined parallax $\pi = 4.51 \pm 0.80$\,mas \citepads{2017AJ....154...11H}, the ALMA continuum equivalent angular radius ($\theta = \rstarB \pm \rstarerrB$\,mas) and the measured $\varv_\mathrm{eq}\,\sin i = \vsiniSolidB \pm \vsiniSoliderrSystB$\,km\,s$^{-1}$ velocity of the $^{28}$SiO molecular shell , we obtain a rotation period of $P / \sin i = \ProtB \pm \ProterrB$\,years.
Considering an inclination of $i = 60^\circ$, the rotation period would be $P = 31$\,years, within the uncertainty range of our $P / \sin i$ estimate.

\subsection{Residual velocity dispersion \label{velresiduals}}

Once the global rotation velocity pattern is subtracted from the velocity map, the residuals have a low amplitude with a typical standard deviation of only $\sigma \approx 1$ to 1.5\,km\,s$^{-1}$ (Fig.~\ref{linemaps}, bottom row; see also Figs.~\ref{12COv0linemaps}, \ref{28SiOv1linemaps}. and \ref{29SiOv0linemaps}).
These small velocity residuals strongly contrast with the large velocity amplitudes of up to 25-30\,km\,s$^{-1}$ observed by \citetads{2011A&A...529A.163O} in the near-infrared for the carbon monoxide MOLsphere of Betelgeuse, and in the comparable red supergiant Antares \citepads{2013A&A...555A..24O, 2017Natur.548..310O}.
These interferometric observations were obtained in the $K$ band ($\lambda \approx 2.2\,\mu$m) and sample the near-infrared, second overtone ($\Delta \varv =2$) emission of the carbon monoxide molecule.
This CO layer is located at approximately $0.94\times$ the radius of the continuum layer of our ALMA observations.
The angular radius of the $^{12}$CO($\varv$=0,\,$J$=3-2) emission line layer is $R_\mathrm{shell} = 39.5 \pm 2.9$\,mas (Table~\ref{betelgeuse-line-params}), that is, $1.3\times$ the radius of the ALMA continuum.
This is a lower excitation line than the CO lines observed using near-infrared interferometry.
The residual velocity map of the solid body rotation model of this line presented in Fig.~\ref{12COv0linemaps} shows a dispersion of only $\sigma = 1.1$\,km\,s$^{-1}$.

For the $^{28}$SiO($\varv$=2,\,$J$=8-7) emission line, we observe residual velocity offsets of $\pm 3$\,km\,s$^{-1}$ peak-to-peak in amplitude over the disk of the star (Fig.~\ref{2models}, bottom panels).
However, the velocity width map (Fig.~\ref{linemaps}, center panel) shows Gaussian velocity dispersions of up to $\sigma \approx 13$\,km\,s$^{-1}$. Such high values cannot be solely due to thermal motions, and indicate the presence of turbulence.
We do not see a comparable amplitude in the radial velocity map (Fig.~\ref{linemaps}, right panel), which indicates that the small-scale rapid motions are (at least partially) smeared out by the limited ALMA angular resolution.
The molecular shell observed with ALMA is therefore likely to be highly turbulent, as is the underlying near-infrared carbon monoxide layer.

A positive velocity residual is visible in Fig.~\ref{2models} (bottom panels) in the northeast quadrant (also in the other studied emission lines), close to the limb of the star, at a position angle compatible with the continuum hot spot.
Its position on the stellar disk implies that geometrical projection effects probably reduce significantly the radial component of the velocity (in addition to the spatial smearing effect).
Other velocity spots are present in the northwest (positive velocity offset) and south (negative offset) regions of the star, and close to the center of the star (positive offset).
They correspond to regions where the width of the emission line is larger (Fig.~\ref{linemaps}, center panel), which suggests that the local turbulence is stronger.
The possible relation between these active regions and mass loss is discussed in Sect.~\ref{naturehotspot} and \ref{massloss}.

\subsection{Nature of the continuum hot spot\label{naturehotspot}}

From the analysis of the continuum of the present ALMA observations, \citetads{2017A&A...602L..10O} detected a large hot spot (major axis $\approx 20$\,mas, $\Delta T \approx 1000$\,K) that is visible in the \emph{spw2} continuum image presented in Fig.~\ref{bet-spw2-cont}.
This spot appears to be closely aligned with the direction of the polar axis (see also Sect.~\ref{massloss} and Fig.~\ref{mass-ejection}).
The combination of its enormous extension on the stellar disk (comparable to the continuum radius of the star) and strong temperature contrast ($\Delta T \approx 1000$\,K) appear too large to be explained by regular convection cells \citepads{2010A&A...515A..12C}.
The stellar limb even appears to be physically distorted due to its presence.

Several authors have reported the observation of bright spots on the disk of Betelgeuse \citepads{2000MNRAS.315..635Y, 2006sf2a.conf..471H, 2011ApJ...740...24R, 2016A&A...591A.119A}, although their presence is not systematic \citepads{1997MNRAS.290L..11B}.
Interferometric observations in the near-infrared domain by \citetads{2016A&A...588A.130M} and \citetads{2009A&A...508..923H} have shown the presence of temperature inhomogeneities on the infrared photosphere of Betelgeuse.
The continuum radius of the star in this wavelength range is approximately $0.7\times$ the continuum equivalent radius measured with ALMA ($R_\mathrm{star}$ in Table~\ref{betelgeuse-line-params}).
The optical interferometric observations therefore correspond to a significantly deeper layer of the atmosphere.
The observed surface structures in this wavelength range are generally consistent with the results from radiative hydrodynamics simulations of convection \citepads{2010A&A...515A..12C, 2014A&A...572A..17M}.
\citetads{2016A&A...591A.119A} demonstrated that spectropolarimetry at visible wavelengths makes it possible to detect the presence of hot spots through their depolarization of the continuum spectrum of the star.
Using this technique, hot spots have been reported by \citetads{2017arXiv170202002T} on Betelgeuse at the epoch of the ALMA observations.
The main spot detected by these authors matches the position of the ALMA hot spot (northeast quadrant of the stellar disk) and has a comparable spatial extension (their Fig.~4).
In the UV domain, \citetads{1996ApJ...463L..29G} detected a hot spot at a position angle $PA = 235^\circ$ (southwest quadrant).
The position of this spot is only $7^\circ$ away from the direction of the polar axis we obtain from ALMA, but it is located in the southwest quadrant, symmetrically to the ALMA hot spot with respect to the disk center.

We propose that the ALMA hot spot is caused by the presence of an exceptionally strong convection cell.
We adopt here the term \emph{rogue cell} to designate such a statistical outlier, by analogy with the abnormally large rogue waves occasionally present in Earth's oceans \citepads{doi:10.1175/JPO-D-15-0017.1}.
Although the existence of such rogue cells is putative, the detection of a faint but significant magnetic field on Betelgeuse by \citetads{2010A&A...516L...2A} indicates the presence of strong convection cells able to generate a local magnetic field through a dynamo effect \citepads{2004A&A...423.1101D,Tessore17b}.
The hot spot could correspond to a rogue cell sufficiently powerful to eject the hot gaseous material up to several times the radius of the star (see also the discussion in Sect.~\ref{massloss}).
The observed trefoil structure of the molecular shell (see also \citeads{2009A&A...504..115K}) suggests that of the order of three rogue cells are currently present on Betelgeuse.
This means that they must be a relatively rare phenomenon, with a few cells (or compact groups of cells) present at the surface of the star at a given time.

\begin{figure}[]
        \centering
        \includegraphics[width=\hsize]{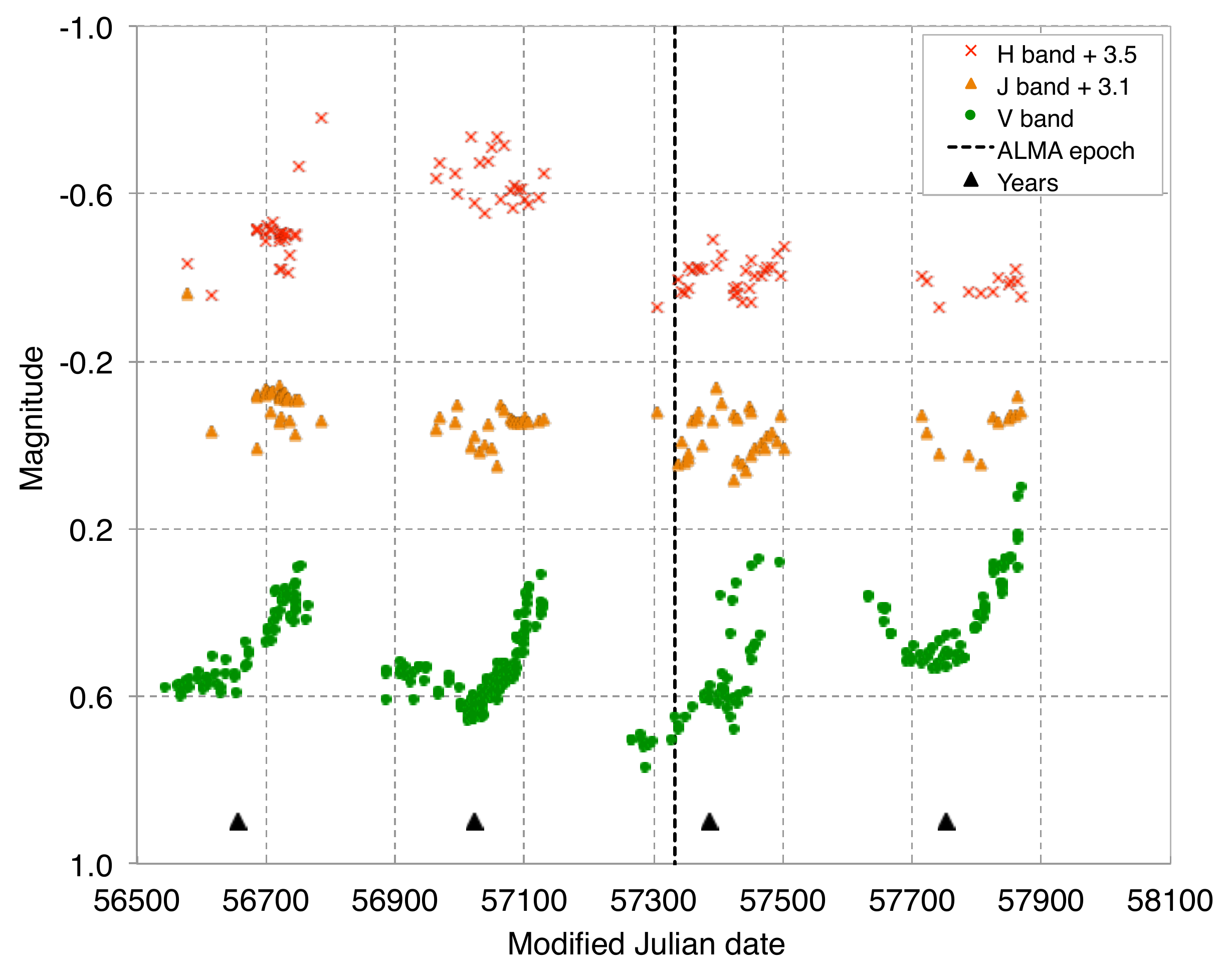}
        \caption{Light curves of Betelgeuse in the $V$, $J$, and $H$ bands from the AAVSO. The epoch of our ALMA observations is marked with a vertical dashed line ($\mathrm{MJD} = 57333$). The $J$ and $H$ light curves are translated respectively by +3.1 and +3.5\,mag for clarity. The black triangles indicate the start of the calendar years 2014, 2015, 2016, and 2017.
        \label{AAVSO-Betelgeuse}}
\end{figure}

\citetads{2016ATel.9503....1W} reported that Betelgeuse has increased its effective temperature over the last 20 years.
According to the AAVSO light curve presented in Fig.~\ref{AAVSO-Betelgeuse}, the star was close to its minimum flux phase in the visible at the time of our ALMA observations; however, it brightened considerably in 2017, reaching its brightest magnitude in the $V$ band in several decades ($m_V \lesssim 0.2$).
This evolution may be due to the development of the ALMA hot spot or to the transit of another rogue convection cell on the stellar disk.
The photometric measurements of Betelgeuse late in its observing season are  affected by high airmass, however, which could bias the photometry.

\subsection{Focused mass loss scenario\label{massloss}}

\begin{figure*}[]
        \sidecaption
        \includegraphics[width=11cm]{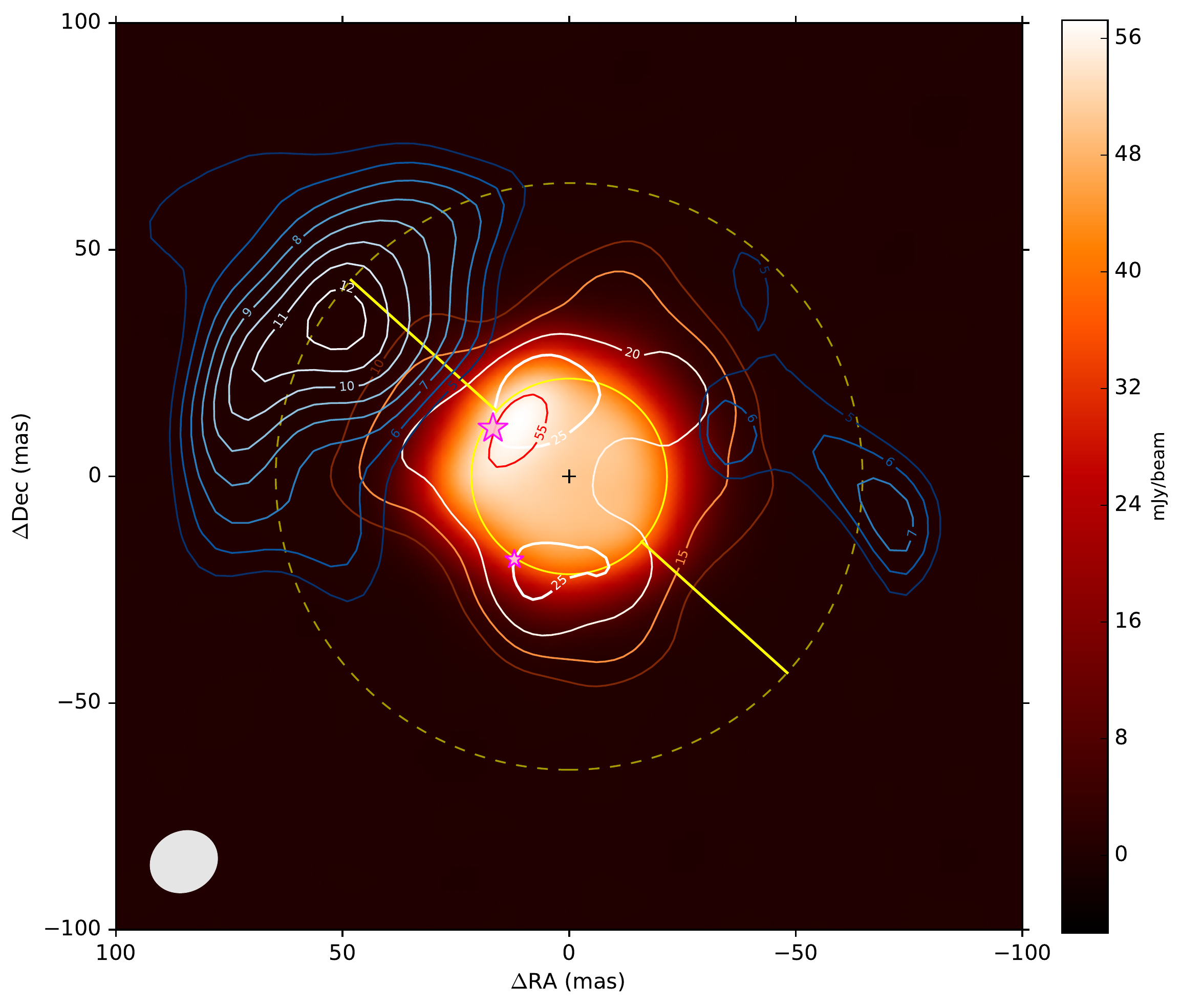}
        \caption{Overview of the structure of the close environment of Betelgeuse.
        The central star image is the ALMA continuum reconstructed image from \citetads{2017A&A...602L..10O} taken in November 2015;  the associated color scale is on the right.
        The red contour level encircles the position of the main continuum hot spot of Betelgeuse (flux level 55 mJy\,beam$^{-1}$).
        The ellipse in the lower left corner of the image shows the ALMA beam size.
        The solid yellow circle marks the size of the infrared photosphere as measured by \citetads{2016A&A...588A.130M} in November 2014, and the two yellow segments represent the direction of the rotation axis.
        The dashed circle is located at 3 times the infrared radius of the star.
        The magenta star symbols show the positions of the two spots of magnetic activity identified in spectropolarimetry by \citetads{2017arXiv170202002T} at epoch 16 October 2015. The size of the star symbol approximates the brightness of each spot.
        The white-orange contours represent the emission amplitude of the $^{28}$SiO($\varv$=2,\,$J$=8-7) line (labels in mJy\,beam$^{-1}$) as determined in Sect.~\ref{molecvelocity} (see also Fig.~\ref{linemaps}).
        The white-blue contours represent the degree of linear polarization measured by \citetads{2016A&A...585A..28K} at epoch 30 March 2015 in the CntH$\alpha$ filter, as a proxy for the presence of dust. The contour labels indicate the degree of linear polarization in percentage from 5 to 12\%.
        \label{mass-ejection}}
\end{figure*}

The principal observed components of the close environment of Betelgeuse are represented in Fig.~\ref{mass-ejection}.
The different observations represented in this figure were all obtained between March and November 2015.
Considering the typical linear scales ($\approx 10$\,au) and velocities ($\approx 5$\,km\,s$^{-1}$) of the environment, the evolution time scale for the close-in envelope of Betelgeuse is of the order of 10\,years.
The morphology of the star's environment is therefore not expected to have evolved significantly between the different observation epochs in 2015.

Several noticeable features are present in the northeast quadrant of the star.
There is a remarkable coincidence of the position angles of the rotation axis of the star, the main hot spot reported by \citetads{2017arXiv170202002T} from spectropolarimetry, the ALMA hot spot observed by \citetads{2017A&A...602L..10O}, and the dust arc detected by \citetads{2016A&A...585A..28K}.
Additionally, the molecular emission map exhibits a strong intensity peak close to the position of the main hot spot.
We represent in Fig.~\ref{mass-ejection} the emission from the $^{28}$SiO($\varv$=2,\,$J$=8-7) line, and this pattern is observed in the amplitude maps of all four emission lines (Figs.~\ref{linemaps}, \ref{12COv0linemaps}, \ref{28SiOv1linemaps}, and \ref{29SiOv0linemaps}).
The scenario that we propose to interpret these converging observations is the ejection of gaseous material from the northeast hot spot.
The hot spot temperature ($T_\mathrm{spot} \approx 3700$\,K;  \citeads{2017A&A...602L..10O}) is identical to the effective temperature of the near-infrared photosphere \citepads{2004A&A...418..675P} located much deeper.
This may indicate that the gas in the hot spot is moving upward rapidly from deeper layers, and does not have time to cool down.
Alternatively, the spot could be a shock interface between the expanding convective cell and surrounding extended atmosphere, or even a local low density gap in the circumstellar gas density allowing to see deeper and hotter layers.
Above the hot spot, the gas can be followed in the northeast quadrant in the ALMA molecular emission maps.
The emission extends along three main directions: in the northeast, south, and northwest quadrants.
This trefoil distribution matches remarkably well the directions of the plumes detected by \citetads{2009A&A...504..115K} using VLT/NACO adaptive optics observations in the near-infrared at epoch 2009.0.
The persistence of these extended plumes over a period of at least 6 years is an indication that they may correspond to long-lived mass loss ejection spots.
Farther out, at a radius of $60-70$\,mas, the ALMA molecular plume unfolds in a partial, light-scattering dust shell detected in visible imaging polarimetry by \citetads{2016A&A...585A..28K}.
The position angle alignment of the ALMA molecular plume and the dust shell, as well as their comparable opening angle, point to an outward moving, radial flow of gas that condenses into dust at a radius of $3\,R_\mathrm{IR}$.
The continuum hot spot has an elongated elliptical geometry, and is located at the limb of the stellar disk. Under the assumption of a radial mass loss flow, this implies that the velocity vector of the ejected material is mostly orthogonal to the line of sight (i.e., within the plane of the sky) and thus has a minimal measurable Doppler component. 
This is consistent with the absence of a specific signature of the region of the continuum hot spot in the radial velocity map (Fig.~\ref{linemaps}), but the confirmation of our focused mass loss scenario will require the simultaneous observation of a continuum hot spot and a clearly coincident radial velocity signature.
From geometrical considerations, such a combination would be more easily observable for a hot spot located near the center of the stellar disk.

The alignment in the northeast quadrant of the dust-forming molecular plume with the direction of the pole of Betelgeuse is particularly intriguing.
The maps presented by \citetads{2017arXiv170202002T} show that a hot spot persisted for almost three years in the northeast quadrant of the disk of Betelgeuse between 2013 and 2016.
The polar region of Betelgeuse therefore appears as a specially active region from which a more intense and continuous mass loss than elsewhere on the stellar disk may occur. 
The hypothesis of an enhanced mass loss at the pole through interactions between the pulsation and rotation was mentioned by \citetads{1998AJ....116.2501U}.
We note that several other regions of the star also host hot spots and molecular mass ejection plumes.

In our ALMA high angular resolution observations, we are sensitive to the hottest and densest gas, and the molecular emission is detected from the surface of the star up to $\approx 50$\,mas ($2.5\times$ the near-infrared photospheric radius).
In a future work, we will investigate the relationship of this inner envelope to larger scale, fainter surface brightness structures using the data obtained on the compact ALMA TC configuration (Sect.~\ref{observations}).
Radio observations of Betelgeuse at centimeter wavelengths with e-MERLIN \citepads{2017AJ....154...11H} showed the presence of spots at larger angular scales than the ALMA structures.

As a side note, the structure of the close-in molecular and dust envelope of Betelgeuse (see also \citeads{2011A&A...531A.117K}) may be relevant for the ongoing research on the environment of supernovae explosions (see, e.g., \citeads{2016MNRAS.458.1253V, 2013MNRAS.433.1745D, 2017arXiv170401697D}).
Finally, we note the coincidence of the position angle of the rotation axis of Betelgeuse with the symmetry axis of its large bow shock and with its velocity vector with respect to the ISM \citepads{1997AJ....114..837N, 2008PASJ...60S.407U, 2012A&A...548A.113D}, which is likely a chance alignment.

\section{Conclusion}

Our ALMA observations show that Betelgeuse is rotating with a projected velocity of $\varv_\mathrm{eq}\,\sin i = \vsiniSolidB \pm \vsiniSoliderrSystB$\ km\,s$^{-1}$ at the equivalent continuum angular radius $R_\mathrm{star} = \rstarB \pm \rstarerrB$\ mas, corresponding to an angular rotation velocity of $\omega\,\sin i = (\omegasini \pm \omegasinierr) \times 10^{-9}$\ rad\,s$^{-1}$.
Observations in the UV domain \citepads{1998AJ....116.2501U} showed that the rotation of the chromosphere at a radius of $\approx 60$\,mas occurs at a similar projected velocity to the value we measure at 30\,mas, and thus at a reduced angular velocity.
This suggests a rotational coupling of the star and its chromosphere up to $\approx 10$\,au or $\approx 1.5\times$ the ALMA continuum radius ($R_\mathrm{star} = 1400 \pm 250\,R_\odot$), possibly of magnetic origin.

As expected for such a very large star, the rotation is slow, but its influence on the mass loss appears significant.
The position angle of the rotation axis coincides with a large hot spot located in the northeast quadrant of the stellar disk detected by \citetads{2017A&A...602L..10O}.
We propose that the principal ALMA continuum hot spot corresponds to a rogue convection cell located close to the pole of Betelgeuse.
It is at the origin of an ongoing focused mass ejection.
The coincidence of the position angle of the hot spot with the polar axis may be a chance alignment.
However, the possible presence of a global magnetic field on Betelgeuse, consistent with the hypothesis of a magnetic rotational coupling of the star and its chromosphere, could have an impact on the local convection efficiency along the magnetic polar axis.
A slight misalignment of the magnetic and rotation axes could periodically hide the north and south magnetic polar caps with a period of half the rotation period, that is, of $\approx 15$ to 20\,years.
Such a scenario could explain the presence of a hot spot in the symmetrical quadrant to the ALMA hot spot in the observations recorded by \citetads{1998AJ....116.2501U} in 1995, twenty years before the present ALMA data.

The combination of the ALMA continuum and line emission maps with the polarimetric imaging in the visible reported by \citetads{2016A&A...585A..28K} allows us to trace the ejection of stellar material up to the condensation of dust.
The ALMA line emission, however, does not show any global, spherically symmetric outflow of molecular gas at the radius of the molecular shell ($\approx 9$\,au).
This suggests that rogue cells are an important ingredient to explain the mass loss of red supergiants.
They seem to be at the origin of focused mass loss plumes, powerful enough to eject material at a sufficiently large distance from the star to allow for the condensation of dust.
The considerable radiation pressure on the formed dust grains can then carry them out of the star's deep gravitational well.
In order to test this scenario, we suggest that  the statistical occurrence of rogue cells in rotating, radiative 3D magneto-hydrodynamical simulations should be studied, possibly including internal differential rotation.

\begin{acknowledgements}
We thank Dr. Malcolm Gray for discussions on the maser properties of the SiO circumstellar emission of Betelgeuse.
LD acknowledges support from the ERC consolidator grant 646758 AEROSOL.
GMH acknowledges support from program HST-AR-\#14566.001 provided by NASA through a grant from the Space Telescope Science Institute, which is operated by the Association of Universities for Research in Astronomy, Inc., under NASA contract NAS 5-26555.
This work was supported by the UK Science and Technology Research Council, under grant number ST/L000768/1.
This project has received funding from the European Union's Horizon 2020 research and innovation program under the Marie Sk\l{}odowska-Curie Grant agreement No. 665501 with the research Foundation Flanders (FWO) ([PEGASUS]$^2$ Marie Curie fellowship 12U2717N awarded to MM).
KO acknowledges the grant from the Universidad Cat{\'o}lica del Norte.
This paper makes use of the following ALMA data:
ADS/JAO.ALMA\#2015.1.00206.S. ALMA is a partnership of ESO (representing
its member states), NSF (USA) and NINS (Japan), together with NRC
(Canada), MOST and ASIAA (Taiwan), and KASI (Republic of Korea), in
cooperation with the Republic of Chile. The Joint ALMA Observatory is
operated by ESO, AUI/NRAO and NAOJ.
We acknowledge with thanks the variable star observations from the AAVSO International Database contributed by observers worldwide and used in this research.
We used the SIMBAD and VIZIER databases at the CDS, Strasbourg (France), and NASA's Astrophysics Data System Bibliographic Services.
This research made use of Astropy\footnote{Available at \url{http://www.astropy.org/}}, a community-developed core Python package for Astronomy \citepads{2013A&A...558A..33A}.
\end{acknowledgements}

\bibliographystyle{aa} 
\bibliography{biblioBetelgeuse}

\begin{thebibliography}{58}
\expandafter\ifx\csname natexlab\endcsname\relax\def\natexlab#1{#1}\fi

\bibitem[{Archinal {et~al.}(2011)Archinal, A'Hearn, Bowell, Conrad,
  Consolmagno, Courtin, Fukushima, Hestroffer, Hilton, Krasinsky, Neumann,
  Oberst, Seidelmann, Stooke, Tholen, Thomas, \& Williams}]{Archinal2011}
Archinal, B.~A., A'Hearn, M.~F., Bowell, E., {et~al.} 2011, Celestial Mechanics
  and Dynamical Astronomy, 109, 101

\bibitem[{{Astropy Collaboration} {et~al.}(2013){Astropy Collaboration},
  {Robitaille}, {Tollerud}, {Greenfield}, {Droettboom}, {Bray}, {Aldcroft},
  {Davis}, {Ginsburg}, {Price-Whelan}, {Kerzendorf}, {Conley}, {Crighton},
  {Barbary}, {Muna}, {Ferguson}, {Grollier}, {Parikh}, {Nair}, {Unther},
  {Deil}, {Woillez}, {Conseil}, {Kramer}, {Turner}, {Singer}, {Fox}, {Weaver},
  {Zabalza}, {Edwards}, {Azalee Bostroem}, {Burke}, {Casey}, {Crawford},
  {Dencheva}, {Ely}, {Jenness}, {Labrie}, {Lim}, {Pierfederici}, {Pontzen},
  {Ptak}, {Refsdal}, {Servillat}, \& {Streicher}}]{2013A&A...558A..33A}
{Astropy Collaboration}, {Robitaille}, T.~P., {Tollerud}, E.~J., {et~al.} 2013,
  \aap, 558, A33

\bibitem[{{Auri{\`e}re} {et~al.}(2010){Auri{\`e}re}, {Donati},
  {Konstantinova-Antova}, {Perrin}, {Petit}, \&
  {Roudier}}]{2010A&A...516L...2A}
{Auri{\`e}re}, M., {Donati}, J.-F., {Konstantinova-Antova}, R., {et~al.} 2010,
  \aap, 516, L2

\bibitem[{{Auri{\`e}re} {et~al.}(2016){Auri{\`e}re}, {L{\'o}pez Ariste},
  {Mathias}, {L{\`e}bre}, {Josselin}, {Montarg{\`e}s}, {Petit}, {Chiavassa},
  {Paletou}, {Fabas}, {Konstantinova-Antova}, {Donati}, {Grunhut}, {Wade},
  {Herpin}, {Kervella}, {Perrin}, \& {Tessore}}]{2016A&A...591A.119A}
{Auri{\`e}re}, M., {L{\'o}pez Ariste}, A., {Mathias}, P., {et~al.} 2016, \aap,
  591, A119

\bibitem[{Benetazzo {et~al.}(2015)Benetazzo, Barbariol, Bergamasco, Torsello,
  Carniel, \& Sclavo}]{doi:10.1175/JPO-D-15-0017.1}
Benetazzo, A., Barbariol, F., Bergamasco, F., {et~al.} 2015, Journal of
  Physical Oceanography, 45, 2261

\bibitem[{{Burns} {et~al.}(1997){Burns}, {Baldwin}, {Boysen}, {Haniff},
  {Lawson}, {Mackay}, {Rogers}, {Scott}, {Warner}, {Wilson}, \&
  {Young}}]{1997MNRAS.290L..11B}
{Burns}, D., {Baldwin}, J.~E., {Boysen}, R.~C., {et~al.} 1997, \mnras, 290, L11

\bibitem[{{Chiavassa} {et~al.}(2011){Chiavassa}, {Freytag}, {Masseron}, \&
  {Plez}}]{2011A&A...535A..22C}
{Chiavassa}, A., {Freytag}, B., {Masseron}, T., \& {Plez}, B. 2011, \aap, 535,
  A22

\bibitem[{{Chiavassa} {et~al.}(2010){Chiavassa}, {Haubois}, {Young}, {Plez},
  {Josselin}, {Perrin}, \& {Freytag}}]{2010A&A...515A..12C}
{Chiavassa}, A., {Haubois}, X., {Young}, J.~S., {et~al.} 2010, \aap, 515, A12

\bibitem[{{Decin} {et~al.}(2012){Decin}, {Cox}, {Royer}, {Van Marle},
  {Vandenbussche}, {Ladjal}, {Kerschbaum}, {Ottensamer}, {Barlow}, {Blommaert},
  {Gomez}, {Groenewegen}, {Lim}, {Swinyard}, {Waelkens}, \&
  {Tielens}}]{2012A&A...548A.113D}
{Decin}, L., {Cox}, N.~L.~J., {Royer}, P., {et~al.} 2012, \aap, 548, A113

\bibitem[{{Dessart} {et~al.}(2017){Dessart}, {Hillier}, \&
  {Audit}}]{2017arXiv170401697D}
{Dessart}, L., {Hillier}, D.~J., \& {Audit}, E. 2017, ArXiv e-prints
  [\eprint[arXiv]{1704.01697}]

\bibitem[{{Dessart} {et~al.}(2013){Dessart}, {Hillier}, {Waldman}, \&
  {Livne}}]{2013MNRAS.433.1745D}
{Dessart}, L., {Hillier}, D.~J., {Waldman}, R., \& {Livne}, E. 2013, \mnras,
  433, 1745

\bibitem[{{Dolan} {et~al.}(2016){Dolan}, {Mathews}, {Lam}, {Quynh Lan},
  {Herczeg}, \& {Dearborn}}]{2016ApJ...819....7D}
{Dolan}, M.~M., {Mathews}, G.~J., {Lam}, D.~D., {et~al.} 2016, \apj, 819, 7

\bibitem[{{Domiciano de Souza} {et~al.}(2004){Domiciano de Souza}, {Zorec},
  {Jankov}, {Vakili}, {Abe}, \& {Janot-Pacheco}}]{2004A&A...418..781D}
{Domiciano de Souza}, A., {Zorec}, J., {Jankov}, S., {et~al.} 2004, \aap, 418,
  781

\bibitem[{{Dorch}(2004)}]{2004A&A...423.1101D}
{Dorch}, S.~B.~F. 2004, \aap, 423, 1101

\bibitem[{{Dupree} \& {Stefanik}(2013)}]{2013EAS....60...77D}
{Dupree}, A.~K. \& {Stefanik}, R.~P. 2013, in EAS Publications Series, Vol.~60,
  EAS Publications Series, ed. P.~{Kervella}, T.~{Le Bertre}, \& G.~{Perrin},
  77--84

\bibitem[{{Ekstr{\"o}m} {et~al.}(2012){Ekstr{\"o}m}, {Georgy}, {Eggenberger},
  {Meynet}, {Mowlavi}, {Wyttenbach}, {Granada}, {Decressin}, {Hirschi},
  {Frischknecht}, {Charbonnel}, \& {Maeder}}]{2012A&A...537A.146E}
{Ekstr{\"o}m}, S., {Georgy}, C., {Eggenberger}, P., {et~al.} 2012, \aap, 537,
  A146

\bibitem[{Endres {et~al.}(2016)Endres, Schlemmer, Schilke, Stutzki, \&
  M{\"o}ller}]{ENDRES201695}
Endres, C.~P., Schlemmer, S., Schilke, P., Stutzki, J., \& M{\"o}ller, H.~S.
  2016, Journal of Molecular Spectroscopy, 327, 95

\bibitem[{{Gilliland} \& {Dupree}(1996)}]{1996ApJ...463L..29G}
{Gilliland}, R.~L. \& {Dupree}, A.~K. 1996, \apjl, 463, L29

\bibitem[{{Gray}(2000)}]{2000ApJ...532..487G}
{Gray}, D.~F. 2000, \apj, 532, 487

\bibitem[{{Gray} {et~al.}(2009){Gray}, {Wittkowski}, {Scholz}, {Humphreys},
  {Ohnaka}, \& {Boboltz}}]{2009MNRAS.394...51G}
{Gray}, M.~D., {Wittkowski}, M., {Scholz}, M., {et~al.} 2009, \mnras, 394, 51

\bibitem[{{Harper} \& {Brown}(2006)}]{2006ApJ...646.1179H}
{Harper}, G.~M. \& {Brown}, A. 2006, \apj, 646, 1179

\bibitem[{{Harper} {et~al.}(2017{\natexlab{a}}){Harper}, {Brown}, {Guinan},
  {O'Gorman}, {Richards}, {Kervella}, \& {Decin}}]{2017AJ....154...11H}
{Harper}, G.~M., {Brown}, A., {Guinan}, E.~F., {et~al.} 2017{\natexlab{a}},
  \aj, 154, 11

\bibitem[{{Harper} {et~al.}(2017{\natexlab{b}}){Harper}, {DeWitt}, {Richter},
  {Greathouse}, {Ryde}, {Guinan}, {O'Gorman}, \& {Vacca}}]{2017ApJ...836...22H}
{Harper}, G.~M., {DeWitt}, C., {Richter}, M.~J., {et~al.} 2017{\natexlab{b}},
  \apj, 836, 22

\bibitem[{{Haubois} {et~al.}(2006){Haubois}, {Perrin}, {Lacour}, {Schuller},
  {Monnier}, {Berger}, {Ridgway}, {Millan-Gabet}, {Pedretti}, \&
  {Traub}}]{2006sf2a.conf..471H}
{Haubois}, X., {Perrin}, G., {Lacour}, S., {et~al.} 2006, in SF2A-2006: Semaine
  de l'Astrophysique Francaise, ed. D.~{Barret}, F.~{Casoli}, G.~{Lagache},
  A.~{Lecavelier}, \& L.~{Pagani}, 471

\bibitem[{{Haubois} {et~al.}(2009){Haubois}, {Perrin}, {Lacour}, {Verhoelst},
  {Meimon}, {Mugnier}, {Thi{\'e}baut}, {Berger}, {Ridgway}, {Monnier},
  {Millan-Gabet}, \& {Traub}}]{2009A&A...508..923H}
{Haubois}, X., {Perrin}, G., {Lacour}, S., {et~al.} 2009, \aap, 508, 923

\bibitem[{Jones {et~al.}(2001)Jones, Oliphant, Peterson,
  {et~al.}}]{Jones:2001aa}
Jones, E., Oliphant, T., Peterson, P., {et~al.} 2001, {SciPy}: Open source
  scientific tools for {Python}, \url{http://www.scipy.org/}

\bibitem[{{Kervella} {et~al.}(2016){Kervella}, {Lagadec}, {Montarg{\`e}s},
  {Ridgway}, {Chiavassa}, {Haubois}, {Schmid}, {Langlois}, {Gallenne}, \&
  {Perrin}}]{2016A&A...585A..28K}
{Kervella}, P., {Lagadec}, E., {Montarg{\`e}s}, M., {et~al.} 2016, \aap, 585,
  A28

\bibitem[{{Kervella} {et~al.}(2011){Kervella}, {Perrin}, {Chiavassa},
  {Ridgway}, {Cami}, {Haubois}, \& {Verhoelst}}]{2011A&A...531A.117K}
{Kervella}, P., {Perrin}, G., {Chiavassa}, A., {et~al.} 2011, \aap, 531, A117

\bibitem[{{Kervella} {et~al.}(2009){Kervella}, {Verhoelst}, {Ridgway},
  {Perrin}, {Lacour}, {Cami}, \& {Haubois}}]{2009A&A...504..115K}
{Kervella}, P., {Verhoelst}, T., {Ridgway}, S.~T., {et~al.} 2009, \aap, 504,
  115

\bibitem[{{Lobel} \& {Dupree}(2000)}]{2000ApJ...545..454L}
{Lobel}, A. \& {Dupree}, A.~K. 2000, \apj, 545, 454

\bibitem[{{Lobel} \& {Dupree}(2001)}]{2001ApJ...558..815L}
{Lobel}, A. \& {Dupree}, A.~K. 2001, \apj, 558, 815

\bibitem[{{Meynet} {et~al.}(2013){Meynet}, {Haemmerl{\'e}}, {Ekstr{\"o}m},
  {Georgy}, {Groh}, \& {Maeder}}]{2013EAS....60...17M}
{Meynet}, G., {Haemmerl{\'e}}, L., {Ekstr{\"o}m}, S., {et~al.} 2013, in EAS
  Publications Series, Vol.~60, EAS Publications Series, ed. P.~{Kervella},
  T.~{Le Bertre}, \& G.~{Perrin}, 17--28

\bibitem[{{Montarg{\`e}s} {et~al.}(2016){Montarg{\`e}s}, {Kervella}, {Perrin},
  {Chiavassa}, {Le Bouquin}, {Auri{\`e}re}, {L{\'o}pez Ariste}, {Mathias},
  {Ridgway}, {Lacour}, {Haubois}, \& {Berger}}]{2016A&A...588A.130M}
{Montarg{\`e}s}, M., {Kervella}, P., {Perrin}, G., {et~al.} 2016, \aap, 588,
  A130

\bibitem[{{Montarg{\`e}s} {et~al.}(2014){Montarg{\`e}s}, {Kervella}, {Perrin},
  {Ohnaka}, {Chiavassa}, {Ridgway}, \& {Lacour}}]{2014A&A...572A..17M}
{Montarg{\`e}s}, M., {Kervella}, P., {Perrin}, G., {et~al.} 2014, \aap, 572,
  A17

\bibitem[{M{\"u}ller {et~al.}(2005)M{\"u}ller, Schl{\"a}der, Stutzki, \&
  Winnewisser}]{MULLER2005215}
M{\"u}ller, H.~S., Schl{\"a}der, F., Stutzki, J., \& Winnewisser, G. 2005,
  Journal of Molecular Structure, 742, 215

\bibitem[{{Neilson} {et~al.}(2016){Neilson}, {Baron}, {Norris}, {Kloppenborg},
  \& {Lester}}]{2016ApJ...830..103N}
{Neilson}, H.~R., {Baron}, F., {Norris}, R., {Kloppenborg}, B., \& {Lester},
  J.~B. 2016, \apj, 830, 103

\bibitem[{{Neilson} {et~al.}(2011){Neilson}, {Lester}, \&
  {Haubois}}]{2011ASPC..451..117N}
{Neilson}, H.~R., {Lester}, J.~B., \& {Haubois}, X. 2011, in Astronomical
  Society of the Pacific Conference Series, Vol. 451, 9th Pacific Rim
  Conference on Stellar Astrophysics, ed. S.~{Qain}, K.~{Leung}, L.~{Zhu}, \&
  S.~{Kwok}, 117

\bibitem[{{Noriega-Crespo} {et~al.}(1997){Noriega-Crespo}, {van Buren}, {Cao},
  \& {Dgani}}]{1997AJ....114..837N}
{Noriega-Crespo}, A., {van Buren}, D., {Cao}, Y., \& {Dgani}, R. 1997, \aj,
  114, 837

\bibitem[{{O'Gorman} {et~al.}(2015){O'Gorman}, {Harper}, {Brown}, {Guinan},
  {Richards}, {Vlemmings}, \& {Wasatonic}}]{2015A&A...580A.101O}
{O'Gorman}, E., {Harper}, G.~M., {Brown}, A., {et~al.} 2015, \aap, 580, A101

\bibitem[{{O'Gorman} {et~al.}(2012){O'Gorman}, {Harper}, {Brown}, {Brown},
  {Redfield}, {Richter}, \& {Requena-Torres}}]{2012AJ....144...36O}
{O'Gorman}, E., {Harper}, G.~M., {Brown}, J.~M., {et~al.} 2012, \aj, 144, 36

\bibitem[{{O'Gorman} {et~al.}(2017){O'Gorman}, {Kervella}, {Harper},
  {Richards}, {Decin}, {Montarg{\`e}s}, \& {McDonald}}]{2017A&A...602L..10O}
{O'Gorman}, E., {Kervella}, P., {Harper}, G.~M., {et~al.} 2017, \aap, 602, L10

\bibitem[{{Ohnaka} {et~al.}(2009){Ohnaka}, {Hofmann}, {Benisty}, {Chelli},
  {Driebe}, {Millour}, {Petrov}, {Schertl}, {Stee}, {Vakili}, \&
  {Weigelt}}]{2009A&A...503..183O}
{Ohnaka}, K., {Hofmann}, K.-H., {Benisty}, M., {et~al.} 2009, \aap, 503, 183

\bibitem[{{Ohnaka} {et~al.}(2013){Ohnaka}, {Hofmann}, {Schertl}, {Weigelt},
  {Baffa}, {Chelli}, {Petrov}, \& {Robbe-Dubois}}]{2013A&A...555A..24O}
{Ohnaka}, K., {Hofmann}, K.-H., {Schertl}, D., {et~al.} 2013, \aap, 555, A24

\bibitem[{{Ohnaka} {et~al.}(2017){Ohnaka}, {Weigelt}, \&
  {Hofmann}}]{2017Natur.548..310O}
{Ohnaka}, K., {Weigelt}, G., \& {Hofmann}, K.-H. 2017, \nat, 548, 310

\bibitem[{{Ohnaka} {et~al.}(2011){Ohnaka}, {Weigelt}, {Millour}, {Hofmann},
  {Driebe}, {Schertl}, {Chelli}, {Massi}, {Petrov}, \&
  {Stee}}]{2011A&A...529A.163O}
{Ohnaka}, K., {Weigelt}, G., {Millour}, F., {et~al.} 2011, \aap, 529, A163

\bibitem[{{Perrin} {et~al.}(2004){Perrin}, {Ridgway}, {Coud{\'e} du Foresto},
  {Mennesson}, {Traub}, \& {Lacasse}}]{2004A&A...418..675P}
{Perrin}, G., {Ridgway}, S.~T., {Coud{\'e} du Foresto}, V., {et~al.} 2004,
  \aap, 418, 675

\bibitem[{{Ravi} {et~al.}(2011){Ravi}, {Wishnow}, {Townes}, {Lockwood},
  {Mistry}, \& {Tatebe}}]{2011ApJ...740...24R}
{Ravi}, V., {Wishnow}, E.~H., {Townes}, C.~H., {et~al.} 2011, \apj, 740, 24

\bibitem[{{Ryde} {et~al.}(2006){Ryde}, {Harper}, {Richter}, {Greathouse}, \&
  {Lacy}}]{2006ApJ...637.1040R}
{Ryde}, N., {Harper}, G.~M., {Richter}, M.~J., {Greathouse}, T.~K., \& {Lacy},
  J.~H. 2006, \apj, 637, 1040

\bibitem[{{Tessore} {et~al.}(2017{\natexlab{a}}){Tessore}, A., {Morin},
  {Mathias}, {Josselin}, \& {Auri{\`e}re}}]{Tessore17b}
{Tessore}, B., A., L., {Morin}, J., {et~al.} 2017{\natexlab{a}}, A\&A, 603,
  A129

\bibitem[{{Tessore} {et~al.}(2017{\natexlab{b}}){Tessore}, {L{\`o}pez-Ariste},
  {Mathias}, {L{\`e}bre}, {Morin}, \& {Josselin}}]{2017arXiv170202002T}
{Tessore}, B., {L{\`o}pez-Ariste}, A., {Mathias}, P., {et~al.}
  2017{\natexlab{b}}, ArXiv e-prints [\eprint[arXiv]{1702.02002}]

\bibitem[{{Tsuji}(2000)}]{2000ApJ...540L..99T}
{Tsuji}, T. 2000, \apjl, 540, L99

\bibitem[{{Ueta} {et~al.}(2008){Ueta}, {Izumiura}, {Yamamura}, {Nakada},
  {Matsuura}, {Ita}, {Tanab{\'e}}, {Fukushi}, {Matsunaga}, \&
  {Mito}}]{2008PASJ...60S.407U}
{Ueta}, T., {Izumiura}, H., {Yamamura}, I., {et~al.} 2008, \pasj, 60, S407

\bibitem[{{Uitenbroek} {et~al.}(1998){Uitenbroek}, {Dupree}, \&
  {Gilliland}}]{1998AJ....116.2501U}
{Uitenbroek}, H., {Dupree}, A.~K., \& {Gilliland}, R.~L. 1998, \aj, 116, 2501

\bibitem[{{Vlasis} {et~al.}(2016){Vlasis}, {Dessart}, \&
  {Audit}}]{2016MNRAS.458.1253V}
{Vlasis}, A., {Dessart}, L., \& {Audit}, E. 2016, \mnras, 458, 1253

\bibitem[{{Wasatonic} \& {Guinan}(2016)}]{2016ATel.9503....1W}
{Wasatonic}, R.~P. \& {Guinan}, E.~F. 2016, The Astronomer's Telegram, 9503

\bibitem[{{Wheeler} {et~al.}(2017){Wheeler}, {Nance}, {Diaz}, {Smith},
  {Hickey}, {Zhou}, {Koutoulaki}, {Sullivan}, \&
  {Fowler}}]{2017MNRAS.465.2654W}
{Wheeler}, J.~C., {Nance}, S., {Diaz}, M., {et~al.} 2017, \mnras, 465, 2654

\bibitem[{{Wong} {et~al.}(2016){Wong}, {Kami{\'n}ski}, {Menten}, \&
  {Wyrowski}}]{2016A&A...590A.127W}
{Wong}, K.~T., {Kami{\'n}ski}, T., {Menten}, K.~M., \& {Wyrowski}, F. 2016,
  \aap, 590, A127

\bibitem[{{Young} {et~al.}(2000){Young}, {Baldwin}, {Boysen}, {Haniff},
  {Lawson}, {Mackay}, {Pearson}, {Rogers}, {St.-Jacques}, {Warner}, {Wilson},
  \& {Wilson}}]{2000MNRAS.315..635Y}
{Young}, J.~S., {Baldwin}, J.~E., {Boysen}, R.~C., {et~al.} 2000, \mnras, 315,
  635

\end{thebibliography}

\begin{appendix}

\section{$^{12}$CO($\varv$=0,\,$J$=3-2) \label{12COv0line}}

\begin{figure*}[ht]
        \centering
        \includegraphics[height=6cm]{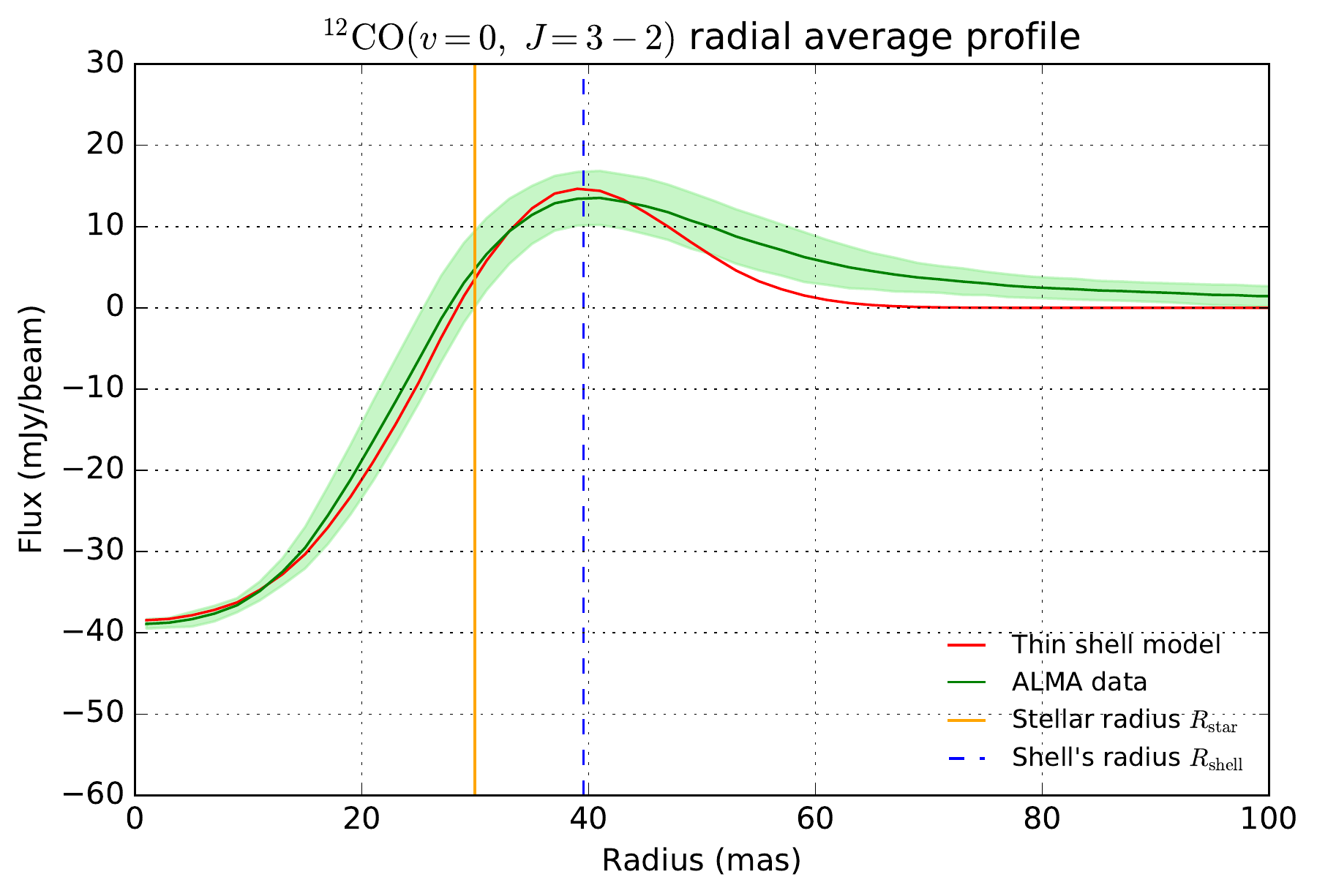}
        \includegraphics[height=6cm]{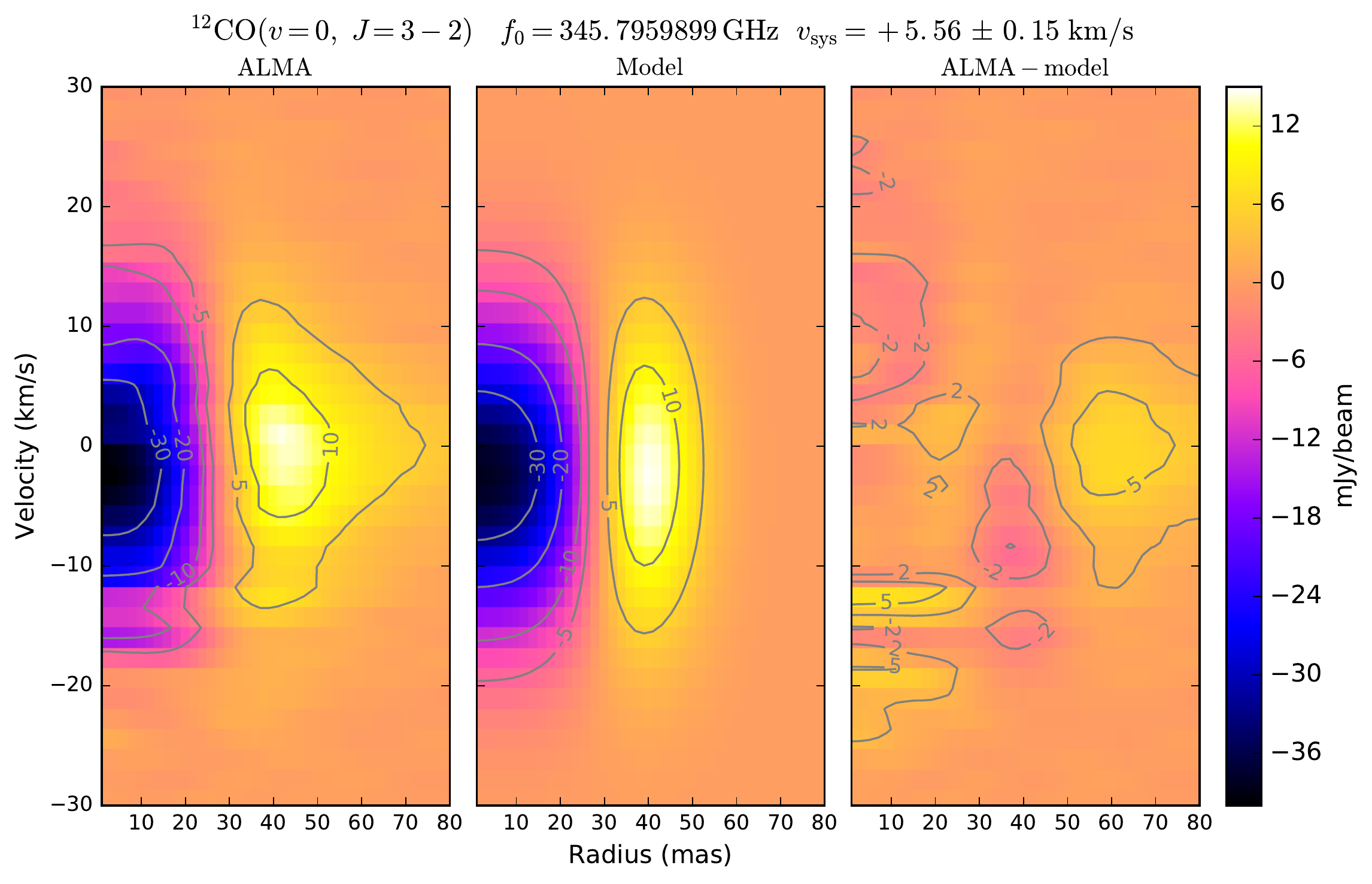}
        \caption{{\it Left panel:} Radial average profile of the ALMA continuum subtracted observation of the $^{12}$CO($\varv$=0,\,$J$=3-2) line (green curve) and best fit model (red curve).
        The shaded green area is the standard deviation of the shell over the considered ring radius.
        The equivalent uniform disk radius of the star is shown with a solid orange line, and the radius of the shell is represented with a dashed blue line.
        {\it Right panels:} Radial average profile as a function of the velocity offset (left panel), best fit model (center panel), and residuals of their subtraction (right panel).
        \label{profileresiduals12COv0}}
\end{figure*}

\begin{figure*}[h]
        \centering
        \includegraphics[width=6cm]{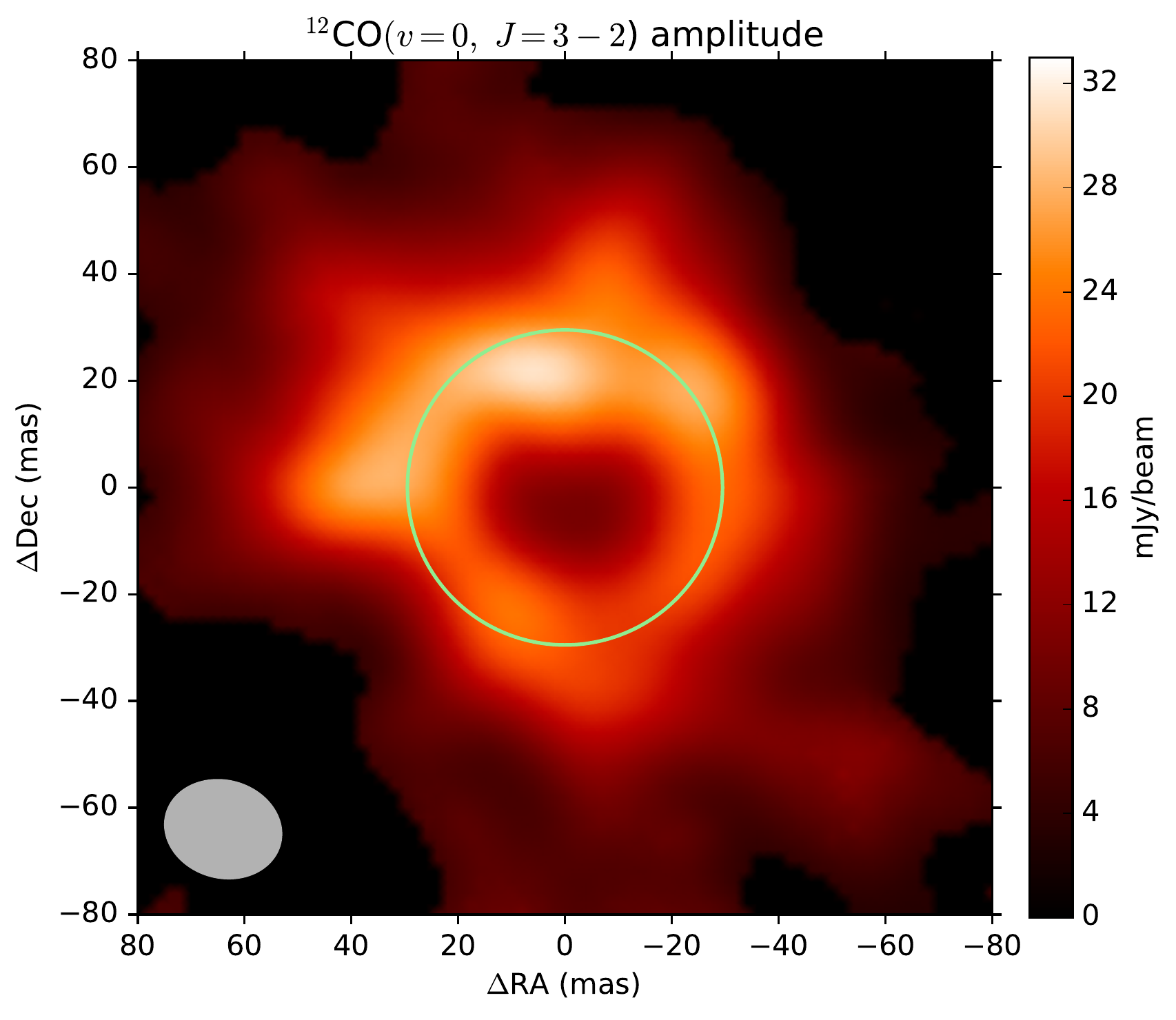}
        \includegraphics[width=6cm]{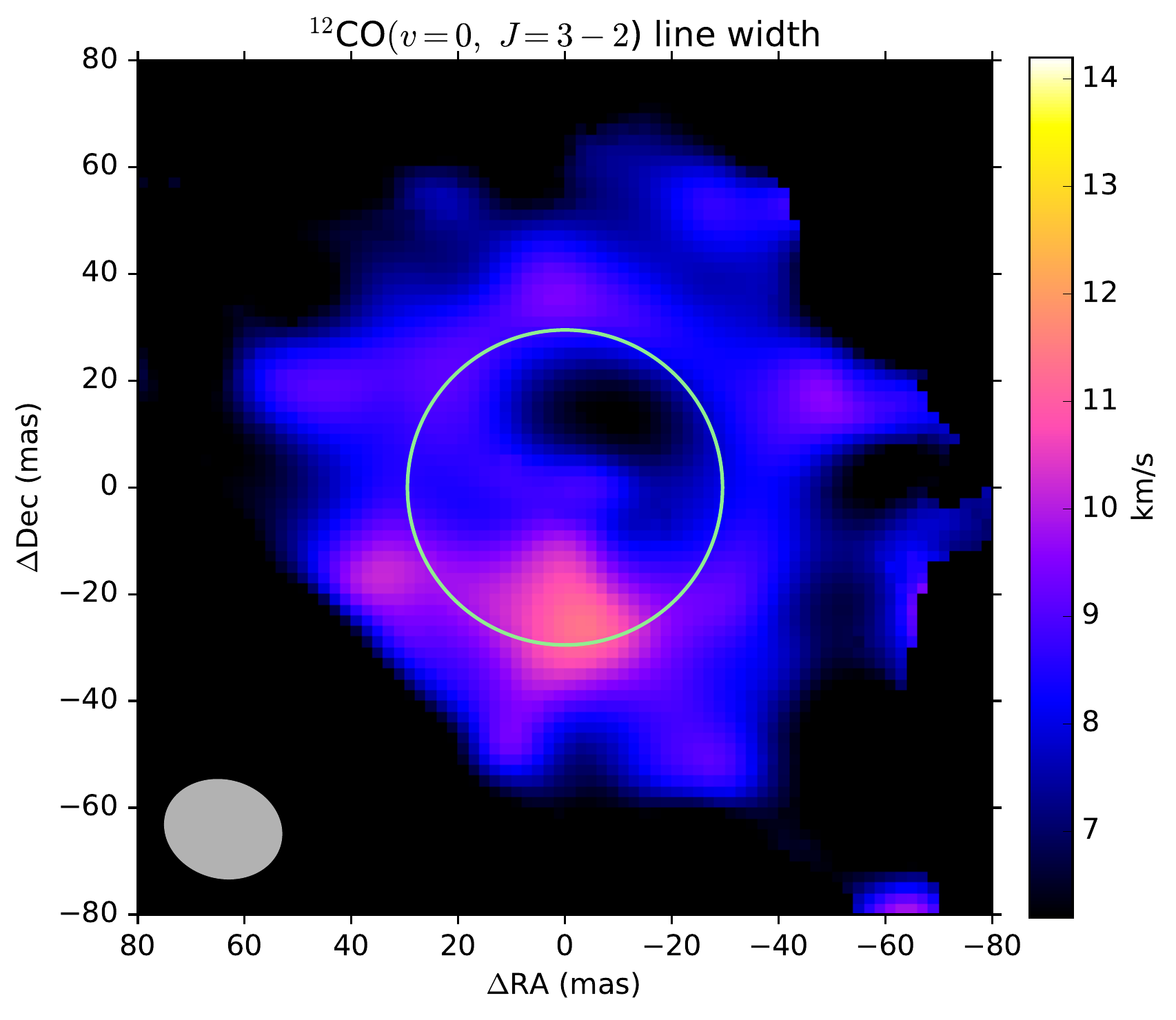}
        \includegraphics[width=6cm]{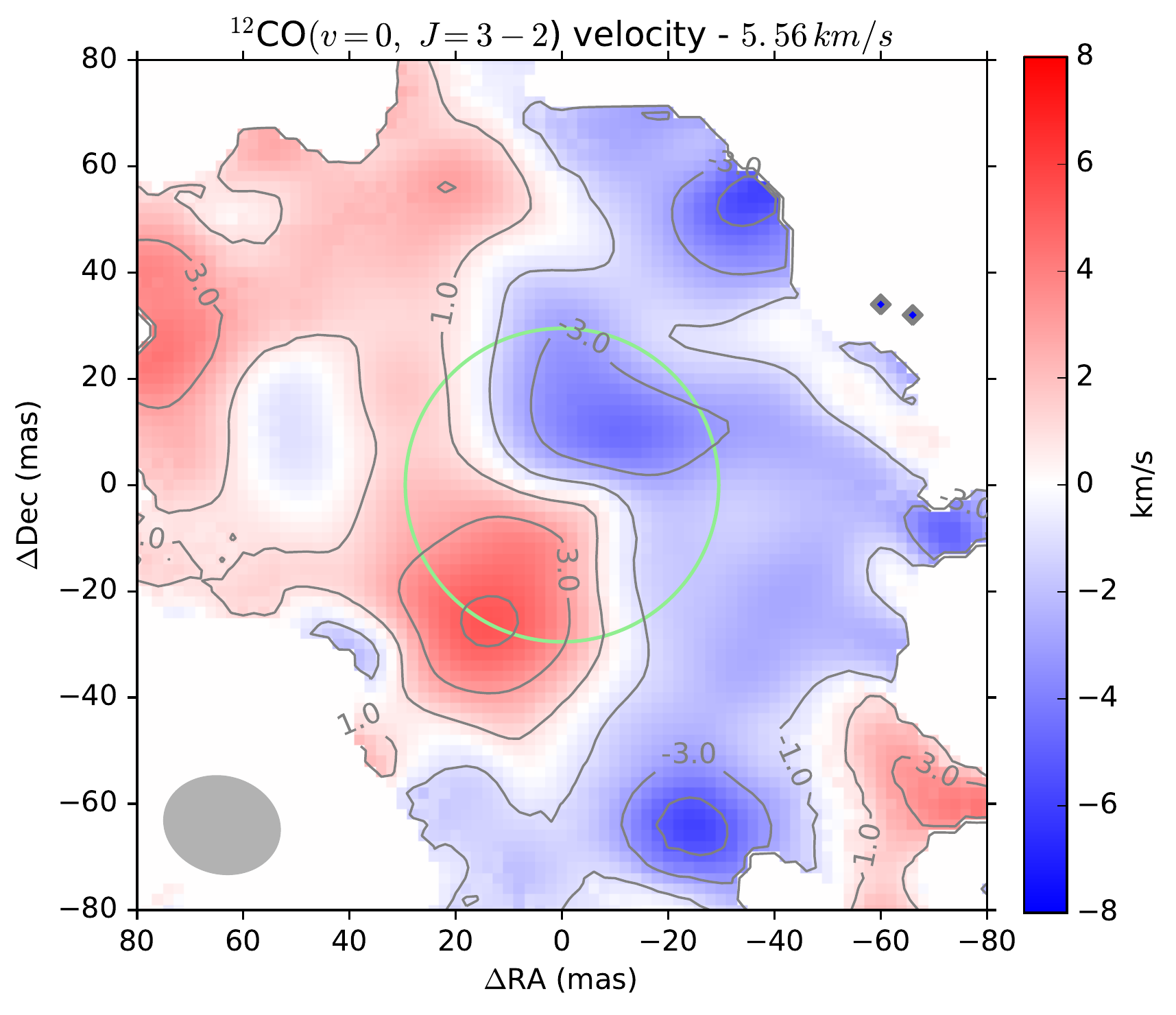}
        \includegraphics[height=5.8cm]{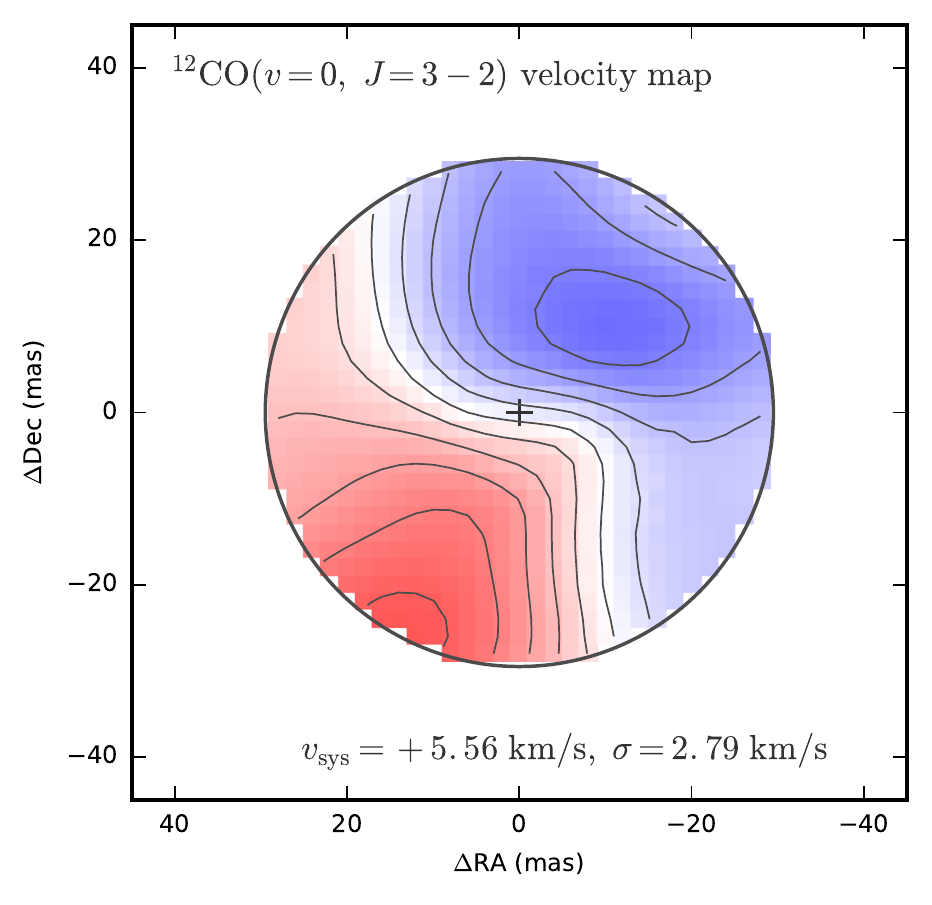}
        \includegraphics[height=5.8cm]{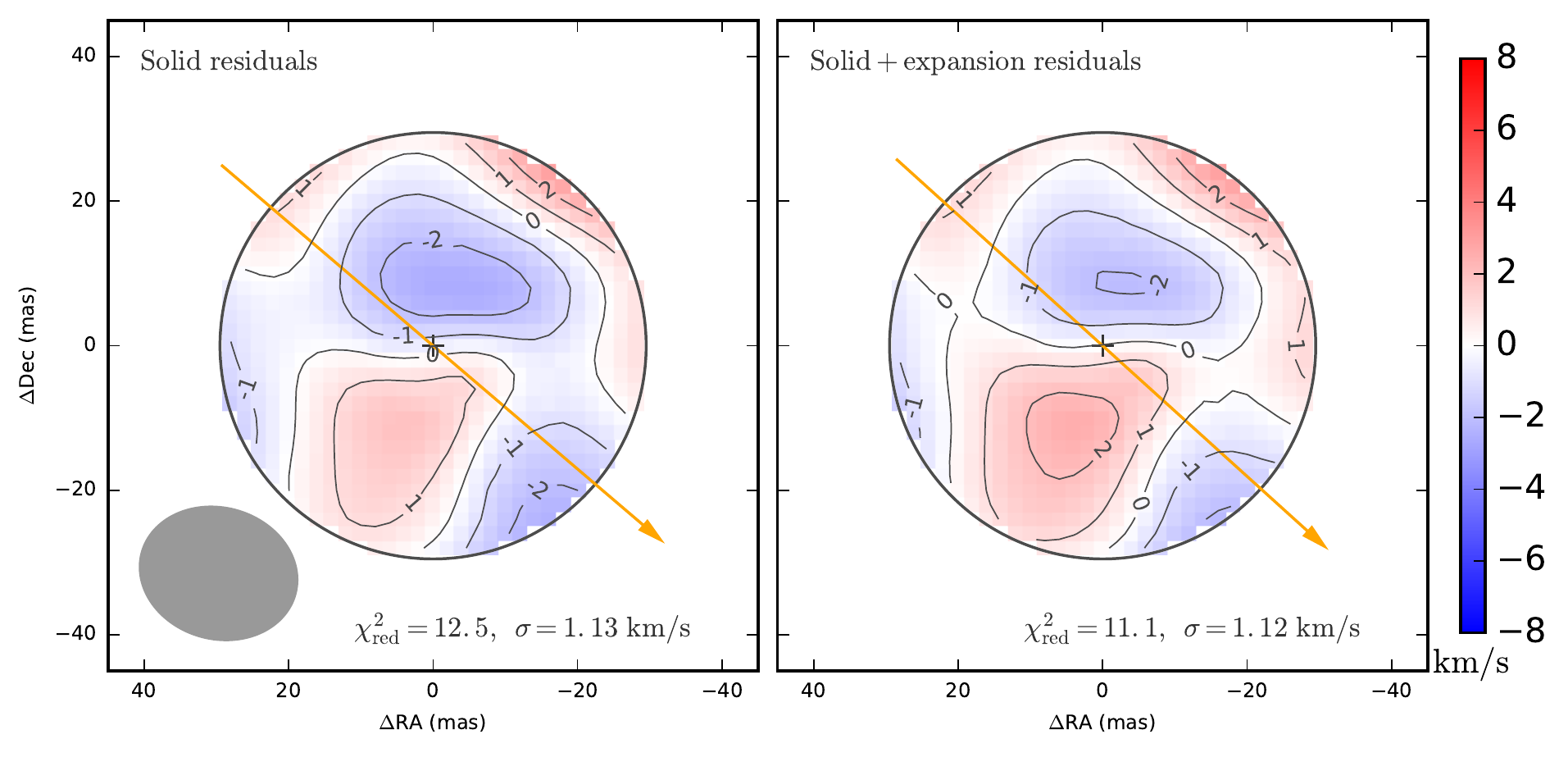}
        \caption{{\it Top row:} Emission line profile maps of Betelgeuse for the $^{12}$CO($\varv=0,\,J$=3-2) line, computed from the absorption corrected data cube (see Sect.~\ref{shellmodel}).
        The top left panel is the map of the amplitude of the best fit Gaussian, the top middle panel the width of the emission line, and the top right panel the velocity.
        The systemic velocity determined from the line profile fitting is subtracted from the velocity map.
        The equivalent uniform disk size of the continuum emission is represented with a green circle, and the beam is shown in the lower left corner.
        {\it Bottom row:} Velocity map represented over the equivalent continuum disk of the star (left panel).
        The residuals of the fit are plotted in the center and right panels.
        The ALMA beam is shown in the lower left corner of the bottom center panel, and the contour levels are separated by 1\,km\,s$^{-1}$ intervals.
        The polar axis is represented with an arrow pointing toward the angular momentum vector for a right-handed rotating coordinate system.
        \label{12COv0linemaps}}
\end{figure*}

\section{$^{28}$SiO($\varv$=1,\,$J$=8-7) \label{28SiOv1line}}

\begin{figure*}[ht]
        \centering
        \includegraphics[height=6cm]{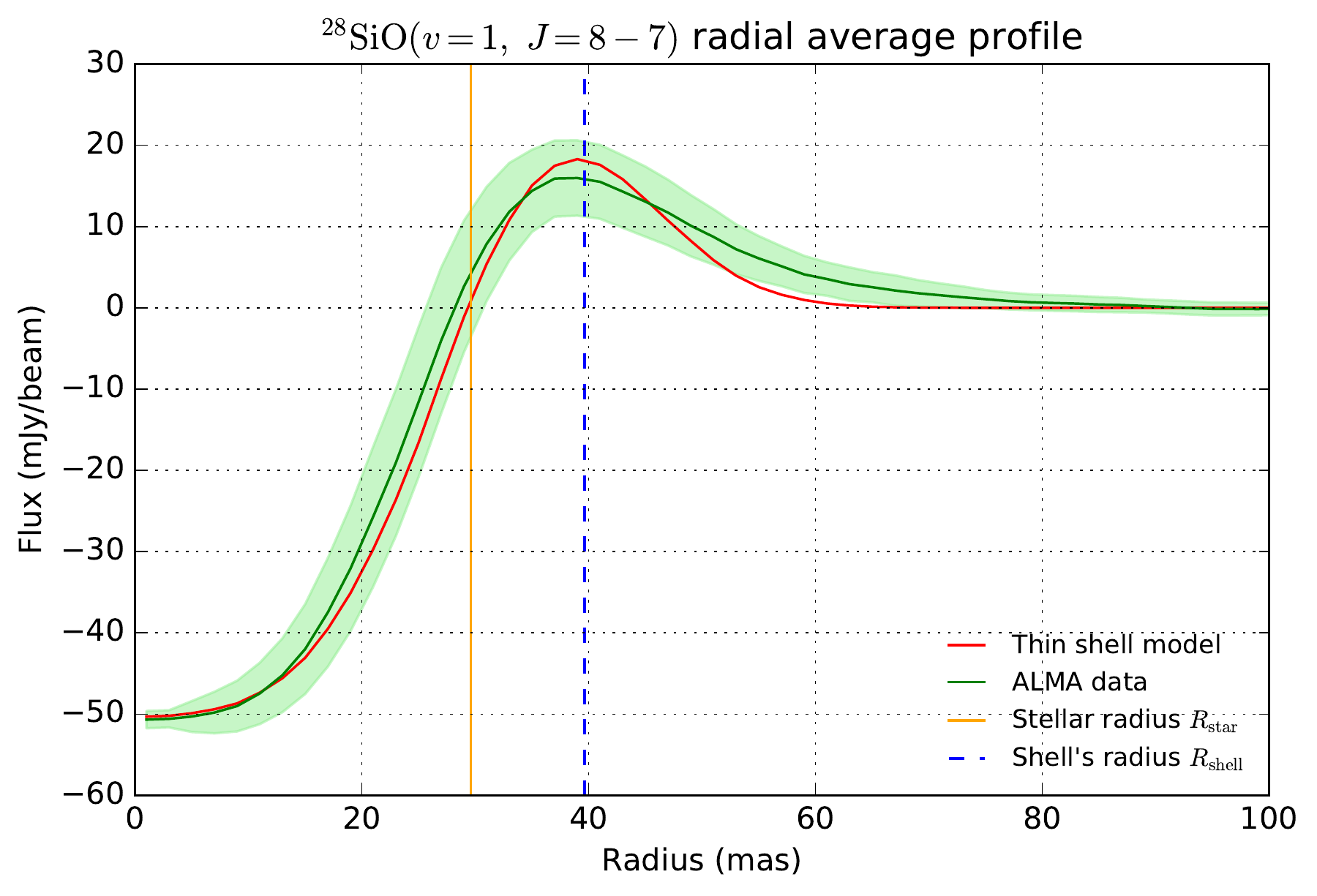}
        \includegraphics[height=6cm]{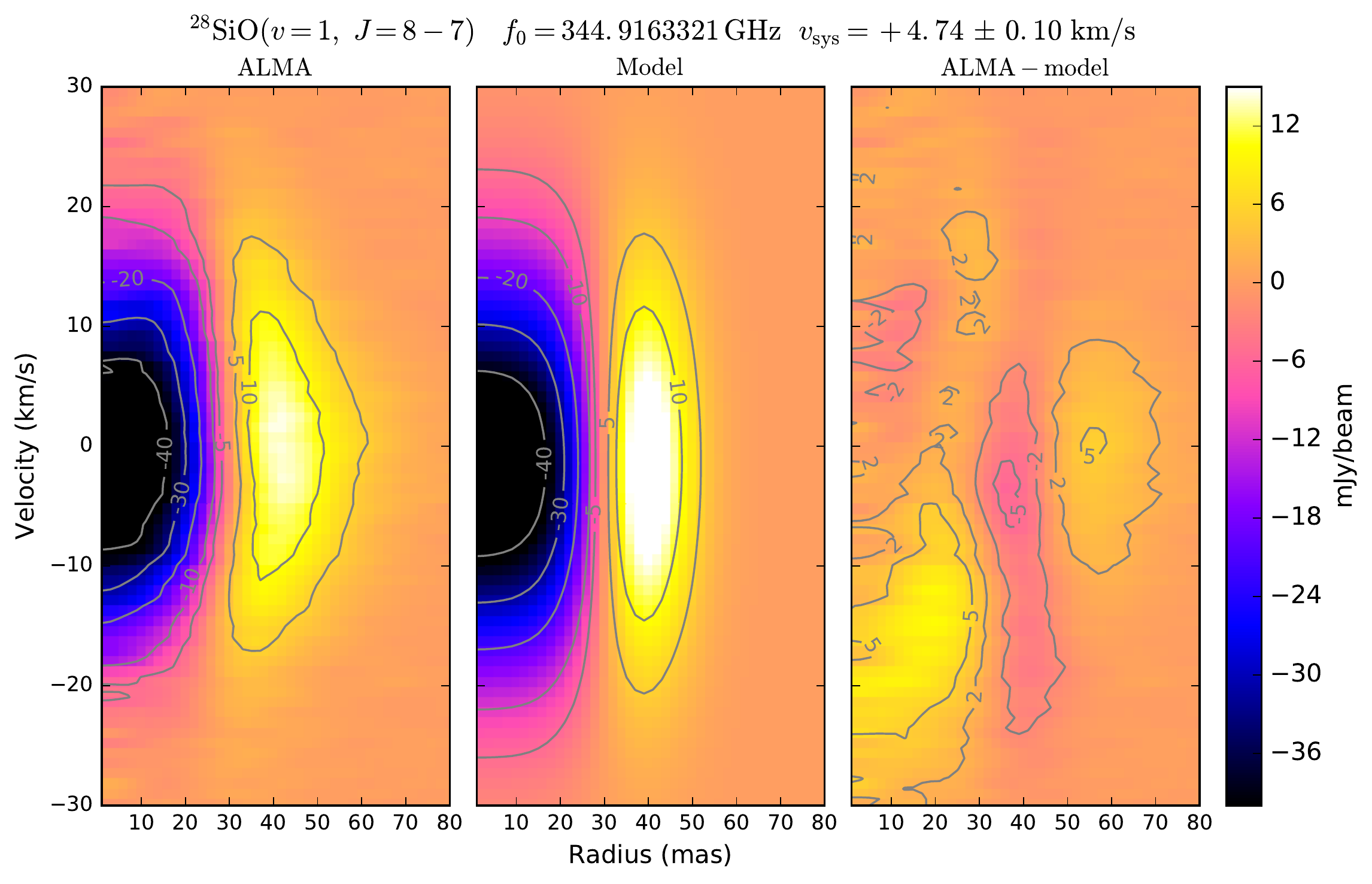}
        \caption{Same plots as in Fig.~\ref{profileresiduals12COv0}, but for the $^{28}$SiO($\varv$=1,\,$J$=8-7) line.
        \label{profileresiduals28SiOv1}}
\end{figure*}

\begin{figure*}[]
        \centering
        \includegraphics[width=6cm]{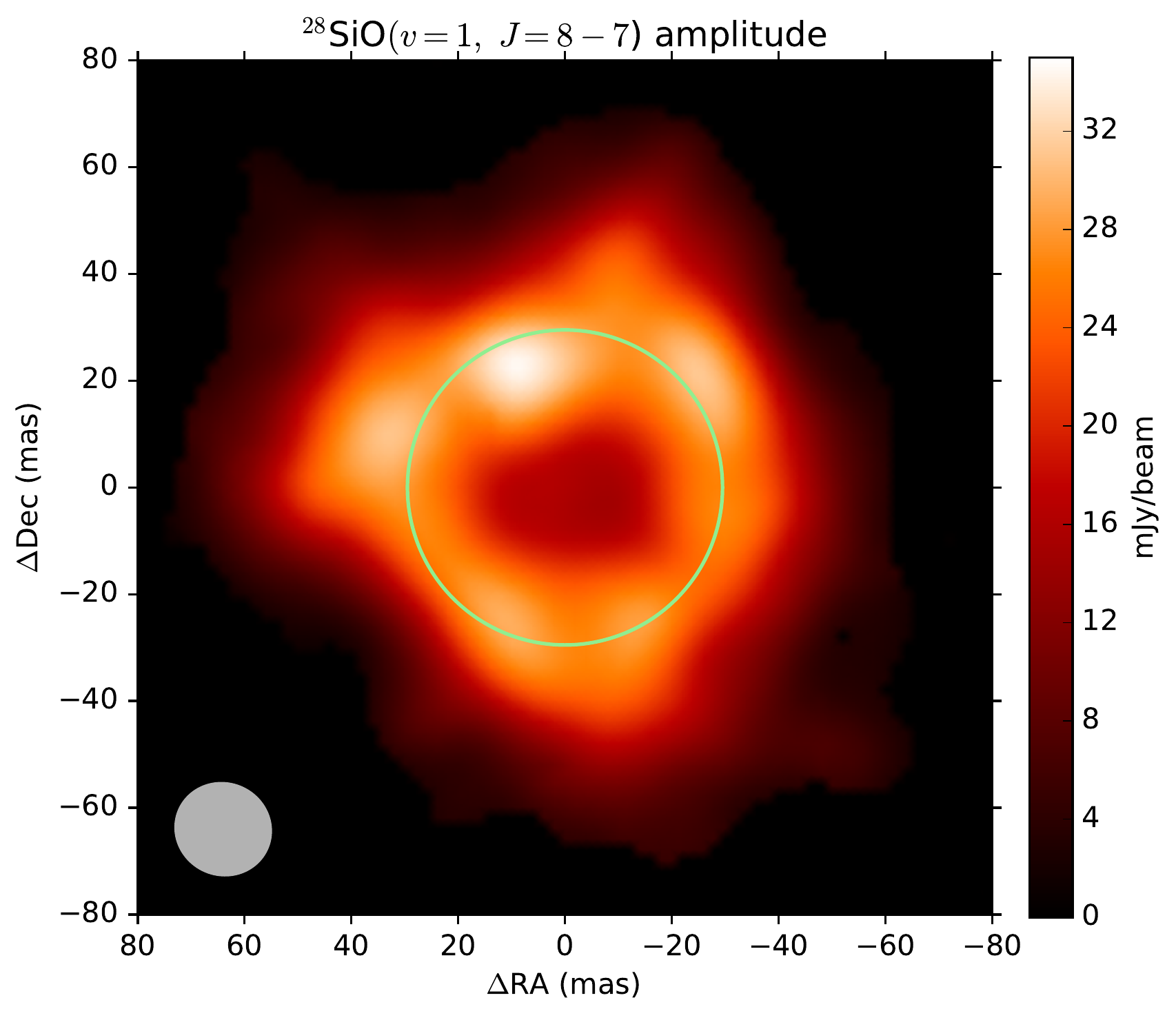}
        \includegraphics[width=6cm]{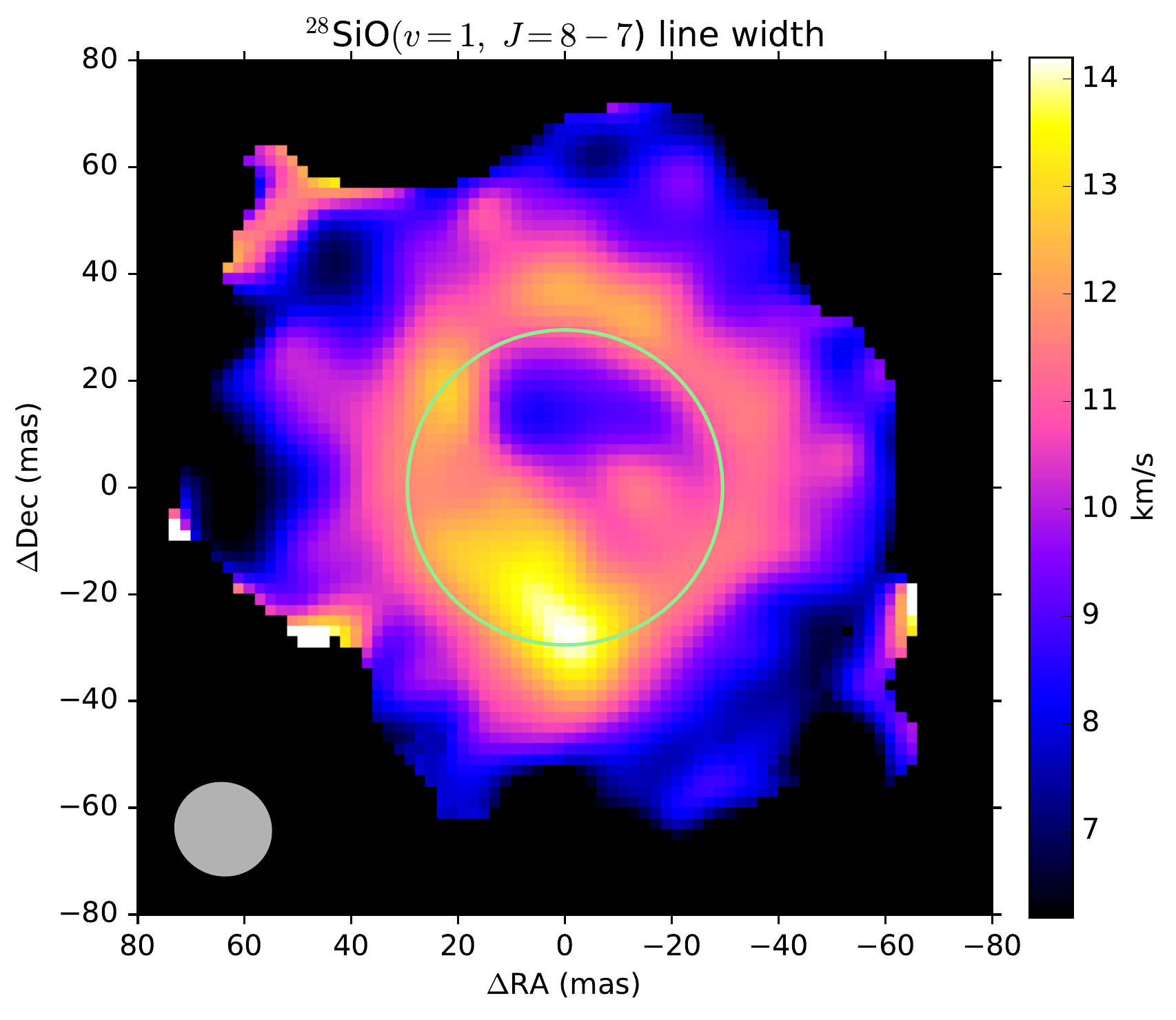}
        \includegraphics[width=6cm]{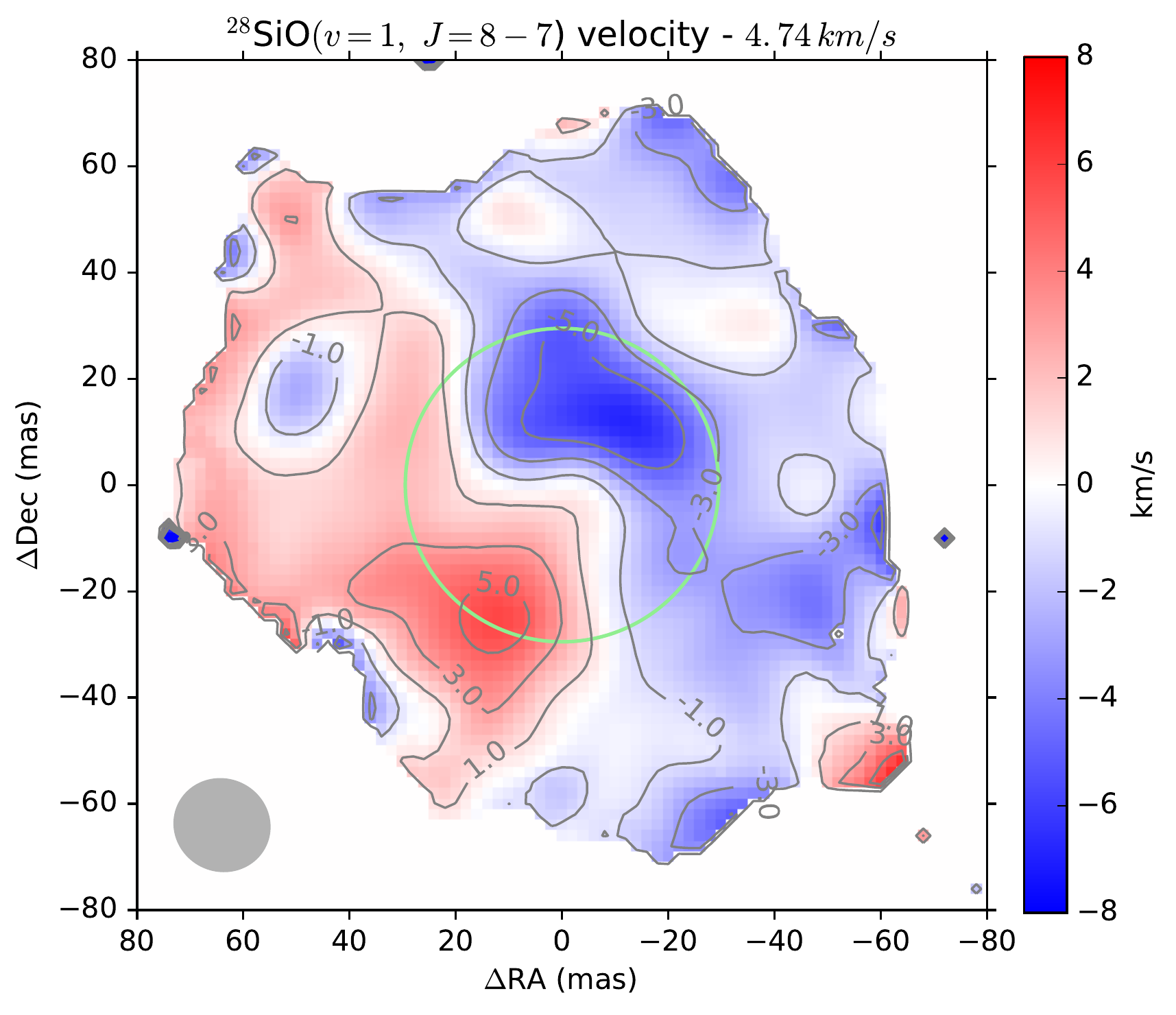}
        \includegraphics[height=5.8cm]{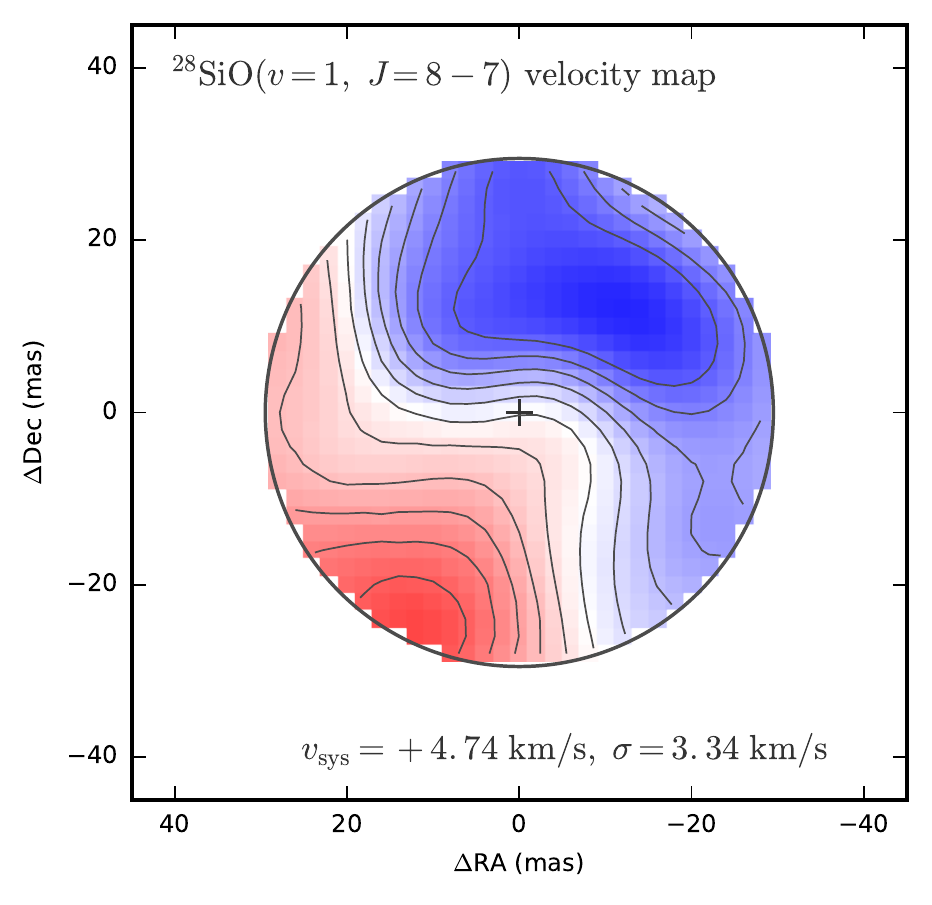}
        \includegraphics[height=5.8cm]{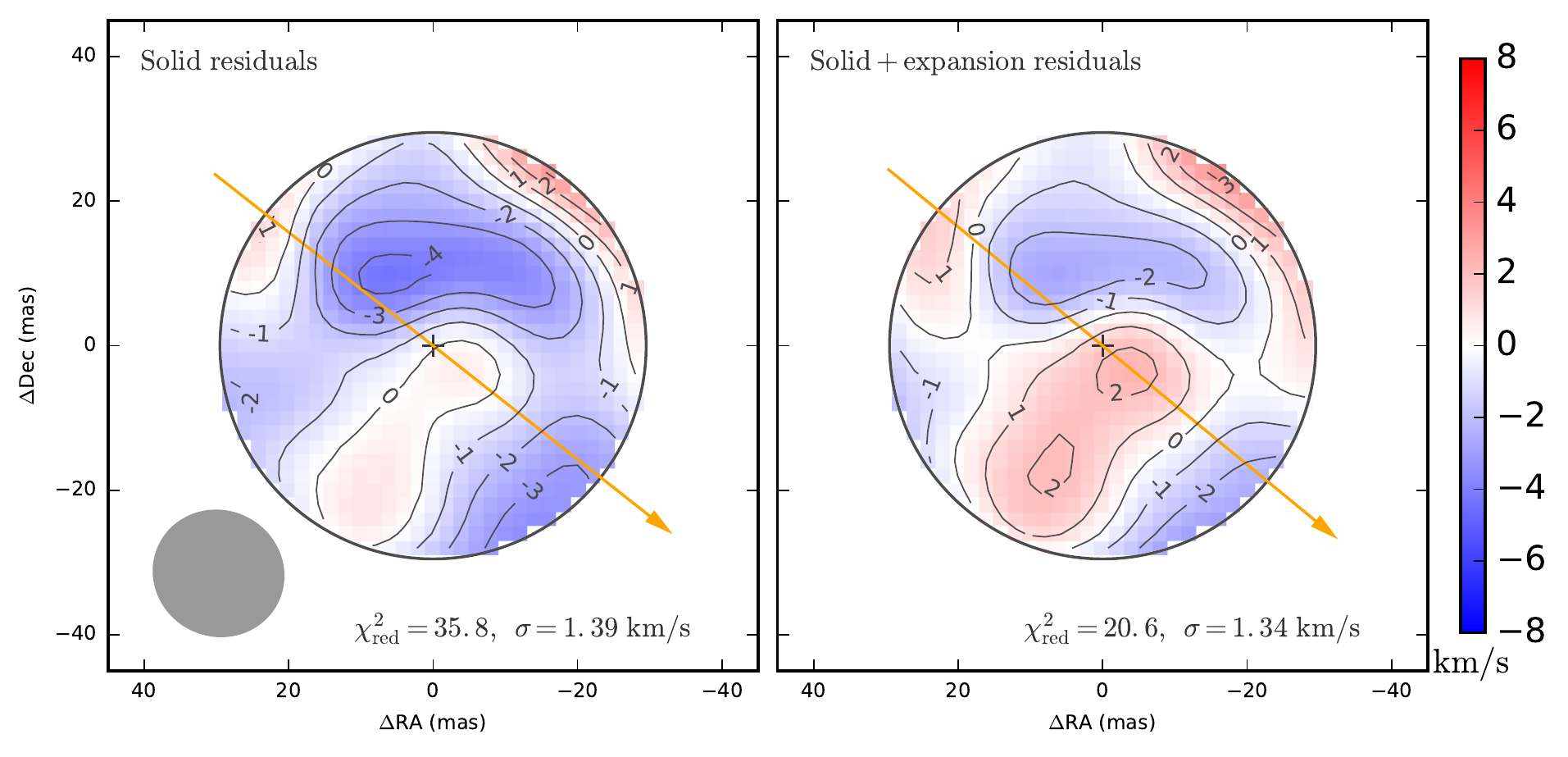}
        \caption{Same maps as in Fig.~\ref{12COv0linemaps}, but for the $^{28}$SiO($\varv$=1,\,$J$=8-7) line.
         \label{28SiOv1linemaps}}
\end{figure*}

\section{$^{29}$SiO($\varv$=0,\,$J$=8-7) \label{29SiOv0line}}

\begin{figure*}[ht]
        \centering
        \includegraphics[height=6cm]{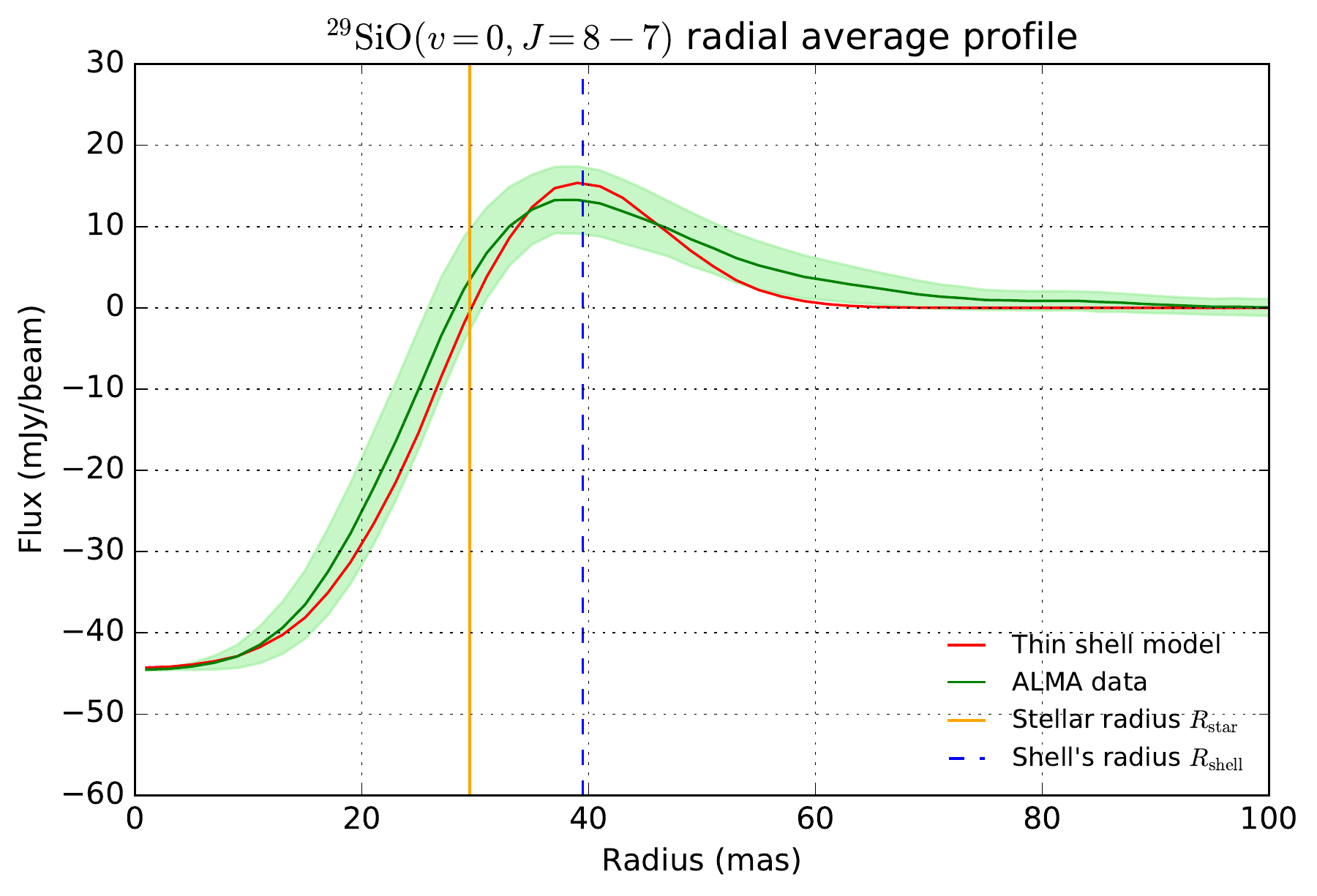}
        \includegraphics[height=6cm]{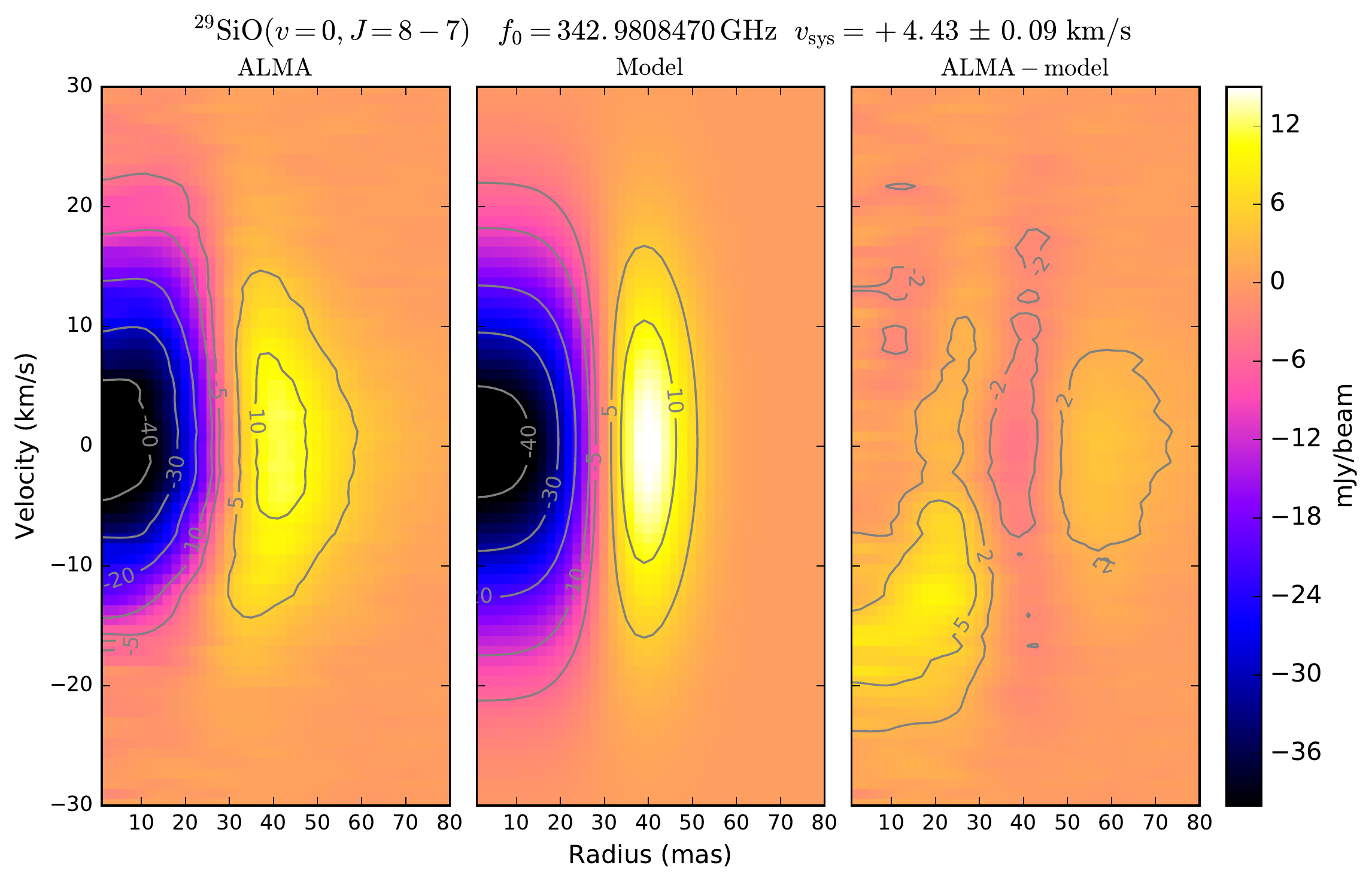}
        \caption{Same plots as in Fig.~\ref{profileresiduals12COv0}, but for the $^{29}$SiO($\varv$=0,\,$J$=8-7) line.
        \label{profileresiduals29SiOv0}}
\end{figure*}

\begin{figure*}[]
        \centering
        \includegraphics[width=6cm]{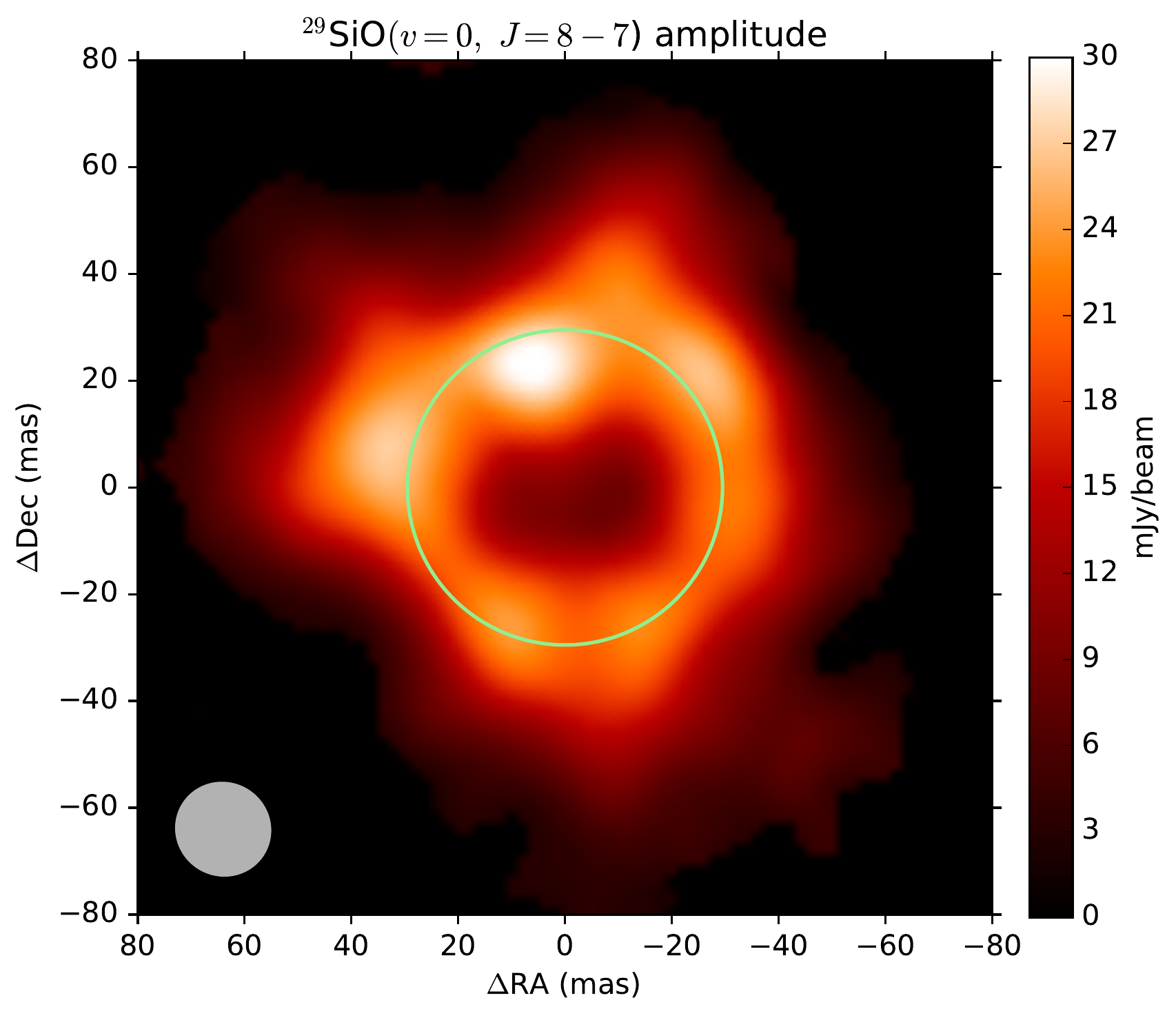}
        \includegraphics[width=6cm]{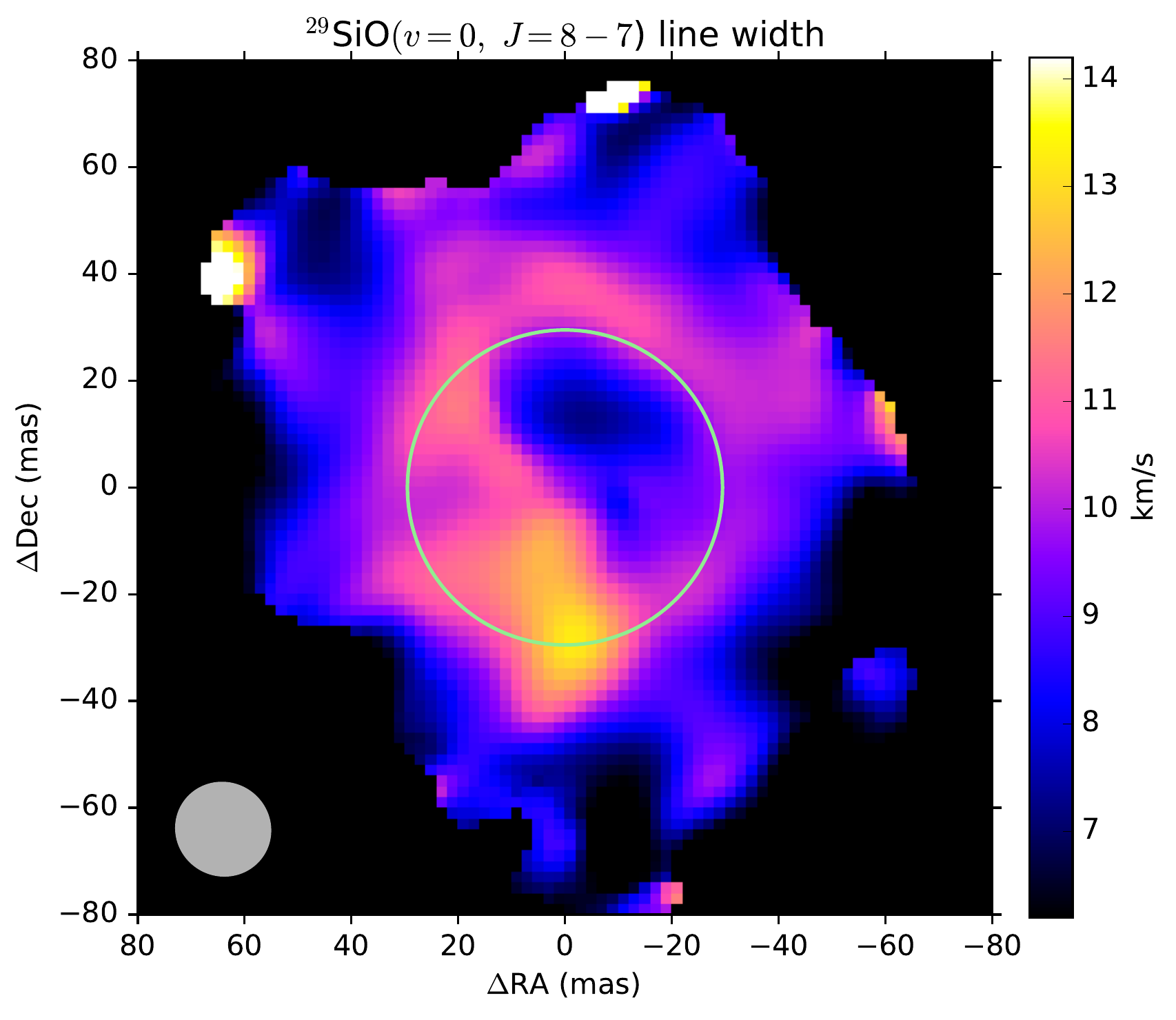}
        \includegraphics[width=6cm]{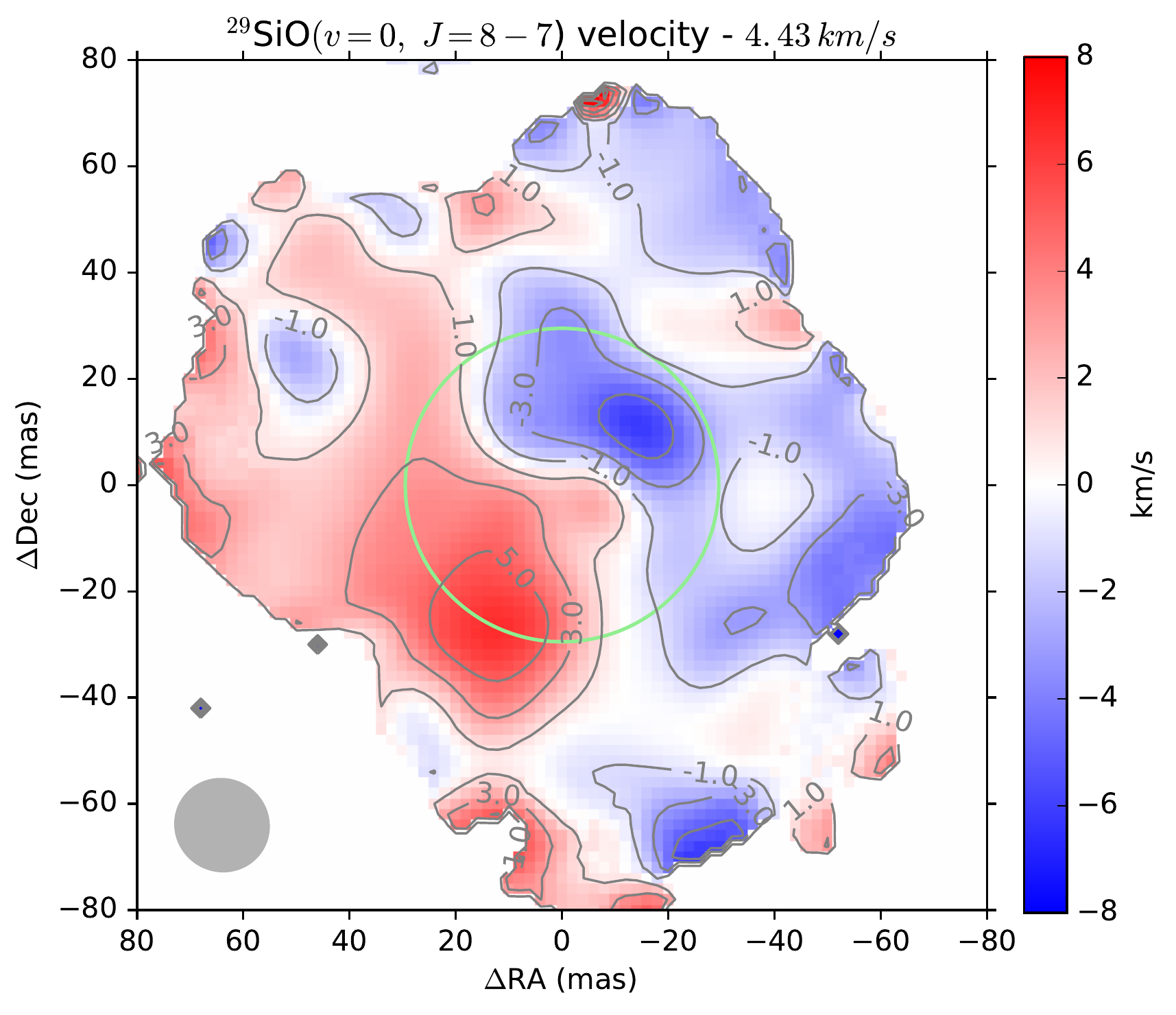}
        \includegraphics[height=5.8cm]{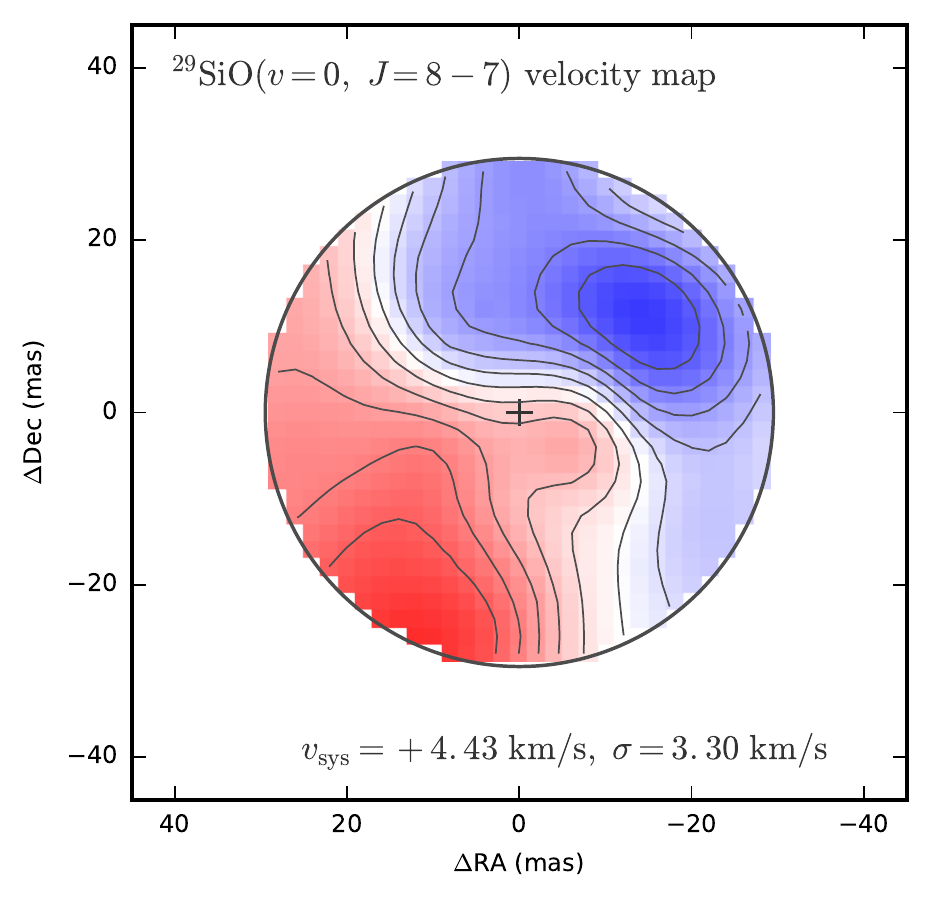}
        \includegraphics[height=5.8cm]{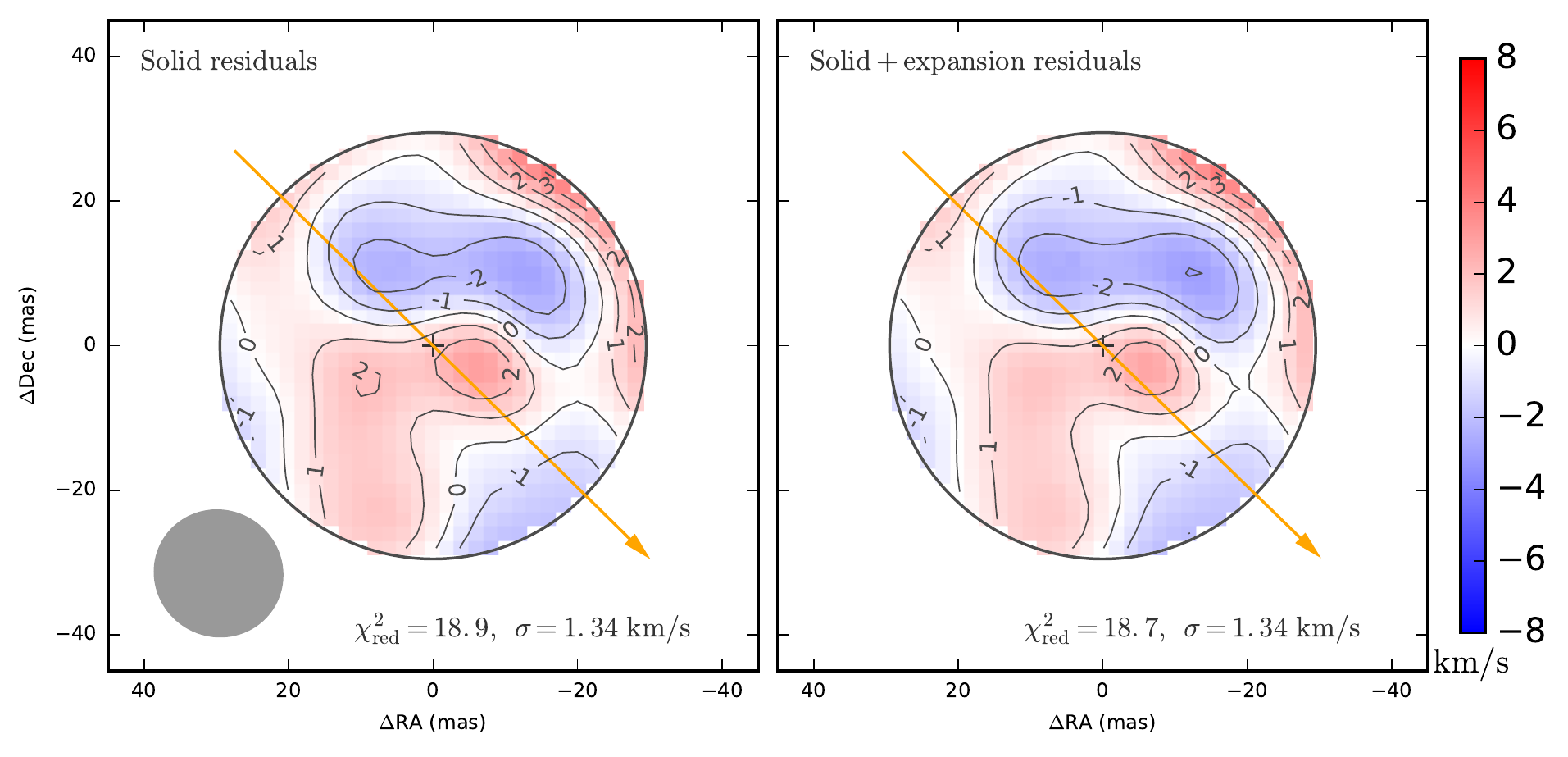}
        \caption{Same maps as in Fig.~\ref{12COv0linemaps}, but for the $^{29}$SiO($\varv$=0,\,$J$=8-7) line.
        \label{29SiOv0linemaps}}
\end{figure*}

\end{appendix}

\end{document}